\documentclass[12pt]{article}

\usepackage{amsfonts}
\usepackage{amsmath}
\usepackage{amssymb}
\usepackage{multicol}
\usepackage{epsfig}
\usepackage{float}
\usepackage{rotating}
\usepackage{booktabs}
\usepackage{threeparttable}
\usepackage{makecell}
\usepackage{multirow}
\usepackage{caption}
\usepackage{mathrsfs}
\usepackage[FIGTOPCAP]{subfigure}
\usepackage[normalem]{ulem}

\usepackage{setspace}
\usepackage[mathscr]{eucal}
\usepackage{url}
\usepackage{tabularx}

\usepackage{graphicx}
\usepackage{xcolor}
\newcommand{\White}[1]{\textcolor{white}{#1}}

\usepackage{titlesec}
\titleformat*{\section}{\large\bfseries}
\titleformat*{\subsection}{\normalsize\bfseries}

\titlespacing*{\section}
{0pt}{1.5ex plus 1ex minus .2ex}{1.3ex plus .2ex}
\titlespacing*{\subsection}
{0pt}{1.5ex plus 1ex minus .2ex}{1.3ex plus .2ex}
\titlespacing*{\paragraph}
{0pt}{1.5ex plus 1ex minus .2ex}{1.3ex plus .2ex}

\usepackage{pdflscape}

\newcommand{\E}{\operatorname{E}}

\newcommand{\Cov}{\operatorname{Cov}}

\renewcommand{\epsilon}{\varepsilon}

\begin{document}
\title{Marshall meets Bartik: \\
Revisiting the mysteries of the trade\thanks{Earlier versions of this paper have been presented at the 17th North American Meeting of the Urban Economics Association, 37th Annual Meeting of the Applied Regional Science Conference, 2024 Spring Meeting of the Japanese Economic Association, and 13th European Meeting of the Urban Economics Association. We thank the discussants and participants at these conferences and other workshops for valuable comments and suggestions.}}
\author{Yasusada Murata\thanks{College of Economics, Nihon University.
E-mail: \texttt{murata.yasusada@nihon-u.ac.jp}}
\quad \quad Ryo Nakajima\thanks{Department of Economics, Keio University.
E-mail: \texttt{nakajima@econ.keio.ac.jp}}}
\date{April 24, 2026}
\maketitle
\begin{abstract}
We identify a causal effect of top inventor inflows on the patent productivity of local inventors by combining the idea-generating process described by Marshall (1890) with the Bartik (1991) instruments involving the state taxes and commuting zone characteristics of the United States. We find that local productivity gains go beyond organizational boundaries and co-inventor relationships, which implies the partially nonexcludable good nature of knowledge in a spatial economy and pertains to the mysteries of the trade in the air. Our counterfactual experiment suggests that the spatial distribution of inventive activity is substantially distorted by the presence of state tax differences.
\end{abstract}

\newpage
\section{Introduction}
\label{sec:sec1}

Knowledge creation has been central to various fields of economics such as trade, growth, and geography. However, little is known about the idea-generating process between individuals, despite Marshall's (1890) simple explanation as follows:
\begin{quote}
\vskip -.13cm
\emph{if one man starts a new idea, it is taken up by others and combined with suggestions of their own; and thus it becomes the source of further new ideas.}
\end{quote}
\vskip -.13cm
While intuitive, verifying this statement has been challenging. The main difficulty lies in the possible endogeneity---those who generate new ideas tend to cluster together.

We address this problem by identifying a causal effect of a top inventor inflow on the patent productivity of local inventors at the commuting-zone level in the United States. In doing so, we use inventor-level data from the PatentsView database, which is an open data platform supported by the United States Patent and Trademark Office (USPTO). Since  top inventor inflows are likely endogenous, we predict those flows by constructing Bartik (1991) instruments: the predicted probability that a top inventor migrates from origin to destination constitutes a share, and the number of top inventors in the origin corresponds to a shift.

To understand the driving forces behind knowledge creation among individuals, we first classify local inventors into internal and external inventors. Local inventors are considered internal if they share the same organization as the migrating top inventors and/or if they are co-inventors of the migrating top inventors. All other local inventors are external because they are not directly linked to the migrating top inventors.

We then examine two types of effects---the productivity gains of all local inventors and those of external inventors. Our baseline results suggest that the former and latter gains from an additional top inventor inflow are 6\% and 4\%, respectively. The former are interpreted as local aggregate gains from both knowledge sharing among internal inventors and knowledge spillovers to external inventors. The latter focus on the gains that go beyond organizational boundaries and co-inventor relationships and pertain to the most frequently quoted passage from Marshall (1890): ``\emph{The mysteries of the trade become no mysteries; but are as it were in the air.}'' We thus disentangle productivity gains due to external knowledge spillovers (``knowledge in the air'') from those due to internal knowledge sharing \mbox{(``knowledge in the lab'').}

Our identification strategy consists of three main steps. We first estimate the impact of spatial and temporal variation in top earners' income tax rates on the migration probability of top inventors for any pair of origin and destination commuting zones while controlling for origin-destination characteristics. We then aggregate, for each destination commuting zone, the predicted bilateral probabilities across origin commuting zones to construct a Bartik instrument for top inventor inflows. We finally employ an instrumental variable (IV) approach, where we use the Bartik instrument in the first-stage regression and estimate a structural equation, with the outcome being local patent productivity. The identifying assumption is that local patent productivity in a destination commuting zone does not directly depend on top earners' income tax rates in other commuting zones located in \emph{different states}.\footnote{For example, this assumption implies that local patent productivity in destination commuting zone 37500 (Santa Clara--Monterey--Santa Cruz, CA) does not directly depend on top 5\% or 1\% earners' income tax rates in origin commuting zones 19600 (Bergen--Essex--Middlesex, NJ), 24300 (Cook--DuPage--Lake, IL), and so forth. In line with this assumption, we show in Section~\ref{sec:4.5} that the main source of identifying variation comes from interstate top inventor migrations. We elaborate on this assumption in Section~\ref{sec:4.3} and Appendix~\ref{app:exo1}.}

Our novelty lies in the construction of the Bartik instrument: The predicted migration probability is derived from a location choice model of top inventors who face spatial and temporal differences in individual income tax rates. Thus, our framework can be used to examine to what extent those tax differences distort the spatial distribution of inventive activity. To illustrate this, we run a counterfactual experiment by setting individual income taxes to their average while allowing for endogenous wage changes in the model and find that the existence of tax differences affects local patent productivity up to $-64.8$\% to $72.3$\%, with considerable spatial heterogeneity. We further decompose those gains and losses into two types---direct gains from tax changes and indirect gains via top inventor migration induced by tax changes. We find that the former share is $0.275$, while the latter share is $0.725$.

The contribution of our paper is threefold. First, we shed new light on the idea-generating process described in Marshall (1890) using Bartik (1991) instruments. Our framework differs from (quasi-)natural experimental approaches to knowledge production in historical contexts (e.g., Borjas and Doran, 2012; Moser et al., 2014; Terry et al., 2026) or exploitation of the sudden death of inventors (e.g., Azoulay et al., 2010; Azoulay et al., 2019). We leverage the variation in tax rates across space and time to show that the tax-induced migration of top inventors leads to local productivity gains in their destination, thereby contributing to the agglomeration and innovation literature \mbox{(e.g., Carlino and Kerr, 2015; Kerr and Robert-Nicould, 2020).}

Second, we disentangle the productivity gains due to external knowledge spillovers from those due to internal knowledge sharing through organizations or co-inventor relationships, which allows us to revisit the mysteries of the trade in the air. The theoretical foundation for separating the nonexcludable part from the excludable part of the gains dates back at least to Griliches (1979) and Romer (1990), whereas the empirical literature typically emphasizes the productivity gains of migrants themselves or those from internal knowledge sharing (e.g., Moretti, 2021;  Prato, 2025). Thus, the productivity gains attributed to external knowledge spillovers among individuals have remained unexplored in a spatial framework using modern causal inference methods. Since this partially nonexcludable good nature of knowledge leads to market failures and constitutes a rationale for spatial agglomeration of inventive activity, our analysis contributes to the innovation policy literature (e.g., Chatterji et al.~2014; Aghion and Jaravel, 2015).

Finally, we derive the Bartik instruments from a location choice model involving policy variables \`{a} la Moretti and Wilson (2017). By construction, our model-based Bartik instruments can be used in any setting where origin-destination flows are affected by changes in location-specific policies. This paper applies these instruments to tax-induced domestic migration and conducts a counterfactual experiment to illustrate a way of bridging the gap between the tax and innovation literature (e.g., Stantcheva, 2021; Akcigit et al., 2022; Akcigit and Stantcheva, 2022; Jones, 2022) and the tax and migration literature (e.g., Kleven et al., 2020). This application provides new insights into these two strands of literature since it allows us to assess the relative importance of direct productivity gains from tax changes and indirect productivity gains through the tax-induced migration of top inventors.

The remainder of the paper is organized as follows. In Section 2, we explain the data and show descriptive statistics. In Section 3, we analyze how tax differences affect the migration of top inventors. Section 4 constructs the Bartik instruments and presents our main results on local patent productivity gains by employing the instrumental variable approach. In Section~5, we check the robustness of the main results. Section 6 discusses the underlying mechanisms through which local productivity gains materialize. We conduct the counterfactual experiment in Section 7 and conclude the paper in Section 8.

\vspace{-.1cm}
\section{Data and descriptive statistics}
\label{sec:sec2}
\vspace{-.1cm}

Our main dataset is the PatentsView database, which is an open data platform supported by the USPTO and provides various administrative data on issued patents and patent applications. The data are based on the disambiguation process and contain, for each issued patent, patent inventors, assignees, residential addresses of patent inventors, and patent citations. Our sample period is from 1977 to 2009, during which there were $3,015,305$ patent applications by 1,282,708 unique inventors (see \ref{sec:secA} for a more detailed description of the data sources and construction, as well as the disambiguation of inventors and assignees).\footnote{The sample period and data construction are dictated by data availability and consistent with those in Moretti and Wilson (2017).}

Since our objective is to estimate the impact of top inventor migration on the productivity of local inventors in the destination, we need to define the productivity of an inventor and determine (i) who qualifies as a top inventor, (ii) under what condition we detect the migration of a top inventor, and (iii) who in the destination potentially gains from top inventor inflows.

To this end, we first define the productivity of an inventor as the number of patents applied for by that inventor.\footnote{If there are multiple inventors for a patent, we allocate an equal fraction of that patent to each of its inventors; i.e., if there are three inventors for a patent, one-third of that patent is allocated to each inventor.}
We then identify, for each year, the top 5\% of inventors based on productivity over the last ten years and refer to them as \emph{top inventors} for short. It follows that  the status of a top inventor varies from year to year. During our sample period, there are $263,259$ top inventor $\times$ year observations,
and the number of unique top inventors is $60,294$. Thus, on average, the total duration of being a top inventor is $4.366$ years.

We detect the migration of a top inventor if the commuting zone of residence of that top inventor in the patent application data differs between two consecutive years.\footnote{We assume that the migration occurs at the end of the first year and use the definition of commuting zones as of 1990. When an inventor applied for more than one patent in a year, the most frequently observed commuting zone is regarded as that inventor's place of residence in that year. In case of a tie, we use the commuting zone observed for the first time in that year. We exclude commuting zones in Alaska and Hawaii. This detection requires the observations of the same inventor for two consecutive years, which may lead to the underestimation of top inventor migrations and measurement error. Thus, the same caveats as those in Moretti and Wilson (2017, pp.1864--1865) apply to our data construction.}
Since the status of a top inventor varies from year to year, we consider the migration of inventors who qualify as top inventors in the first year.\footnote{As a robustness check, we consider the migration of inventors who qualify as top inventors in both years in Figure~\ref{fig:fig3}. The details are provided in Murata and Nakajima (2025, Appendix C.1).}
In our sample, the total number of top inventor migrations is $9,178$, and the number of unique top inventors who migrated at least once is $5,725$. Thus, on average, each top inventor moved $1.603$ times, \mbox{conditional on moving at least once.}\footnote{\label{fn6}These numbers differ from those in Moretti and Wilson (2017), who use the pre-disambiguation data in the baseline case (Table 2A) and Li et al.'s (2014) disambiguation data as a robustness check (Table Appendix~14), because we use the PatentView database based on a more recent disambiguation algorithm.}

\begin{figure}[hbtp]
\caption{Geographic distribution of top inventor inflows.}
\vspace{-.2cm}
\hskip 2.3cm
\includegraphics[trim = 0 0 0 0, height=6cm, width=12cm, clip]{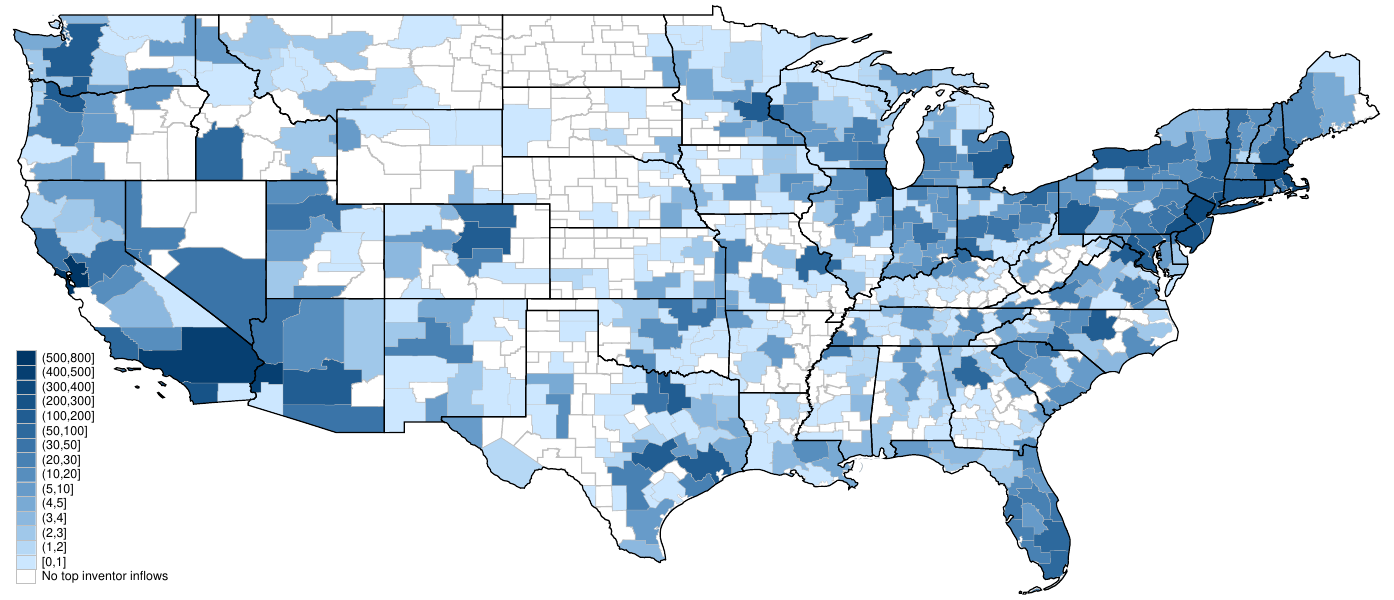} \\
\footnotesize{\noindent
\emph{Notes:} Inflows are defined as the number of top inventors who migrated into each commuting zone from 1977 to 2009.}
\label{fig:fig1}
\end{figure}

Since we analyze the impact of top inventor migration on the productivity of local inventors in the destination, we aggregate migration flows at the destination level. Figure~\ref{fig:fig1} depicts the geographic distribution of all $9,178$ top inventor inflows by commuting zone, and Table~\ref{tab:tab2} summarizes top 10 commuting zones by top inventor inflows.\footnote{We aggregate migration flows at the commuting zone level because it captures stronger commuting ties and thus more inventor interactions within labor market areas and because knowledge spillovers tend to be localized at short distances (see, e.g., Murata et al.~2014). We check the robustness of the result regarding geographic space in Section~\ref{sec:5.2}.}

\vskip -.05cm
\vspace{-.9cm}
\begin{table}[hbtp]
\caption{Top 10 commuting zones by top inventor inflows.}
\label{tab:tab2}
\vspace{-.1cm}
{\footnotesize
\hskip 2.2cm \begin{tabular}{rrlcr}
\hline \hline
rank & cz number & counties & state & inflows \\
\hline
1	&	37500	&	Santa Clara--Monterey--Santa Cruz	&	CA	&	724	\\
2	&	37800	&	Alameda--Contra Costa--San Francisco	&	CA	&	557	\\
3	&	38300	&	Los Angeles--Orange--San Bernardino	&	CA	&	408	\\
4	&	19600	&	Bergen--Essex--Middlesex	&	NJ	&	372	\\
5	&	20500	&	Middlesex--Worcester--Essex	&	MA	&	335	\\
6	&	38000	&	San Diego	&	CA	&	266	\\
7	&	19400	&	Kings--Queens--New York	&	NY	&	240	\\
8	&	19700	&	Philadelphia--Montgomery--Delaware	&	PA	&	220	\\
9	&	24300	&	Cook-DuPage--Lake	&	IL	&	219	\\
10	&	20901	&	Hartford--Fairfield--New Haven	&	CT	&	195	\\
\hline
\hline
\end{tabular}}
\begin{tablenotes}[flushleft]
\footnotesize
\item \hskip -.1cm \emph{Notes:} Inflows are defined as the number of top inventors who migrated into each commuting zone from 1977 to 2009.
\vspace{-.2cm}
\end{tablenotes}
\end{table}

\clearpage
\newpage

\begin{figure}[hbtp]
\caption{Geographic distribution of local patent productivity (in logs).}
\label{fig:fig2}
\hskip 2.4cm
\includegraphics[trim = 0 0 0 0, height=6cm, width=12cm, clip]{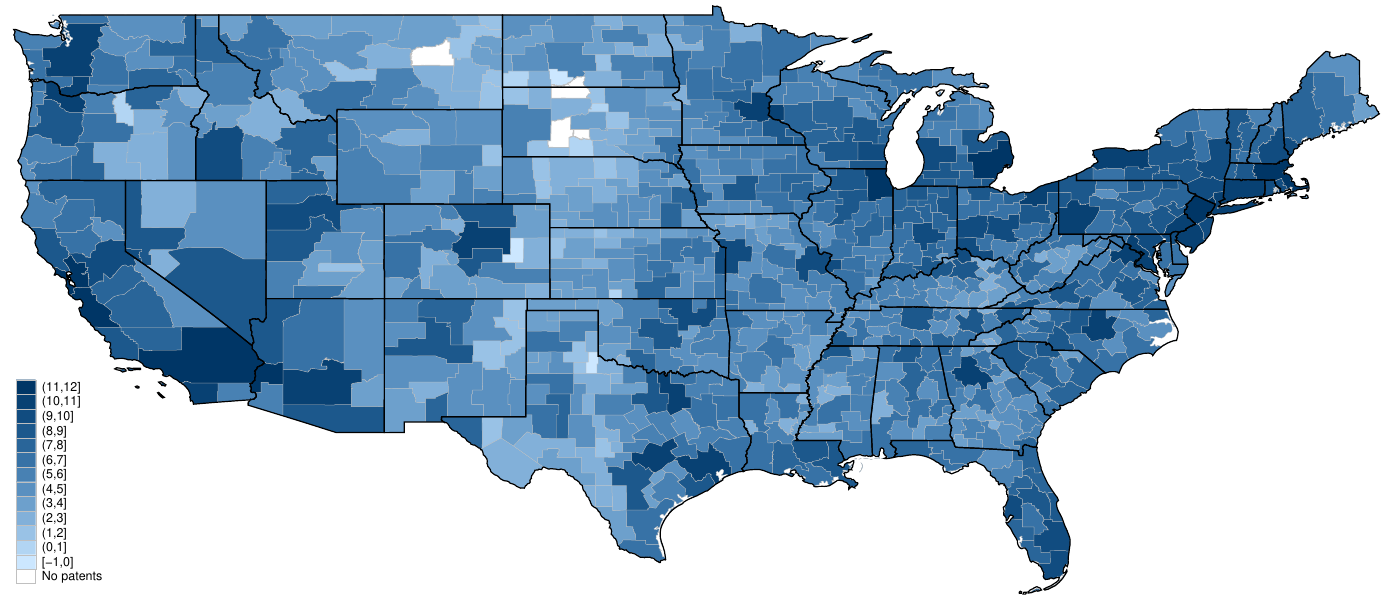} \\
\footnotesize{\noindent
\emph{Notes:} The productivity in each commuting zone is defined as the total number of patents applied for by local inventors between 1977 and 2009. If there are multiple inventors for a patent, then we allocate an equal fraction of that patent to each of its inventors.}
\vskip -.5cm
\end{figure}

When assessing the impact of top inventor migration into a commuting zone, we focus on the \emph{local inventors} who lived in that commuting zone at that time while excluding the top inventors who had already moved in that commuting zone. In our sample, the number of those local inventors is $1,274,192$, and they applied for $2,027,777$ patents from 1977 to 2009. Figure~\ref{fig:fig2} illustrates the geographic distribution of local patent productivity (in logs), and Table \ref{tab:tab3} summarizes the top 10 commuting zones by local patent productivity.

\begin{table}[hbtp]
\caption{Top 10 commuting zones by local patent productivity.}
\label{tab:tab3}
{\footnotesize
\hskip 1.85cm \begin{tabular}{rrlcr}
\hline \hline
rank	&	cz number	&	counties	&	state	& productivity
\\
\hline
1	&	37500	&	Santa Clara-Monterey-Santa Cruz	&	CA	&	143,069.321	\\
2	&	38300	&	Los Angeles-Orange-San Bernardino	&	CA	&	116,303.629	\\
3	&	37800	&	Alameda-Contra Costa-San Francisco	&	CA	&	81,828.603	\\
4	&	20500	&	Middlesex-Worcester-Essex	&	MA	&	80,486.767	\\
5	&	19600	&	Bergen-Essex-Middlesex	&	NJ	&	77,093.518	\\
6	&	24300	&	Cook-DuPage-Lake	&	IL	&	75,095.386	\\
7	&	11600	&	Wayne-Oakland-Macomb	&	MI	&	63,437.324	\\
8	&	19400	&	Kings-Queens-New York	&	NY	&	54,570.527	\\
9	&	21501	&	Hennepin-Ramsey-Dakota	&	MN	&	50,587.528	\\
10	&	39400	&	King-Pierce-Snohomish	&	WA	&	49,760.915	\\
\hline
\hline
\end{tabular}}
\begin{tablenotes}[flushleft]
\footnotesize
\item  \hskip -.1cm \emph{Notes:}
The productivity in each commuting zone is defined as the total number of patents applied for by local inventors between 1977 and 2009.
If there are multiple inventors for a patent, then we allocate an equal fraction of that patent to each of its inventors.
\end{tablenotes}
\vskip -.25cm
\end{table}

As seen from Figures~\ref{fig:fig1} and \ref{fig:fig2}, their spatial patterns are quite similar. The correlation between the top inventor inflows and local patent productivity (the log of local patent productivity) is $0.97$ ($0.52$) and their rank correlation is $0.85$. However, since correlation does not necessarily imply causation, we take an instrumental variable approach to examine the causal effect of the top inventor inflows on the productivity of local inventors. To this end, we use the variation in the individual income average tax rates for top earners (ATRs) by state and year to construct predicted flows of top inventors by commuting zone and year.\footnote{We assume that top inventors are taxpayers at the ninety-fifth (ninety-ninth) percentile of the U.S. income distribution as a baseline (as a robustness check). We further check the robustness of our results using statutory marginal tax rates (MTRs). As discussed in Moretti and Wilson (2017), ATRs and MTRs are highly correlated, and indeed, we obtain similar results regardless of the choice of tax rates. See Figure~\ref{fig:fig3} for those robustness checks. The details are provided in Murata and Nakajima (2025, Appendix C.1).}
We also consider corporate income tax rates (CITRs), investment tax credits (ITCs), and R\&D tax credits (RTCs) that can affect top inventor migration.\footnote{Data on state taxes and tax credits for the years 1977 to 2009 are provided by Moretti and Wilson (2017).} 
Table \ref{tab:taby} presents summary statistics for the top inventor inflows and the local patent productivity at the commuting zone $\times$ year level. We show in \ref{sec:secA} the summary statistics for other commuting zone-level variables, as well as state taxes and tax credits, that we use in the subsequent  analysis.

\vspace{.1cm}
\begin{table}[hbtp]
\caption{Summary statistics (main variables).}
\label{tab:taby}
{\footnotesize
\hskip 1.1cm \begin{tabular}{lrrrrr}
\hline\hline
 & \multicolumn{1}{c}{total} & \multicolumn{1}{c}{mean} & \multicolumn{1}{c}{sd} & \multicolumn{1}{c}{min} & \multicolumn{1}{c}{max} \\
\hline
Local patent productivity (overall) & 2,027,776.570 &  85.821   & 386.089  & 0.000  & 10,205.625  \\
Local patent productivity (internal) & 1,061,199.866 & 44.913  & 252.589   & 0.000  & 8,232.745  \\
Local patent productivity (external) & 966,576.704 & 40.908   & 152.466  & 0.000  & 3,108.842  \\
Top inventor inflows (overall) & 9,178.000 & 0.388   &  2.139  & 0.000  & 81.000  \\
Top inventor inflows (intrastate) & 2,271.000  & 0.096   &  0.929 &  0.000 & 37.000   \\
Top inventor inflows (interstate) & 6,907.000 & 0.292  & 1.453  & 0.000  & 48.000  \\
\hline
Number of observations& \multicolumn{5}{r}{23,628}  \\
Number of commuting zones  & \multicolumn{5}{r}{716} \\
Number of years &  \multicolumn{5}{r}{33}  \\
\hline\hline
\end{tabular}}
\begin{tablenotes}[flushleft]
\footnotesize \item \hskip -.1cm \emph{Notes:} Summary statistics are based on the data described in Section 2 for the years 1977 to 2009. The local patent productivity can be decomposed into two: One is by the internal inventors who share the same assignee as the migrating top inventors and/or who are co-inventors of the migrating top inventors; and the other is by the external inventors. The top inventor inflows can be decomposed into  intrastate and interstate migration. Of the 722 commuting zones, four have no patents and two have only one patent during the sample period. We thus use 716 commuting zones in our regression analysis with fixed effects.
\end{tablenotes}
\vspace{-.1cm}
\end{table}

We further classify local inventors into \textit{internal} and \textit{external} inventors. Local inventors are internal if they share the same assignee as the migrating top inventors and/or if they are co-inventors of the migrating top inventors. All the other local inventors are external because they are not directly linked to the migrating top inventors. In our sample, $42.20$\% of local inventors are internal, whereas the remaining $57.80$\% are external.\footnote{The number of internal inventors is $537,717$, and that of external inventors is $736,475$. Internal inventors are classified into three groups: $89,637$ inventors share the same assignee as the migrating top inventors and are co-inventors of the migrating top inventors, $419,378$ inventors share the same assignee as the migrating top inventors but are not co-inventors of the migrating top inventors, and $28,702$ inventors do not share the same assignee as the migrating top inventors but are co-inventors of the migrating top inventors.}

The knowledge of the migrating top inventors can be shared with internal inventors within the same organization and/or through co-inventors relationships (``knowledge in the lab'') or can spill over to external inventors within the same commuting zone (``knowledge in the air''). We call the former \textit{internal knowledge sharing} and the latter \textit{external knowledge spillovers}.

\abovedisplayskip=5pt
\belowdisplayskip=5pt

\section{Tax differences and the migration of top inventors}
\label{sec:sec3}

In this section, we estimate the impact of tax differences across states on the migration of top inventors from origin commuting zone $o$ to destination commuting zone $d$, which corroborates the results in Moretti and Wilson (2017). In the next section, we use the predicted migration of top inventors to develop a new method of constructing a Bartik instrument.

Let $\sigma (o)$ and $\sigma (d)$ denote the states to which origin and destination commuting zones \linebreak
belong, respectively. In the beginning of period $t$, top inventors in $o$, whose number~is~denoted \linebreak
by $I_{ot}$, observe individual income tax rates in origin and all possible destination commuting zones, $\tau_{\sigma(o)t}$ and $\{\tau_{\sigma(d)t}\}_{d \neq o}$. By the end of period $t$, they decide whether to migrate to $d$ or to stay in $o$. The number and share of top inventors who migrate from $o$ to $d$ in period~$t$ is defined as $M_{odt}$ and $P_{odt}=M_{odt}/I_{ot}$, respectively. Similarly, the number and share of top inventors who stay in $o$ in period $t$ is defined as $M_{oot}$ and $P_{oot}=M_{oot}/I_{ot}$.

\subsection{Inventors}
In each period, top inventors choose the location that gives them the highest utility. The utility of top inventor~$i$, who lived in commuting zone $o$ in the previous period and moves to commuting zone $d$ in the current period $t$, is given by
$U_{iodt} = \alpha \ln (1-\tau_{\sigma(d)t})  +  \alpha \ln w_{dt} + Z_d -C_{od} + \varepsilon_{iodt},$
where $\tau_{\sigma(d)t}$ and $w_{dt}$ are the individual income tax rate and wage in $d$, respectively; $\alpha$ is the coefficient on the log of after-tax income; $Z_d$ captures consumption amenities and the cost of living in $d$; $C_{od}$ is the cost of migration measured in utility; and $ \varepsilon_{iodt}$ represents time-varying idiosyncratic preferences for locations. The utility of top inventor $i$ who stays in $o$ is given by
$U_{ioot} = \alpha \ln (1-\tau_{\sigma(o)t}) + \alpha \ln w_{ot}  + Z_o -C_{oo} +  \varepsilon_{ioot},$
where we assume that $C_{oo}=0$. Taking the difference between $U_{iodt}$ and $U_{ioot}$ yields the utility change for top inventor $i$, conditional on moving from $o$ to $d$. Assume that $ \varepsilon_{iodt}$ is independent and identically Gumbel distributed. Let $P_{odt}/P_{oot}$ denote the share of top inventors who move from $o$ to $d$ relative to the share of top inventors who stay in~$o$. The log odds ratio for top inventors is then given by
\begin{equation}
\ln (P_{odt}/P_{oot}) =  \alpha [ \ln (1-\tau_{\sigma(d)t})  -  \ln (1-\tau_{\sigma(o)t})  ]
+\alpha [ \ln w_{dt} - \ln w_{ot} ]
+ [ Z_d -Z_o ] - C_{od}.
\label{eq:inventors}
\end{equation}

\subsection{Firms}
In each period, firms choose a location and hire a top inventor to maximize profit. The profit of firm~$j$, which was located in commuting zone $o$ in the previous period and moves to commuting zone $d$ in the current period $t$, is given by $\ln \pi_{jodt} =  \beta \ln (1-\tau'_{\sigma(d)t}) - \ln w_{dt}+ Z'_d -C'_{od} + \varepsilon'_{jodt},$ where $\tau'_{\sigma(d)t}$ stands for state policies such as the CITR, ITC, and RTC in $\sigma(d)$; $Z'_d$ captures production amenities in $d$; $C'_{od}$ is the cost of migration for a firm; and $ \varepsilon'_{jodt}$ represents time-varying idiosyncratic firm productivity shocks. As in the case with inventors, assume that $C'_{oo}=0$, and that $ \varepsilon'_{jodt}$ is independent and identically Gumbel distributed. Let  $P'_{odt}/P'_{oot}$ denote the share of firms that move from $o$ to $d$ relative to the share of firms that stay in~$o$. The log odds ratio for firms is then given by
\begin{equation}
\ln (P'_{odt}/P'_{oot}) =  \beta [ \ln (1-\tau'_{\sigma(d)t})  -  \ln (1-\tau'_{\sigma(o)t})  ]
- [ \ln w_{dt} - \ln w_{ot} ]
+ [ Z'_d -Z'_o ] - C'_{od}.
\label{eq:firms}
\end{equation}

\subsection{Equilibrium}
In equilibrium, the demand for top inventors must equal the supply of top inventors in each commuting zone in each year. To derive an equilibrium relationship between tax differences and the migration of top inventors, we first solve \eqref{eq:firms} for $\ln w_{dt} - \ln w_{ot}$. We then plug the resulting expression into \eqref{eq:inventors} and set $\ln (P'_{odt}/P'_{oot}) = \ln (P_{odt}/P_{oot})$ as in Moretti and Wilson (2017), which yields the equation we estimate as follows  (see Appendix~\ref{sec:secB1} for the derivation):
\begin{eqnarray}
\ln (P_{odt}/P_{oot})
&=&  \eta [ \ln (1-\tau_{\sigma(d)t})  -  \ln (1-\tau_{\sigma(o)t}) ] +  \eta' [ \ln (1-\tau'_{\sigma(d)t})  -  \ln (1-\tau'_{\sigma(o)t})  ]  \notag \\
&&  + \gamma_d  + \gamma_o +\gamma_{od}+ u_{odt},
\label{eq:eqm}
\end{eqnarray}
where $\eta =  \frac{\alpha}{1+\alpha} $ and $\eta'= \frac{\alpha \beta}{1+\alpha}$ are parameters governing inventor and firm mobility, respectively; $\gamma_d= \frac{1}{1+\alpha} [ Z_d +\alpha Z'_d ]$ and $\gamma_o= - \frac{1}{1+\alpha} [  Z_o  + \alpha  Z'_o ]$ are destination and origin fixed effects that account for consumption and production amenities; $\gamma_{od}= -\frac{1}{1+\alpha} [ C_{od} + \alpha C'_{od}  ]$ denotes fixed effects that are specific to each pair of commuting zones to capture the cost of migration for inventors and firms; and $u_{odt}$ is an error term. We consider different combinations of fixed effects in the next subsection.

\subsection{Estimation}
\vspace{-.2cm}
When estimating \eqref{eq:eqm}, we proxy $\tau_{\sigma(d)t}$ by the ATR for a hypothetical taxpayer at the ninety-fifth or ninety-ninth percentile of the U.S. income distribution because, as in Moretti and Wilson (2017), we do not observe top inventors' income.\footnote{We report the results at the ninety-fifth percentile as a baseline and include the results at the ninety-ninth percentile in Figure~\ref{fig:fig3} as a robustness check. In Figure~\ref{fig:fig3} we further check the robustness of our results using statutory marginal tax rates (MTRs) and average property tax rates (APTRs). The details are provided in Murata and Nakajima (2025, Appendix C.1).}
We regard $\tau'_{\sigma(d)t}$ as consisting of the CITR, ITC, and RTC. We use different combinations of fixed effects in \eqref{eq:eqm}, as well as year fixed effects or region pair $\times$ year fixed effects and report robust standard errors that allow for three-way clustering by commuting zone pair, origin-state $\times$ year, \mbox{and destination-state $\times$ year.}

\begin{table}[hbtp]
\caption{The impact of tax differences on the migration of top inventors.}
\label{tab:tab4}
\vspace{-.1cm}
{\footnotesize
\hskip 2.25cm
\begin{tabular}{lrrrr}
\hline \hline
 & \multicolumn{1}{c}{(1)} & \multicolumn{1}{c}{(2)} & \multicolumn{1}{c}{(3)} & \multicolumn{1}{c}{(4)} \\  \hline
$\Delta \ln (1 - {\rm ATR})$ & $7.357$	& $6.902$	& $6.406$	& $6.586$	\\
	&	(1.611)	&	(1.420)	&	(1.292)	&	(1.124)	\\
$\Delta \ln (1 - {\rm CITR})$ &	-0.435	&	-0.195	&	-0.300	&	-0.140	\\
	&	(1.058)	&	(0.999)	&	(0.812)	&	(0.717)	\\
$\Delta \ln (1 + {\rm ITC})$ &	0.172	&	-0.083	&	0.118	&	-0.034	\\
	&	(0.737)	&	(0.688)	&	(0.993)	&	(0.689)	\\
$\Delta \ln (1 + {\rm RTC})$ &	0.323	&	0.311	&	0.377	&	0.178	\\
	&	(0.443)	&	(0.395)	&	(0.321)	&	(0.281)	\\[1mm]
CZ pair FE	& Yes	&	Yes	&	No &	No	\\ 
Origin CZ FE and destination CZ FE & No	&	No	&	Yes &	Yes \\
Year FE	& Yes	&	No	&	Yes &	No	\\ 
Region pair $\times$ year FE     & No	&	Yes	&	No &	Yes \\[1mm]
Observations	&	4,866	&	4,866	&	7,226	&	7,225	\\
$\overline R^2$ (total)	&	0.893	&	0.904	&	0.907	&	0.917	\\
$\overline R^2$ (within)	&	0.400	&	0.458	&	0.411	&	0.013	\\
\hline
\hline
\end{tabular}}
\begin{tablenotes}[flushleft]
\footnotesize
\item \hskip -.1cm \emph{Notes:} The dependent variable in each column is the log odds ratio in equation \eqref{eq:eqm}. ATR, CITR, ITC, and RTC stand for the individual income average tax rate at the ninety-fifth percentile, corporate income tax rate, investment tax credit, and R\&D tax credit, respectively. $\Delta \ln (1 - {\rm ATR})$ is defined as $\ln(1-{\rm ATR}_{\sigma(d)t}) - \ln(1-{\rm ATR}_{\sigma(o)t})$, where ${\rm ATR}_{\sigma(d)t}$ and ${\rm ATR}_{\sigma(o)t}$ denote the ATRs in states $\sigma (d)$ and $\sigma (o)$ in year $t$. $\Delta \ln (1 - {\rm CITR})$, $\Delta \ln (1 + {\rm ITC})$, and $\Delta \ln (1 + {\rm RTC})$ are defined analogously. Cluster-robust standard errors are in parentheses.
\end{tablenotes}
\vspace{-.4cm}
\end{table}

Table~\ref{tab:tab4} shows that the elasticity of the migration of top inventors with respect to the difference in ATRs between origin and destination is positive and significant in all cases, thus corroborating Moretti and Wilson's (2017) interstate migration results using the commuting zone level data.\footnote{In addition to the difference in geographic units, there is another important difference that explains why our estimates are not exactly the same as those in Moretti and Wilson (2017). As detailed in footnote~\ref{fn6}, our dataset is based on a different disambiguation algorithm from theirs.}
In what follows, we use the result in Column 2 of Table~\ref{tab:tab4} since the specification is most closely related to their baseline case.

\begin{figure}[h!]
\caption{Binned scatter plot of the relationship between top inventor migrations and ATRs.}
\hskip 4cm \includegraphics[width=0.5\textwidth]{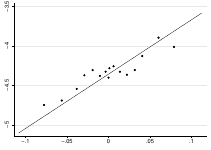} \\
\footnotesize{\noindent
\emph{Notes:} This figure illustrates a binned scatter plot, where the vertical axis is the log odds ratio of top inventor migrations, $\ln(P_{odt}/P_{oot})$, and the horizontal axis is the difference in ATRs between destination and origin commuting zones, $\Delta \ln (1 - {\rm ATR})=\ln(1-{\rm ATR}_{\sigma(d)t}) - \ln(1-{\rm ATR}_{\sigma(o)t})$. It is depicted using the Stata package \texttt{binsreg} (see Cattaneo et al., 2024), where we incorporate commuting zone pair fixed effects and region pair $\times$ year fixed effects, as well as CITRs, ITCs, and RTCs, as covariate adjustments.}
\label{fig:binscatter_line}
\vspace{-.75cm}
\end{figure}

Figure \ref{fig:binscatter_line} illustrates a binned scatter plot, where the vertical axis is the log odds ratio of top inventor migrations, $\ln(P_{odt}/P_{oot})$, and the horizontal axis is the difference in ATRs between destination and origin commuting zones, $\Delta \ln (1 - {\rm ATR}) = \ln(1-{\rm ATR}_{\sigma(d)t}) - \ln(1-{\rm ATR}_{\sigma(o)t})$.\footnote{${\rm ATR}_{\sigma(d)t}$ and ${\rm ATR}_{\sigma(o)t}$ denote the ATRs in states $\sigma (d)$ and $\sigma (o)$ in year $t$. We use similar notation for other state policy variables.} The figure reveals a pronounced tendency for top inventors to migrate from commuting zones with higher individual income tax rates to those with lower rates.

\vspace{.15cm}

\section{The migration of top inventors and local patent productivity}
\label{sec:sec4}
\vspace{-.15cm}

\noindent
To analyze the impact of top inventor inflows on local patent productivity, we first present a new method to construct a Bartik instrument based on the estimated flows of top inventors obtained in Section~\ref{sec:sec3}. We then show the main results using a static framework. To check the robustness of our main results, we further consider a dynamic setting in the next section. In both cases, we estimate two types of effects: (a) the productivity gains of all local inventors and (b) those of external inventors. The former effect can be interpreted as local aggregate productivity gains from both internal knowledge sharing and external knowledge spillovers. The latter effect can be viewed as the gains from external knowledge spillovers that go beyond organizational boundaries and co-inventor relationships. Our main focus is on the latter gains since they pertain to what Marshall (1890) referred to as \mbox{the mysteries of trade in the air.}

\subsection{Bartik instrument}

To construct a Bartik instrument, we start with the identity regarding top inventor inflows, $M_{dt} =\sum_{o \neq d} M_{odt}$; i.e., the total number of top inventors migrating to destination commuting zone $d$ in period $t$ equals the sum of the number of top inventors migrating from origin commuting zone $o$ to destination commuting zone $d$ across all $o \neq d$ in period $t$. Recalling that $P_{odt} = M_{odt} / I_{ot}$, we have
\begin{equation}
\textstyle M_{dt} = \sum_{o \neq d} P_{odt} I_{ot},
\label{eq:identity}
\end{equation}
where the right-hand side consists of the share $P_{odt}$ of top inventors migrating from $o$ to $d$ in period $t$ and the shift $I_{ot}$, i.e., the number of top inventors in $o$ at the beginning of period $t$.

As shown by Goldsmith-Pinkham et al.~(2020) and Borusyak et al.~(2022), there are two approaches to ensuring the exogeneity of a Bartik instrument. The former consider share exogeneity, whereas the latter employ shift exogeneity. We rely on the former in this section and check the robustness of the results using the latter in Section~\ref{sec:sec5.6}.

Our novelty lies in combining the share exogeneity approach in Goldsmith-Pinkham et al.~(2020) with the estimates obtained from the location choice model \`{a} la Moretti and Wilson (2017). Let $\{\widehat \eta, \widehat \eta', \widehat \gamma_d, \widehat \gamma_o, \widehat \gamma_{od}\}$ denote the estimates from \eqref{eq:eqm}. Using these estimates, we first compute the predicted probability that a top inventor, who lived in $o$ at the beginning of period $t$, moves to $d$ by the end of period~$t$ as follows (see Appendix~\ref{sec:secB2} for the derivation):
\vspace{.1cm}
\begin{equation}
\widehat P_{odt} = \frac{ \exp \{ \widehat   \eta  \ln (1-\tau_{\sigma(d)t})  +   \widehat  \eta'  \ln (1-\tau'_{\sigma(d)t})    +  \widehat \gamma_d
+\widehat  \gamma_{od} \} }{\sum_{c \in {\cal C}}
 \exp \{ \widehat  \eta  \ln (1-\tau_{\sigma(c)t}) + \widehat   \eta'  \ln (1-\tau'_{\sigma(c)t})  +\widehat   \gamma_c  + \widehat \gamma_{oc} \}},
\label{eq:migprob}
\vspace{.1cm}
\end{equation}
where ${\cal C}$ is the set of \emph{all} commuting zones including origin commuting zone $o$ and destination commuting zone $d$.\footnote{\label{fn12}When estimating \eqref{eq:eqm}, we do not simultaneously use the set of fixed effects $\{\widehat \gamma_d, \widehat \gamma_o, \widehat \gamma_{od}\}$. Recall that in Table~\ref{tab:tab4}, we consider $\widehat \gamma_{od}$ in Columns 1-2 and adopt $\{\widehat \gamma_d, \widehat \gamma_o \}$ in Columns 3-4. In what follows, we thus modify the way we incorporate fixed effects into \eqref{eq:migprob} according to empirical specifications.}
We then construct a Bartik instrument by replacing the share $P_{odt}$ in \eqref{eq:identity} with the predicted share $\widehat P_{odt}$ in \eqref{eq:migprob} as follows:
\begin{equation}
\textstyle B_{dt} = \sum_{o \neq d} \widehat{P}_{odt} I_{ot}.
\label{eq:Bartik}
\end{equation}
We use the Bartik instrument \eqref{eq:Bartik} in the first-stage regression in Section~\ref{sec:sec4.1} and discuss the share exogeneity in Section~\ref{sec:4.3} and the relevance and validity of the instrument in Section~\ref{sec:4.5}.

\subsection{Empirical specifications}
\label{sec:sec4.1}
\vspace{-.1cm}
We start with the fixed effect (FE) model:
\vspace{-.1cm}
\begin{equation}
\ln Y_{dt} = \phi M_{dt} +\xi X_{dt} +\varepsilon_{dt},
\label{eq:ols}
\vspace{-.1cm}
\end{equation}
where $Y_{dt}$ is the patent productivity of local inventors in commuting zone $d$ in period~$t$ (defined as the number of patents by all local inventors or by external inventors), $M_{dt}=\sum_{o \neq d} M_{odt}$ is the number of top inventors who migrate to destination commuting zone~$d$ in period~$t$, $\phi$ stands for the impact of a top inventor inflow on local patent productivity, $X_{dt}$ is a vector of control variables and fixed effects, as detailed below, and $\varepsilon_{dt}$ is an i.i.d.~shock.

However, the top inventor inflows $M_{dt}$ may be endogenous due to reverse causality or the existence of omitted variables that have direct impacts on both top inventor inflows and local patent productivity. Reverse causality arises when greater local patent productivity attracts top inventors, whereas omitted variables exist when there are unobserved consumption and production amenities that have been studied since Roback (1982).

To address these endogeneity issues, we consider an instrumental variable (IV) regression, which consists of the structural equation
\vspace{-.02cm}
\begin{equation}
\ln Y_{dt} = \phi^s M_{dt} +\xi^s X_{dt} +\varepsilon_{dt}^s,
\label{eq:iv}
\vspace{-.3cm}
\end{equation}
and the first-stage regression
\vspace{-.3cm}
\begin{equation}
\textstyle
M_{dt} = \psi^f B_{dt} +\xi^f X_{dt} +\varepsilon_{dt}^f,
\label{eq:fs}
\vspace{-.1cm}
\end{equation}
where $B_{dt}$ is the Bartik instrument given by \eqref{eq:Bartik}. We further consider two variants of the Bartik instrument to assess the sensitivity of our results. One is the prediction of between-state top inventor flows, $B_{dt}^\sigma = \sum_{o \notin \sigma(d)} \widehat P_{odt} I_{o t}$, to highlight state tax differences. The other is the predicted top inventor flows into commuting zone $\nu(d)$, which is the nearest neighborhood of~$d$, $B_{dt}^\nu =  \sum_{o \neq d, \nu(d)} \widehat P_{o \nu(d) t} I_{o t}$, to take a spatial lag of $B_{dt}$.

In both the FE regression \eqref{eq:ols} and the IV regression \eqref{eq:iv} and \eqref{eq:fs}, $X_{dt}$ includes commuting zone fixed effects, $\delta_d$, year fixed effects, $\delta_t$, and the ATR at the ninety-fifth percentile of the U.S. income distribution, $\tau_{\sigma(d)t}$. As robustness checks, we consider the ATR at the ninety-ninth percentile, MTR, and APTR. Depending on specifications, we include in $X_{dt}$ manufacturing employment as a time-varying control in commuting zone $d$, as well as other taxes and tax credits such as the CITR, ITC, RTC, and ATR at the fiftieth percentile in state $\sigma(d)$.\footnote{See in Figure~\ref{fig:fig3} for these robustness checks. The details are provided in Murata and Nakajima (2025, Appendices~C.1 and C.2). We further incorporate other time-varying employment variables at the commuting zone level such as ``finance and insurance,'' ``professional, scientific, and technical services,'' and ``management of companies and enterprises'' into $X_{dt}$ in Figure~\ref{fig:fig3}.}$^,$\footnote{In both the FE and IV regressions, we replace $\ln Y_{dt}$ with $\ln (1+ Y_{dt})$ in the baseline case to accommodate commuting zone $\times$ year observations with no patents. As a robustness check, we drop such observations and retain $\ln Y_{dt}$. The IV regression estimates using $\ln Y_{dt}$ are shown in Figure~\ref{fig:fig3}, while the corresponding FE estimates and the details of these analyses are reported in \mbox{Murata and Nakajima (2025, Appendix C.3).}}
In both the FE and IV regressions, we cluster standard errors at the commuting zone level.

\subsection{Potential threats to identification}
\label{sec:4.3}

Recall that the Bartik instrument $B_{dt}$ consists of the shares $\widehat P_{odt}$ and the shifts $I_{o t}$.\footnote{In what follows, we refer to the predicted shares $\widehat P_{odt}$ as the shares for simplicity when there is no confusion.} It is based on the migration identity and predicts the top inventor flows to destination $d$ by the sum of the products of these two elements. In the main analysis, we follow the shares perspective; i.e., it is the shares $\widehat P_{odt}$ that provide the exogenous variation satisfying $E(\varepsilon_{dt}^s \widehat P_{odt}|X_{dt})=0$, and the shifts $I_{ot}$ do not affect the identification of $\phi^s$ provided that the shares are exogenous (Goldsmith-Pinkham et al., 2020; Borusyak et al., 2025).

We show in Appendix~\ref{app:exo1} that the share exogeneity holds, i.e., $E(\varepsilon_{dt}^s \widehat P_{odt}|X_{dt})=0$ for $o \in {\cal C}$, under the following two assumptions: (i) $\varepsilon_{dt}^s$ is mean zero conditional on $X_{dt}$, i.e., $E(\varepsilon_{dt}^s|X_{dt})=0$, and (ii) $\varepsilon_{dt}^s$ and other state taxes $\{ \tau_{\sigma (c)t}  \}_{c \notin \sigma (d)}$ are independent conditional on $X_{dt}$, i.e., $\varepsilon_{dt}^s \perp \tau_{\sigma(c)t} | X_{dt}$ for $c \notin \sigma (d)$ and $c,d \in {\cal C}$, where we let $X_{dt}=\{ \tau_{\sigma(d)t}, \delta_d, \delta_t\}$ in the main analysis.\footnote{${\mathfrak A}\perp {\mathfrak B}| {\mathfrak C}$ denotes the independence of ${\mathfrak A}$ and ${\mathfrak B}$ conditional on ${\mathfrak C}$.}
Note that the exclusion restriction in conventional IV regressions is stated at the aggregate level as $E(\varepsilon_{dt}^s B_{dt} |X_{dt})=0$. In contrast, following Goldsmith-Pinkham et al.~(2020), we refer to the orthogonality at the share level, $E(\varepsilon_{dt}^s \widehat P_{odt}|X_{dt})=0$, as the exclusion restriction, which is the key requirement for identifying $\phi^s$ from the shares perspective.

One may worry that the first assumption, $E(\varepsilon_{dt}^s|X_{dt}) = 0$, may not hold due to a possible correlation between $\varepsilon_{dt}^s$ and $\tau_{\sigma(d)t}$ through unobserved state-specific time-varying factors. To alleviate potential concerns that state taxes may respond to local economic conditions or be correlated with local economic policies affecting innovation, we follow Akcigit et al.~(2022) and employ alternative specifications with state $\times$ year fixed effects $\delta_{\sigma(d)t}$. Specifically, we replace $X_{dt}=\{ \tau_{\sigma(d)t}, \delta_d, \delta_t \}$ and $\varepsilon_{dt}^s$ in the structural equation \eqref{eq:iv} with $X_{dt}'=\{ \delta_{\sigma(d)t}, \delta_d, \delta_t \}$ and $\zeta_{dt}^s$, respectively, so that $\delta_{\sigma(d)t}$ subsumes $\tau_{\sigma(d)t}$. We can then recover the share exogeneity $E(\zeta_{dt}^s \widehat P_{odt}|X_{dt}')=0$ by replacing the assumption $E(\varepsilon_{dt}^s |X_{dt})=0$ with $E(\zeta_{dt}^s |X_{dt}')=0$ (see Section~\ref{sec:sec4.2} and Appendix~\ref{app:exo_temp}).

One may also worry that the second assumption, $\varepsilon_{dt}^s \perp \{\tau_{\sigma(c)t} \}_{c \notin \sigma (d)} | \{ \tau_{\sigma(d)t}, \delta_d, \delta_t\}$ for $c,d \in {\cal C}$, may not hold. Recall that the shares $\widehat P_{odt}$ in \eqref{eq:migprob} depend on the tax rates in all states $\{ \tau_{\sigma (c)t} \} _{c \in \cal C}$. Thus, in the destination specific IV regression model \eqref{eq:iv} and \eqref{eq:fs}, the second assumption implies that the state taxes, $\{ \tau_{\sigma (c)t}  \}_{c \notin \sigma (d)}$, other than that in the destination, $\tau_{\sigma (d)t}$, have an effect on the local patent productivity $Y_{dt}$ only indirectly via $\widehat P_{odt}$ in the Bartik instrument $B_{dt}$ (or the predicted inflows of top inventors). To address the concern that other state taxes $\{ \tau_{\sigma (c)t}  \}_{c \notin \sigma (d)}$ may influence the inflows of non-top inventors in $d$, thereby enhancing $Y_{dt}$, we estimate \eqref{eq:iv} and \eqref{eq:fs} by directly excluding \mbox{such potential inflows (see Section~\ref{sec:5.3}).}\footnote{To alleviate another concern that state tax competition may induce correlation between $\varepsilon_{dt}^s$ and $\{ \tau_{\sigma (c)t}  \}_{c \notin \sigma (d)}$, we examine the possibility of strategic interactions among state governments by estimating a reaction function, where the income tax in one state responds to the income taxes in other states (see, e.g., Brueckner, 2003). In line with the second assumption, we do not find strong evidence for state tax competition \mbox{(see \ref{app:taxc}).}}

Finally, to show that our results are not driven by the particular construction of the Bartik instruments, we use the subsample from 1990 to 2009 and compare the results based on $B_{dt}$ with those based on the canonical Bartik instrument $B_{dt}^0 = \sum_{o \neq d} P_{od}^0 I_{o t}$, where $P_{od}^0$ is the initial shares constructed by the observed top inventor flows from 1977 to 1989~(see~Section~\ref{sec:sec4.2}).

\subsection{Main results}
\label{sec:sec4.2}

Table~\ref{tab:tab5} presents the estimation results for the FE and IV regressions. Column 1 reports the FE case. Columns 2-7 are the results for different IV regressions. Column 2 considers the Bartik instrument $B_{dt} = \sum_{o \neq d} \widehat P_{odt} I_{o t}$ in \eqref{eq:Bartik}. Column 3 adds to Column 2 its variant $B_{dt}^\sigma = \sum_{o \notin \sigma(d)} \widehat P_{odt} I_{o t}$, which captures top inventor flows only from other states. Column~4 further adds to Column~3 the other variant $B_{dt}^\nu = \sum_{o \neq d, \nu(d)} \widehat P_{o \nu(d) t} I_{o t}$. It involves top inventor flows from origin commuting zones $o \neq d$ to commuting zone $\nu(d)$, which is the nearest neighborhood of $d$. Columns 5-7 replace $\ln (1-{\rm ATR})$ in Columns 2-4 with {\rm state} $\times$ {\rm year} fixed effects, $\delta_{\sigma(d)t}$, to control for time-varying state-specific unobservables, which alleviates the concern that there may be a correlation between $\varepsilon_{dt}^s$ and $\tau_{\sigma(d)t}$. Finally, Columns 8 and 9 compare $B_{dt}$ with $B_{dt}^0 = \sum_{o \neq d} P_{od}^0 I_{o t}$ using the subsample from 1990 to 2009. In both Panels (a) and (b), we exclude the patents by top  inventors \mbox{who moved in from the dependent variable.}

\begin{table}[tp]
\caption{The impact of top inventor inflows on local patent productivity.}
\label{tab:tab5}
\centering
{\footnotesize
\hskip 1.2cm
\begin{threeparttable}
\begin{tabular}{lrrrrrrrrr}
\hline \hline
& \multicolumn{1}{c}{(1)} & \multicolumn{1}{c}{(2)} & \multicolumn{1}{c}{(3)} &
\multicolumn{1}{c}{(4)} & \multicolumn{1}{c}{(5)} & \multicolumn{1}{c}{(6)} &
\multicolumn{1}{c}{(7)} & \multicolumn{1}{c}{(8)} & \multicolumn{1}{c}{(9)} \\ \hline

\multicolumn{10}{l}{(a) All local inventors} \\[0.2em]
Top inventor inflows
 & 0.043 & 0.062 & 0.060 & 0.059 & 0.066 & 0.059 & 0.060 & 0.050 & 0.039 \\
 & (0.006) & (0.013) & (0.012) & (0.011) & (0.014) & (0.013) & (0.012) & (0.010) & (0.007) \\
 $\ln (1-{\rm ATR})$
 & 6.041 & 5.915 & 5.899 & 6.017 & & & & 5.427 & 5.323 \\
 & (1.038) & (1.040) & (1.038) & (1.038) & & & & (1.843) & (1.846) \\[0.3em]

 Effective $F$ statistic
 & & 37.755 & 33.377 & 33.040 & 51.824 & 35.130 & 34.997 & 30.528 & 24.300 \\
 $\tau=5\%$
 & & 37.418 & 31.930 & 34.734 & 37.418 & 31.214 & 32.989 & 37.418 & 37.418 \\
 $\tau=10\%$
 & & 23.109 & 19.892 & 21.389 & 23.109 & 19.473 & 20.364 & 23.109 & 23.109 \\
 $\tau=20\%$
 & & 15.062 & 13.094 & 13.901 & 15.062 & 12.839 & 13.272 & 15.062 & 15.062 \\
 $\tau=30\%$
 & & 12.039 & 10.531 & 11.093 & 12.039 & 10.336 & 10.610 & 12.039 & 12.039 \\[0.6em]

\multicolumn{10}{l}{(b) External inventors} \\[0.2em]
Top inventor inflows
 & 0.027 & 0.042 & 0.040 & 0.041 & 0.041 & 0.036 & 0.038 & 0.038 & 0.032 \\
 & (0.004) & (0.010) & (0.009) & (0.009) & (0.011) & (0.010) & (0.010) & (0.008) & (0.007) \\
 $\ln (1-{\rm ATR})$
 & 4.781 & 4.684 & 4.641 & 4.616 & & & & 4.636 & 4.574 \\
 & (0.848) & (0.851) & (0.848) & (0.842) & & & & (1.739) & (1.741) \\[0.3em]
 
Effective $F$ statistic
 & & 37.755 & 33.377 & 33.040 & 51.824 & 35.130 & 34.997 & 30.528 & 24.300 \\
 $\tau=5\%$
 & & 37.418 & 31.921 & 34.738 & 37.418 & 31.203 & 32.988 & 37.418 & 37.418 \\
 $\tau=10\%$
 & & 23.109 & 19.887 & 21.392 & 23.109 & 19.467 & 20.364 & 23.109 & 23.109 \\
 $\tau=20\%$
 & & 15.062 & 13.091 & 13.902 & 15.062 & 12.835 & 13.272 & 15.062 & 15.062 \\
 $\tau=30\%$
 & & 12.039 & 10.529 & 11.094 & 12.039 & 10.333 & 10.610 & 12.039 & 12.039 \\[0.6em]

CZ FE
 & Yes & Yes & Yes & Yes & Yes & Yes & Yes & Yes & Yes \\
 Year FE
 & Yes & Yes & Yes & Yes & No & No & No & Yes & Yes \\
 State $\times$ year FE
 & No & No & No & No & Yes & Yes & Yes & No & No \\
 Years: 1977--2009
 & Yes & Yes & Yes & Yes & Yes & Yes & Yes & No & No \\
 Years: 1990--2009
 & No & No & No & No & No & No & No & Yes & Yes \\[0.4em]

Observations
 & 23,628 & 23,628 & 23,628 & 23,463 & 23,562 & 23,562 & 23,397 & 14,320 & 14,320 \\
\hline \hline
\end{tabular}

\begin{tablenotes}[flushleft]
\footnotesize
\item \hskip -0.1cm \emph{Notes:}
The coefficient on top inventor inflows is converted to semi-elasticity.
ATR stands for the individual income average tax rate at the ninety-fifth percentile.
The coefficient on $\ln (1-{\rm ATR})$ is converted to elasticity.
Column 1 does not control for the endogeneity of top inventor inflows.
Column 2 uses $B_{dt}$ as an instrument.
Column 3 uses $B_{dt}$ and $B_{dt}^{\sigma}$ as instruments.
Column 4 uses $B_{dt}$, $B_{dt}^{\sigma}$, and $B_{dt}^{\nu}$ as instruments.
Columns 5--7 replace $\ln (1-{\rm ATR})$ in Columns 2--4 with state $\times$ year fixed effects.
Columns 8 and 9 use the subsample from 1990 to 2009 and present the results based on $B_{dt}$ and those based on the canonical Bartik instrument $B_{dt}^0$, respectively.
Cluster-robust standard errors are in parentheses.
\end{tablenotes}
\end{threeparttable}
\vspace{-.25cm}
}
\end{table}

As seen from Table~\ref{tab:tab5}, the semi-elasticities of local patent productivity with respect to top inventor inflows, as well as the elasticities of local patent productivity with respect to $1-{\rm ATR}$, are nearly identical for all IV regressions in Columns 2-7 within each panel.\footnote{The semi-elasticities of local patent productivity with respect to top inventor inflows are somewhat smaller for the FE regression. One possible explanation for this is the presence of urban costs such as land rents and commuting costs that are specific to commuting zones and can vary over time (see Duranton and Puga, 2020).}
Panel (a) shows that an inflow of a top inventor raises the patent productivity of all local inventors by approximately 6\%, which can be interpreted as local aggregate gains from both knowledge sharing among internal inventors and knowledge spillovers to external inventors. Panel (b) shows the main result that a top inventor inflow raises local patent productivity by approximately 4\% when we focus on external inventors who are not directly connected to the migrating top inventors.
The latter result can be interpreted as evidence for the existence of the mysteries of trade in the air as the number reflects neither knowledge flows within the same assignee nor those between co-inventors.
We thus disentangle productivity gains due to external knowledge spillovers (``knowledge in the air'') from those due to internal knowledge sharing (``knowledge in the lab''). We further verify that this main result is not sensitive to the construction of the Bartik instruments and the sample period \mbox{(see Columns 8-9 in Panel (b)).}

Our analysis at the commuting zone level differs from the national level analysis in Borjas and Doran (2012), who find a negative impact of the inflows of Soviet mathematicians on the productivity of U.S. mathematicians, and from Moser~et~al.~(2014), who summarize that knowledge spillovers from German Jewish \'{e}migr\'{e}s to incumbent U.S. inventors are unlikely to have been the main driver of the U.S. patent productivity gains.\footnote{We elaborate on incumbents' productivity gains in Section~\ref{sec:5.3}.}
Our main finding---the existence of gains from knowledge in the air (i.e., external inventors' productivity gains due to external knowledge spillovers)---also differs from De la Roca and Puga (2017), Moretti (2021), and Prato (2025) in that these studies highlight the impacts on those who migrate themselves or consider internal knowledge sharing through organizations or co-inventor relationships.

Furthermore, our 4-6\% patent productivity gains for local inventors due to an additional inventor inflow could be compared with the 5\% reduction in patenting due to hypothetically removing the rise in immigration to the US after the 1965 Immigration National Act in Terry et al. (2026). Our results also could be related to the 12\% increase in incumbent plants' TFP due to a new plant opening in Greenstone et al.~(2010) or with the 62\% increase in local patent productivity due to the establishment of a new college in Andrews (2023). However, given the difference between top inventor arrival on the one hand and firm entry and college establishment on the other hand,
it is not surprising that the former is smaller than the latter.

\begin{figure}[tp]
\caption{Specification curve analysis.}
\subfigure[all local inventors]
{\includegraphics[clip, width=0.5\columnwidth]{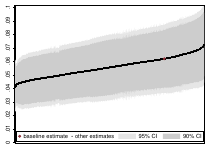}}
\subfigure[external inventors]
{\includegraphics[clip, width=0.5\columnwidth]{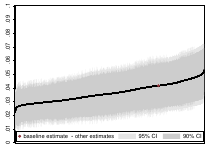}} \\
\footnotesize{\noindent
\emph{Notes:} Panels (a) and (b) illustrate the impacts of a top inventor inflow on the patent productivity of all local inventors and that of external inventors, respectively. Each specification curve is depicted using $31, 969$ alternative specifications, as explained in footnote~\ref{fn_spec1}. The vertical axis is the estimated value of $\widehat \phi^s$.}
\label{fig:fig3}
\end{figure}

We relegate extensive robustness checks of the main result to Section 5. However, before concluding this subsection, let us briefly illustrate the specification curve analysis as in Simonsohn et al.~(2020). We employ different specifications of the IV regressions by considering various dimensions.\footnote{\label{fn_spec1}We consider (i) whether to use the ATR at the ninety-fifth or ninety-ninth percentile; (ii) whether to use $\ln (1+Y_{dt})$ or drop commuting zones with $Y_{dt}=0$; (iii) whether to use $\{B_{dt} \}$, $\{B_{dt}, B_{dt}^\sigma \}$ or $\{B_{dt}, B_{dt}^\sigma, B_{dt}^\nu   \}$ as instruments; (iv) whether to include state $\times$ year fixed effects;
(v) whether to use the baseline or alternative detection of top inventor migrations;
(vi) whether to use the baseline or alternative definition of local inventors (see the analysis on incumbents' productivity gains in Section~\ref{sec:5.3});
(vii) whether to use ATRs or MTRs;
(viii) whether to include APTRs;
and (ix) whether to include each of the other controls (ATR50, CITR, ITC, RTC, manufacturing employment, and other employment variables such as ``finance and insurance,'' ``professional, scientific, and technical services,'' and ``management of companies and enterprises'' to capture other omitted variables, e.g., access to venture capital). Since the usual caveat on weak instruments is applicable here, we adopt only specifications for which the null hypothesis of weak instruments is rejected. The details are provided in Murata and Nakajima (2025, Appendix C).}
Figure~\ref{fig:fig3} plots the specification curve for $\widehat \phi^s$ with 90\% and 95\% confidence intervals. As seen from Panels (a) and (b), the productivity gains of approximately 6\% and 4\% are fairly robust for $31,969$ alternative specifications, thus verifying that our main estimates do not come from data mining. In what follows, we use the specification in Column 3 of Table~\ref{tab:tab5} as a baseline unless otherwise stated.

\subsection{Relevance and validity of Bartik instruments}
\label{sec:4.5}

Our empirical strategy relies on the relevance and validity of the Bartik instruments, which we discuss in what follows.

\begin{figure}[tp]
\caption{Actual versus predicted top inventor flows.}
\subfigure[all flows]
{\includegraphics[clip, width=0.325\columnwidth]{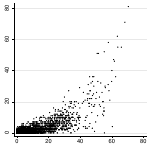}}
\subfigure[interstate flows]
{\includegraphics[clip, width=0.325\columnwidth]{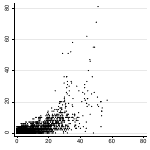}} 
\subfigure[flows to neighboring cz]
{\includegraphics[clip, width=0.325\columnwidth]{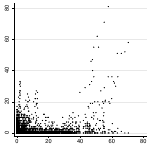}}  \\
\footnotesize{\noindent
\emph{Notes:} In each panel, the vertical and horizontal axes are the actual top inventor flows and the Bartik instruments constructed from the predicted top inventor flows, respectively. The actual flows in Panels (a), (b), and (c) are defined as $M_{dt}$.~The Bartik instruments in Panels (a), (b), and (c) are $B_{dt}$, $B_{dt}^\sigma$, and $B_{dt}^\nu$, respectively.}
\label{fig:fig4}
\end{figure}

To assess the relevance of the Bartik instruments, we first plot in Figure~\ref{fig:fig4} the relationship between the actual top inventor flows $M_{dt} = \sum_{o \neq d} M_{odt}$ and the Bartik instruments constructed from the predicted top inventor flows. For the latter, we consider
$B_{dt} = \sum_{o \neq d} \widehat P_{odt} I_{o t}$,
$B_{dt}^\sigma = \sum_{o \notin \sigma(d)} \widehat P_{odt} I_{o t}$, and
$B_{dt}^\nu =  \sum_{o \neq d, \nu(d)} \widehat P_{o \nu(d) t} I_{o t}$,
in Panels (a), (b), and (c) and obtain the correlation coefficients, $0.78$, $0.74$, and $0.38$, respectively. For the subsample from 1990 to 2009, the correlation coefficient between $M_{dt}$ and $B_{dt}$ is $0.83$, whereas that between $M_{dt}$ and $B_{dt}^0 = \sum_{o \neq d} P_{od}^0 I_{o t}$  is $0.79$.

We further apply a test for weak instruments developed by Montiel Olea and Pflueger (2013) to these Bartik instruments. The test is robust to heteroskedasticity, autocorrelation, and clustering (see also Andrews et al.,~2019). The bottom of each panel in Table~\ref{tab:tab5} reports the effective $F$ statistic, which is a scaled version of the nonrobust first-stage $F$ statistic. Following their baseline, we set the threshold at $\tau=10$\% and the significance at $5$\%. In all cases, the effective $F$ statistic exceeds the critical value reported at $\tau=10$\%, thus rejecting the null hypothesis of weak instruments.

To address the validity of the Bartik instruments, we follow the shares perspective and focus on the exogeneity of the shares $\widehat{P}_{odt}$. We assess the plausibility of the exogeneity assumption in two steps. We first use the decomposition result in Goldsmith-Pinkham~et~al.~(2020) to rewrite the overall estimate of the productivity effect as $\widehat{\phi}^{s} = \sum_{o \in {\cal C}} \widehat{\omega}_{o} \widehat{\phi}^{s}_{o}$, which consists of the origin-specific weight $\widehat{\omega}_{o}$ and the origin-specific productivity effect $\widehat{\phi}^{s}_{o}$. The former is referred to as the Rotemberg weight (Rotemberg, 1983) and measures to what extent the bias originating from commuting zone $o$ contributes to the overall bias. For each of the top five origin commuting zones by Rotemberg weight, we then check the correlation between the predicted probabilities of top inventor migrations to the destination commuting zones and key pre-period destination characteristics. This approach allows us to assess the exogeneity of the shares~$\widehat{P}_{odt}$.

We summarize the results in \ref{app:rotem}. Table \ref{tab:tabF1} presents the summary of the Rotemberg weights. The origin commuting zones with the top five highest weights are Bergen-Essex-Middlesex, Cook-DuPage-Lake, Kings-Queens-NewYork, Philadelphia-Montgomery-Delaware, and Allegheny-Westmoreland-Washington.\footnote{The name of each commuting zone shown here is a list of three counties with the largest numbers of inventors (in descending order) in that commuting zone.}
Table~\ref{tab:tabF2} reports the destination commuting zones to which top inventors moved from these five origin commuting zones. The result that origin and destination states differ in almost all cases is in line with the assumption that the main source of identifying variation comes from interstate top inventor migrations induced by personal income tax differences between states. Table \ref{tab:tabF3} reports, for each of the top five origin commuting zones, the correlation between the predicted shares of top inventor migrations to the destination commuting zones and the log of lagged employment in four sectors: manufacturing; finance and insurance; professional, scientific, and technical services; and management of companies and enterprises. These sectors are chosen as they may correlate with unobservable confounders. The correlations are consistently low, and the coefficients obtained from regressing the predicted shares on the log of lagged sectoral employment are not significant at the 5\% level in all specifications, thus supporting \mbox{the validity of our Bartik instruments.}

\section{Robustness}
\label{sec:sec5}
In this section, we examine the robustness of our main results in terms of time, space, aggregation, an alternative high-income occupation, and an alternative exogeneity assumption. We first extend our static framework to a dynamic setting, which allows us to assess the impacts on local patent productivity before and after top inventor inflows. We then check the robustness in terms of the geographic extent of productivity gains. We further analyze two disaggregated cases: where local inventors are classified by their patent productivity; and where local inventors differ in terms of how many years they have engaged in patenting when top inventors arrive. The latter case assesses whether the productivity gains come from the entry of new local inventors (``extensive margins'') or from incumbent local inventors (``intensive margins''). We also perform a falsification exercise using an alternative high-income occupation. Moreover, we consider the shifts perspective (Borusyak et al., 2022) instead of the shares perspective (Goldsmith-Pinkham et al., 2020).

Furthermore, we conduct some additional robustness checks (see Figure~\ref{fig:fig3} and footnote~\ref{fn_spec1}). Specifically, we consider the alternative individual income average tax rate at the ninety-ninth percentile of the U.S. income distribution (ATR99), statutory marginal tax rates (MTRs), average property tax rates (APTRs), and an alternative way of detecting top inventor migrations. We also include other controls (ATR50, CITR, ITC, RTC, and manufacturing employment), or drop commuting zone $\times$ year observations with no patents. We further use the alternative measure of patent productivity based on patent quality in Kogan et al. (2017). All results are consistent with our main findings and the details are provided in Murata and Nakajima (2025, Tables C2, C4, C6, C8, C9, C10, C12 in Appendices C.1-C.4).

\subsection{Dynamic analysis}
As a first robustness check, we conduct event study analysis to examine whether our main results are sensitive to the inclusion of lead and lag effects of top inventor inflows. To this end, we rewrite \eqref{eq:iv} as $\ln Y_{dt} = \sum_{j=\underline j+1}^{\overline j} \phi_j^\ell M_{dt-j} + \xi^\ell X_{dt} +\varepsilon_{dt}^\ell,$ which is a distributed lag model in levels with a binning window $[\underline j+1, \overline j]$. Thus, when $\underline j=-1$ and $\overline j =0$, this extended model degenerates into the static model \eqref{eq:iv}. Schmidheiny and Siegloch (2023) show that the foregoing equation is equivalent to the event study model given by\footnote{Unlike in standard event study models with a single treatment of identical intensity, we consider a more general case with multiple treatments of varying intensities. See Schmidheiny and Siegloch (2023) for the detailed classification of event study models.}
\vspace{.05cm}
\begin{eqnarray}
\textstyle \ln Y_{dt} &=& \textstyle  \sum_{j=\underline j}^{\overline j} \mu_j^{es} \Delta M_{dt}^{(j)} + \xi^{es} X_{dt}
+\varepsilon_{dt}^{es},  \ \ \mbox{where} \label{eq:es} \\[5pt]
\textstyle \Delta M_{dt}^{(j)} &=&
\left\{ 
\begin{array}{ll}
\sum_{k=-\infty}^{\underline j} \left( M_{dt-k} - M_{dt-k-1} \right) & \mbox{if \ \ $j=\underline j<0$} \\
M_{dt-j} -  M_{dt-j-1}  & \mbox{if \ \ $\underline j<j < \overline j$} \\
\sum_{k=\overline j}^{\infty} \left( M_{dt-k} - M_{dt-k-1} \right) & \mbox{if \ \ $j= \overline j>0$}
\end{array}
\right..
\label{eq:defMs}
\end{eqnarray}
\vskip .3cm
\noindent
Our aim is to estimate $\{\mu_{\underline j}^{es}, \mu_{\underline j+1}^{es}, ... , \mu_{\overline j-1}^{es}, \mu_{\overline j}^{es}\}$ with normalization $\mu_{-1}^{es}=0$. The event study coefficients capture the cumulative effect of the event of top inventor inflows, i.e., $\mu_j^{es}  =\mu_{j-1}^{es}+ \phi_j^\ell = \sum_{h=0}^j \phi_h^{\ell}$ for $j=0,1,...,\overline j$ and $\mu_j^{es}  =\mu_{j+1}^{es}- \phi_{j+1}^\ell = - \sum_{h=j+1}^{-1} \phi_h^{\ell}$ for $j=-2,-3,...,\underline j$. Thus, the coefficients for $j \ge  0$ denote cumulative productivity effects from  event year $0$ (when there are top inventor inflows) to year $j$. Since the static model abstracts from the lead and lag effects, the baseline model may produce biased estimates of productivity gains.

As in the static analysis, we incorporate the Bartik instruments into the event study model. Let $\boldsymbol{\Delta B}_{dt}= [\Delta B_{dt}^{(\underline j)} \cdots  \Delta B_{dt}^{(\overline j)}]'$ denote a $(\underline j + \overline j + 1) \times 1$ vector of the first time difference of the IVs, where $\Delta B_{dt}^{(j)}$ is defined in a similar way as in \eqref{eq:defMs}. The IV event study model consists of the structural equation \eqref{eq:es} and the first-stage regression analogous to \eqref{eq:fs} as follows\footnote{This robustness check abstracts from the possibility that treatment effects can be heterogeneous. Although several recent papers have explored under what conditions event study models provide valid average treatment effects in the presence of heterogeneous treatment effects (e.g., de Chaisemartin and D'Haultf{\oe}uille, 2023), they are not readily applicable to our IV event study model with
multiple treatments of varying intensities.}
\vspace{.1cm}
\begin{equation}
{\Delta M}_{dt}^{(j)} = {\boldsymbol{\psi}^{ef}}^{(j)} \boldsymbol{\Delta B}_{dt} +{\xi^{ef}}^{(j)} X_{dt} +{\varepsilon_{dt}^{ef}}^{(j)},
\label{eq:fses}
\end{equation}
where ${\boldsymbol{\psi}^{ef}}^{(j)}= [{{\psi}^{ef}}^{(j, \underline j)} \cdots {{\psi}^{ef}}^{(j, \overline j)}]$ is a vector of coefficients.\footnote{When estimating the event study models with multiple instruments, we set $\boldsymbol{\Delta B}^{\sigma}_{dt}=[\Delta B_{dt}^{\sigma(\underline j)} \cdots \Delta B_{dt}^{\sigma(\overline j)}]'$ and $\boldsymbol{\Delta B}^{\nu}_{dt}=[\Delta B_{dt}^{\nu(\underline j)} \cdots \Delta B_{dt}^{\nu(\overline j)}]'$ and use $[{\boldsymbol{\Delta B}_{dt}}' \  {\boldsymbol{\Delta B}^{\sigma}_{dt}}' ]'$ or  $[{\boldsymbol{\Delta B}_{dt}}' \   {\boldsymbol{\Delta B}^{\sigma}_{dt}}' \  {\boldsymbol{\Delta B}^{\nu}_{dt}}' ]'$ as instruments in (\ref{eq:fses}).}

\begin{figure}[tp]
\caption{IV event study regressions.}
\vspace{-.25cm}
\subfigure[all local inventors]
{\includegraphics[clip, width=0.5\columnwidth]{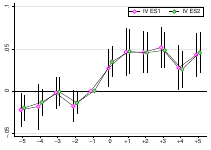}}
\subfigure[external inventors]
{\includegraphics[clip, width=0.5\columnwidth]{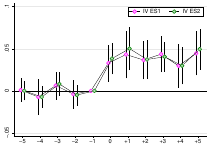}} \\
\footnotesize{\noindent
\emph{Notes:} Panels (a) and (b) illustrate the dynamic impacts of a top inventor inflow on the patent productivity of all local inventors and that of external inventors, respectively. In each panel, IV~ES1 uses $B_{dt}$ and $B_{dt}^\sigma$ as instruments and IV~ES2 uses $B_{dt}$, $B_{dt}^\sigma$, and $B_{dt}^\nu$ as instruments. In \ref{sec:G}, we assess the robustness to possible violations of the parallel trends assumption in both Panels (a) and (b) using the method developed by Rambachan and Roth (2023) and confirm that such violations are unlikely.}
\label{fig:fig5}
\vspace{-.35cm}
\end{figure}

Figure~\ref{fig:fig5} shows the results for the IV event study regressions.\footnote{We report the numbers used in Figure~\ref{fig:fig5} and the associated first-stage statistics in Murata and Nakajima (2025, Tables C13 and C14 in Appendix C.5).}
Panel (a) illustrates the dynamic impacts of a top inventor inflow on the patent productivity of all local inventors, which include not only the gains from internal knowledge sharing within the same assignee and between co-inventors but also the gains from external knowledge spillovers. Panel~(b) corresponds to the dynamic productivity gains of external inventors, which go beyond organizational boundaries and co-inventor relationships. In both cases, we observe a substantial increase in local patent productivity in event year 0 when there are top inventor inflows. The post-event semi-elasticities go up to approximately 0.05, which ensures our main results in Section~\ref{sec:sec4.2}. In contrast, the pre-event semi-elasticities are close to zero in any pre-event year, thus suggesting no productivity gains prior to the event of top inventor migration. In \ref{sec:G}, we further assess the robustness to possible violations of the parallel trends assumption in both Panels (a) and (b) using the method developed by Rambachan and Roth (2023) and confirm that such violations are unlikely.

\subsection{Geographic space}
\label{sec:5.2}
\vspace{-.2cm}
To check the robustness of our main results in terms of the geographic extent of productivity gains, we replace the structural equation \eqref{eq:iv} with
\vspace{-.1cm}
\begin{equation}
\textstyle \ln Y_{dt} = \sum_{r(d)=1}^6 \phi^{sr}_{r(d)} M_{r(d)t} +\xi^{sr} X_{dt} +\varepsilon_{dt}^{sr},
\label{eq:ivdist}
\vspace{-.1cm}
\end{equation}
where $r(d)$ is the distance ring defined for each destination commuting zone $d$ and $M_{r(d)t}$ is the flows of top inventors in the $r(d)$-th ring. The first ring $r(d)=1$ stands for destination~$d$ itself, whereas $r(d) = 2,...,6$ correspond to commuting zones that are $0$-$50$, $50$-$100$, $100$-$150$, $150$-$200$, and $200$-$250$ miles away from commuting zone $d$.\footnote{The distance between any pair of two commuting zones is calculated using the great circle formula.}

\begin{figure}[tp]
\caption{Distance-ring regression.}
\vspace{-.2cm}
\subfigure[all local inventors]
{\includegraphics[clip, width=0.5\columnwidth]{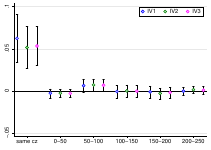}}
\subfigure[external inventors]
{\includegraphics[clip, width=0.5\columnwidth]{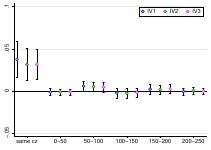}} \\
\footnotesize{\noindent
\emph{Notes:} Panels (a) and (b) illustrate the impacts of a top inventor flow to each distance ring $r(d)$ on the patent productivity of all local inventors in $d$ and that of external inventors in $d$, respectively, where $d$ is the destination commuting zone. In each panel, IV1 uses $B_{dt}$ as an instrument, IV2 uses $B_{dt}$ and $B_{dt}^\sigma$ as instruments, and IV3 uses $B_{dt}$, $B_{dt}^\sigma$, and $B_{dt}^\nu$ as instruments.}
\label{fig:fig6}
\vspace{-.1cm}
\end{figure}

Figure~\ref{fig:fig6} illustrates the estimated coefficients $\{  \widehat \phi^{sr}_{r(d)} \}_{r(d)=1}^6$. Panels (a) and (b) illustrate the impacts of a top inventor flow to each distance ring $r(d)$ on the patent productivity of all local inventors in $d$ and that of external inventors in $d$, respectively. In each panel, the impacts are significant only in the first ring for all three different IVs, which implies that top inventor inflows affect patent productivity only in the commuting zone where they enter. Such localized productivity gains are reminiscent of localized knowledge spillovers in Jaffe et al.~(1993) and Murata et al.~(2014). We will discuss the mechanism of localized productivity gains in terms of \mbox{localized knowledge spillovers as evidenced by patent citations in Section~\ref{sec:sec6.1}.}

\begin{figure}[tp]
\caption{Placebo.}
\vspace{-.2cm}
\subfigure[all local inventors]
{\includegraphics[clip, width=0.5\columnwidth]{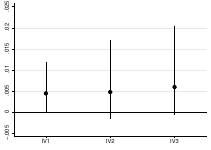}}
\subfigure[external inventors]
{\includegraphics[clip, width=0.5\columnwidth]{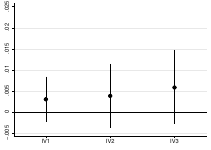}} \\
\footnotesize{\noindent
\emph{Notes:} Panels (a) and (b) illustrate the impacts of a top inventor flow to $d$ on the patent productivity of all local inventors in $R(d)$ and that of external inventors in $R(d)$, respectively, where $d$ is the destination commuting zone and $R(d) \neq d$ is a commuting zone that is randomly drawn from state $\sigma (d)$ to which commuting zone $d$ belongs. In each panel, IV1 uses $B_{dt}$ as an instrument, IV2 uses $B_{dt}$ and $B_{dt}^\sigma$ as instruments, and IV3 uses $B_{dt}$, $B_{dt}^\sigma$, and $B_{dt}^\nu$ as instruments.}
\label{fig:fig7}
\vspace{-.1cm}
\end{figure}

We also conduct a permutation-based placebo test to assess the plausibility of our findings that productivity gains are localized within each commuting zone. This is done by examining the impact of top inventor migration into commuting zone $d$ on productivity gains in a randomly drawn commuting zone $R(d) \neq d$ in state $\sigma (d)$. We thus replace \eqref{eq:iv} with
\begin{equation}
\ln Y_{R(d)t} = \phi^{sR} M_{dt} +\xi^{sR} X_{dt} +\varepsilon_{dt}^{sR}
\label{eq:ivrandom}
\end{equation}
and estimate \eqref{eq:ivrandom} for each IV specification $1000$ times with replacement to obtain the distribution of $\{\phi^{sR}_i\}_{i=1}^{1000}$. We then check if the null hypothesis of no productivity gains, $\phi^{sR}=0$, is rejected. Figure~\ref{fig:fig7} depicts the 95\% confidence interval and the mean of the distribution for each IV specification. The results in both Panels (a) and (b) show that top inventor flows into commuting zone $d$ do not significantly change patent productivity in commuting zone $R(d) \neq d$ randomly drawn from state $\sigma (d)$, thus implying that the extent of productivity gains is geographically limited within each commuting zone. 

\subsection{Productivity gains by local inventor types}
\label{sec:5.3}
\vspace{-.2cm}
We have thus far shown that top inventor inflows enhance patent productivity only for local inventors. We further explore the foregoing result by addressing who gain more from top inventor inflows by allowing for heterogeneity among local inventors.

\begin{figure}[tp]
\caption{Productivity gains by local inventors' productivity.}
\subfigure[all local inventors]
{\includegraphics[clip, width=0.5\columnwidth]{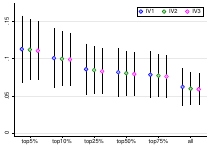}}
\subfigure[external inventors]
{\includegraphics[clip, width=0.5\columnwidth]{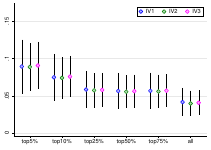}} \\
\footnotesize{\noindent
\emph{Notes:} Both panels illustrate the impacts of a top inventor inflow on local patent productivity, where the impacts differ by local inventors' patent productivity. Panels (a) and (b) are the results for all local inventors and for external inventors, respectively. In each panel, IV1 uses $B_{dt}$ as an instrument, IV2 uses $B_{dt}$ and $B_{dt}^\sigma$ as instruments, and IV3 uses $B_{dt}$, $B_{dt}^\sigma$, and $B_{dt}^\nu$ as instruments.}
\label{fig:fig8}
\end{figure}

First, local inventors differ in terms of their patent productivity. We thus consider the top 5\%, 10\%, 25\%, 50\%, and 75\% of local inventors according to their patent productivity and estimate the causal effect for each productivity group. Panel (a) of Figure~\ref{fig:fig8} illustrates the case with all local inventors. We observe that there are productivity gains for each productivity group and that more-productive local inventors tend to gain more from top inventor inflows. We find a similar pattern in Panel (b). Hence, even when focusing on external inventors who are not directly connected to the migrating top inventors, our result suggests that more-productive local inventors have a tendency to benefit more from their inflows.

\begin{figure}[tp]
\caption{Productivity gains by local stayers' duration.}
\vspace{-.1cm}
\subfigure[all local stayers]
{\includegraphics[clip, width=0.5\columnwidth]{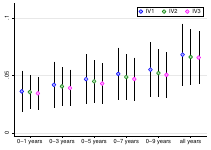}}
\subfigure[external stayers]
{\includegraphics[clip, width=0.5\columnwidth]{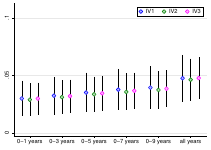}} \\
\footnotesize{\noindent
\emph{Notes:} Both panels illustrate the impacts of a top inventor inflow on the patent productivity of local stayers, where the impacts differ by their duration. Panels (a) and (b) are the results for all local stayers and for external stayers, respectively. In each panel, IV1 uses $B_{dt}$ as an instrument, IV2 uses $B_{dt}$ and $B_{dt}^\sigma$ as instruments, and IV3 uses $B_{dt}$, $B_{dt}^\sigma$, and $B_{dt}^\nu$ as instruments.}
\label{fig:stayers}
\vspace{-.4cm}
\end{figure}

Second, local inventors differ in terms of how many years they have engaged in patenting when top inventors arrive.~To highlight this difference, we focus on local inventors who apply for patents only in a single commuting zone during the sample period. Since these local inventors show no signs of migration during the sample period, we refer to them as \emph{local stayers}, who can be considered as having never moved, and classify them into \emph{internal stayers} and \emph{external stayers}.\footnote{The number of local stayers is 1,195,335, which accounts for 93.81\% of all 1,274,192 local inventors. Thus, in this robustness check, we exclude 78,857 local inventors (6.19\% of all local inventors) who apply for patents in multiple commuting zones. The internal stayers are local stayers who share the same organization as the migrating top inventors and/or if they are co-inventors of the migrating top inventors. All the other local stayers are referred to as external stayers.}
We then disaggregate those local stayers by the length of their inventive activity and estimate the causal effect for each duration.~In each panel of Figure \ref{fig:stayers}, the leftmost bars show the impacts of a top inventor inflow on local stayers with a duration of less than one year, thus capturing the impacts on those who had just started patenting upon the arrival of top inventors (``extensive margins''). The next bars stand for up to three years of patenting, up to five years of patenting, and so on, thus encompassing not only the extensive margins but also the impacts on those who had already engaged in patenting before the top inventor arrival (``intensive margins''). The rightmost bars show the overall effect on local stayers, regardless of their duration. These results suggest that the intensive margins tend to contribute more to the overall productivity gains from top inventor inflows.

Recall the concern in Section~\ref{sec:4.3}, namely that the state taxes, $\{ \tau_{\sigma (c)t}  \}_{c \notin \sigma (d)}$, other than that in the destination, $\tau_{\sigma (d)t}$, may also influence the inflows of non-top inventors, thereby contributing to local patent productivity. This concern is mitigated since the results in Figure~\ref{fig:stayers}, which excludes the role of such potential non-top inventor inflows in local patent productivity by focusing on those who apply for patents in a single commuting zone, are quite similar to our main results in Table~\ref{tab:tab5}.\footnote{Like the definition of a top inventor migration, detecting a non-top inventor migration requires observations in at least two different commuting zones.}

\subsection{Falsification: The case of top baseball players}
\label{sec:fal}
\vspace{-.2cm}
To provide further validation of our analysis, we conduct a falsification exercise using an alternative occupation. Following Kleven et al.~(2013), we focus on high-income professional sports players who should also respond to tax differentials but should not affect patent productivity. In what follows, we examine to what extent the impact of top inventor inflows on local patent productivity is modified by the inflows of top Major League Baseball (MLB) players or ``top players'' for short (see \ref{sec:H} for the institutional settings and data sources).\footnote{We define ``top players'' as those players whose salaries exceeded the ninety-fifth percentile of the U.S. income distribution in the year they declared free agency after six years of MLB service.}

Let ${\cal I} =\{ 1, \dots,I\}$, ${\cal J} = \{1, \dots, J\}$, and ${\cal K} = \{1, \dots, K\}$ denote the set of top players, that of origin teams, and that of destination teams, respectively. The utility of top player $i \in {\cal I}$, who played on team $j \in {\cal J}$ in period $t-1$ and chooses team $k \in {\cal K}$ in period $t$, is given by $U_{i j k t} = V_{i j k t} + \varepsilon_{i j k t}$, where $V_{i j k t}  = \alpha \ln [( 1 - \tau_{\sigma(k) t} ) w_{i k t} ]  + \gamma^h {\rm Home}_{i j k t-1}    + \gamma_k^x X_{i t} - \gamma^c C_{jk}  + Z_k$, $\varepsilon_{i j k t}$ is independent and identically Gumbel distributed, and $k=1$ denotes retirement. Top player $i$'s after-tax salary is given by $( 1 - \tau_{\sigma(k) t} ) w_{i k t}$, where we proxy $\tau_{\sigma(k) t}$ by the ATR at the ninety-fifth percentile in state $\sigma(k)$ in year $t$ as before.\footnote{Two remarks are in order. First, since salary data are available only for a subset of all player-year observations, we use a random forest algorithm to estimate salaries based on various player characteristics and performance. The resulting R-squared values are 0.975 for the training set and 0.813 for the test set. Second, since we observe actual salaries only for chosen teams, we need to construct unobserved potential salaries for other teams. Following Kleven et al.~(2013), we consider several formulations for counterfactual salaries in the context of professional sports labor markets (see \ref{sec:H}).}
${\rm Home}_{i j k t-1}$, which takes a value of $1$ if $j=k$ and $0$ otherwise, is the indicator that captures top player $i$'s preferences for the team in the previous period. $X_{i t}$ is a vector of top player $i$'s characteristics and performance. $C_{jk}$ is the cost of migration measured by the geographical distance between teams $j$ and $k$, and $Z_k$ denotes team fixed effects. The predicted probability that top player $i$, who played on team~$j$ in period $t-1$, chooses team $k$ in period $t$ is then given by
\vspace{.15cm}
\begin{equation}
\widehat P_{ijkt}^{\rm ply} 
= \frac{\exp\{
\widehat{\alpha} \ln [( 1 - \tau_{\sigma(k) t} ) w_{i k t} ]  + \widehat{\gamma}^h {\rm Home}_{i j k t-1}    + \widehat{\gamma}_k^x X_{i t} - \widehat{\gamma}^c C_{jk}  + \widehat{Z}_k 
\}}{\sum_{k'=1}^K \exp\{ \widehat{\alpha} \ln [( 1 - \tau_{\sigma(k') t} ) w_{i k' t} ]  + \widehat{\gamma}^h {\rm Home}_{i j k' t-1}    + \widehat{\gamma}_{k'}^x X_{i t} - \widehat{\gamma}^c C_{jk'}  + \widehat{Z}_{k'} \}}.
\label{eq:plymob}
\end{equation}
\vskip .1cm
\noindent
Like top inventors, we find that top players respond to tax differences within a country (see Table~\ref{tab:mlogit} in \ref{sec:H}), thus showing that the tax-induced international mobility of top players in Kleven et al.~(2013) can be applied to a domestic context. These individual-level predicted probabilities are then aggregated to form the predicted top player flows from origin commuting zone $o$ to destination commuting zone $d$ as $\widehat{P}^{\mathrm{ply}}_{o d t} = \sum_{k \in {\cal K}^d} \sum_{j \in {\cal J}^o} \sum_{i \in {\cal I}^j} \widehat{P}^{\mathrm{ply}}_{ijkt},$ where ${\cal I}^j$ represents the set of top players on team $j$ in period $t-1$; and ${\cal J}^o$ and ${\cal K}^d$ denote the set of teams located in origin $o$ and destination $d$, respectively. Letting $M^{\mathrm{ply}}_{odt}$ and $I^{\mathrm{ply}}_{ot}$ denote the number of top players who migrate from $o$ to $d$ in period $t$ and the number of top players located in origin $o$ in the year prior to $t$, respectively, we define the Bartik instrument for $M^{\mathrm{ply}}_{dt} = \sum_{o\neq d} M^{\mathrm{ply}}_{odt} $  as $B_{dt}^{\mathrm{ply}} = \sum_{o \neq d} \widehat{P}^{\mathrm{ply}}_{odt} I^{\mathrm{ply}}_{ot}$ to estimate the structural equation as follows:
\begin{equation}
\ln Y_{d t}=\phi^{sF} M_{d t}+\check \phi^{sF} M^{\mathrm{ply}}_{d t}+\xi^{sF} X_{d t}+\varepsilon_{d t}^{sF}.
\end{equation}

\begin{table}[htbp]
\caption{The impact of top inventor and top player inflows on local patent productivity.}
\label{tab:regression_results}
{\footnotesize
\hskip 2.1cm   \begin{tabular}{lrrrrrr}
\hline \hline
          	&	\multicolumn{1}{c}{(1)}	&	\multicolumn{1}{c}{(2)}	&	\multicolumn{1}{c}{(3)}	&	\multicolumn{1}{c}{(4)}	&	\multicolumn{1}{c}{(5)}	&	\multicolumn{1}{c}{(6)}	\\ \hline
(a) All local inventors & & & & & &  \\
    Top inventor inflows     & 0.070   & 0.065   & 0.064   & 0.076   & 0.062   & 0.063   \\
                            & (0.018) & (0.015) & (0.015) & (0.021) & (0.016) & (0.016) \\
    Top player inflows       & -0.040  & -0.029  & -0.029  & -0.048  & -0.017  & -0.017  \\
                            & (0.032) & (0.027) & (0.027) & (0.036) & (0.027) & (0.027) \\
       $\ln(1- {\rm ATR})$             & 5.946   & 5.919   & 6.026   &         &         &         \\
                                & (1.041) & (1.038) & (1.039) &         &         &         \\[2mm]
                            \multicolumn{7}{l}{(b) External inventors}      \\
    Top inventor inflows     & 0.049   & 0.045   & 0.047   & 0.050   & 0.039   & 0.041   \\
                            & (0.014) & (0.012) & (0.012) & (0.017) & (0.013) & (0.013) \\
    Top player inflows       & -0.037  & -0.028  & -0.031  & -0.040  & -0.018  & -0.018  \\
                            & (0.026) & (0.022) & (0.022) & (0.029) & (0.021) & (0.022) \\
  $\ln(1 -{\rm ATR})$             & 4.713   & 4.662   & 4.627   &         &         &         \\
  & (0.850) & (0.847) & (0.842) &         &         &         \\[2mm]
    CZ FE                   & Yes     & Yes     & Yes     & Yes     & Yes     & Yes     \\
    Year FE                 & Yes     & Yes     & Yes     & No     & No     & No     \\
    State $\times$ year FE   & No      & No      & No      & Yes     & Yes     & Yes     \\[1mm]
    Observations            & 23,628  & 23,628  & 23,463  & 23,562  & 23,562  & 23,397 \\
\hline
\hline
\end{tabular}
}
\begin{tablenotes}[flushleft]
\footnotesize
\item \hskip -.1cm \emph{Notes:}
The coefficients on top inventor inflows and top player inflows are converted to semi-elasticities. ${\rm ATR}$ stands for the individual income average tax rate at the ninety-fifth percentile. The coefficient on $\ln(1-{\rm ATR})$ is converted to elasticity. Columns 1 and 4 use $B_{dt}$ as an instrument for top inventor inflows $M_{dt}$. Columns 2 and 5 use $B_{dt}$ and $B_{dt}^{\sigma}$ as instruments for $M_{dt}$. Columns 3 and 6 use $B_{dt}$, $B_{dt}^{\sigma}$, and $B_{dt}^{\nu}$ as instruments for $M_{dt}$. In all cases, $B_{dt}^{\mathrm{ply}}$ is used as an instrument for top player inflows $M^{\mathrm{ply}}_{dt}$. Columns 4-6 replace $\ln(1 - {\rm ATR})$ in Columns 1-3 with state $\times$ year FE. Cluster-robust standard errors are in parentheses.
\end{tablenotes}
\vspace{-.4cm}
\end{table}

Table~\ref{tab:regression_results} presents the results for different specifications.\footnote{We report the associated first-stage statistics in Table~\ref{table:sw_tests}.}
Columns 1-6 show that even in the presence of top player inflows, the impact of top inventor inflows on both all local inventors and external inventors remain consistent with our baseline estimates in Table~\ref{tab:tab5}, thus suggesting the robustness of our main results. Indeed, the top player inflows have a negligible effect on local patent productivity, indicating that while high-income top players may respond to tax differences, their presence does not significantly alter the geographic distribution of inventive activity across commuting zones. Hence, we may conclude that although both top inventors and top players take into account the variation in tax rates when making their location decisions, the impacts of their inflows on \mbox{local patent productivity differ substantially.}

\subsection{Shift exogeneity}
\label{sec:sec5.6}
\vspace{-.2cm}
We have so far considered the share exogeneity of the Bartik instruments as in Goldsmith-Pinkham et al.~(2020). We now conduct robustness checks from the shifts perspective by reformulating the destination specific structural equation \eqref{eq:iv} in terms of origin commuting zones (see \ref{sec:I}). Borusyak et al.~(2022) show that the estimator for $\phi^s$ obtained from this reformulation converges in probability to the productivity effect $\phi^s$ in the shares perspective equation \eqref{eq:iv}. Moreover, the standard errors computed in this framework are valid in the presence of exposure-based clustering, as demonstrated by Ad\~{a}o et al. (2019).

\begin{table}[htbp]
\caption{The impact of top inventor inflows on local patent productivity (shift exogeneity).}
\label{tab:shift}
{\footnotesize
\hskip 4.9cm  \begin{tabular}{lrr}
\hline \hline
        & \multicolumn{1}{c}{(1)} & \multicolumn{1}{c}{(2)} \\
        \hline
        (a) All local inventors \\
Top inventor inflows & 0.041\White{)} & 0.039\White{)} \\
& (0.008) & (0.008)  \\
        [0.25em]
        First-stage $F$ statistic &84.781 & 75.042 \\ 
        [0.5em]
        (b) External inventors \\
         Top inventor inflows & 0.032\White{)} & 0.028\White{)} \\ 
         & (0.005) & (0.006) \\
        [0.25em]
        First-stage $F$ statistic & 84.781 & 75.042 \\
        [0.5em]
        Destination controls \\
        \quad $\ln (1 - {\rm ATR})$   & Yes & No \\
        \quad CZ FE and year FE & Yes &Yes \\
        \quad State $\times$ year FE & No &Yes  \\
        [0.5em]
        Observations & 2,760 & 2,760 \\
        Number of origin CZs & 292 & 292 \\
        \hline \hline
\end{tabular}
}
\begin{tablenotes}[flushleft]
\footnotesize
\item \hskip -.1cm \emph{Notes:} This table reports the estimates $\widehat \phi^s$ of the impact of a top inventor inflow on local patent productivity in destination commuting zone $d$, where we obtain these estimates from equivalent IV regressions regarding origin commuting zone $o$, instrumented by $I_{ot}$. The coefficient on top inventor inflows is converted to semi-elasticity. Panel (a) presents the patent productivity gains for all local inventors, whereas Panel (b) focuses on those for external inventors. Column 1 controls for $\ln (1-{\rm ATR})$, commuting zone fixed effects, and year fixed effects. Column 2 replaces $\ln(1 - {\rm ATR})$ in Column 1 with state $\times$ year FE. Exposure-robust standard errors, clustered at the commuting zone level, are reported in parentheses. All the first-stage $F$-statistics, which are derived from the equivalent IV regressions instrumented by $I_{ot}$, exceed 10, thus addressing concerns regarding weak instruments.
\end{tablenotes}
\end{table}

Table~\ref{tab:shift} presents the estimates $\widehat \phi^s$ of the impact of a top inventor inflow on local patent productivity in destination commuting zone $d$, where we obtain these estimates from equivalent IV regressions regarding origin commuting zone $o$, instrumented by $I_{ot}$. Panel (a) reports the patent productivity gains for all local inventors, whereas Panel (b) focuses on those for external inventors. Column 1 controls for $\ln (1-{\rm ATR})$, commuting zone fixed effects, and year fixed effects, and Column 2 replaces $\ln(1 - {\rm ATR})$ in \mbox{Column 1 with state $\times$ year fixed effects.}

\begin{figure}[tp]
\caption{Specification curve analysis (shift exogeneity).}
\subfigure[all local inventors]
{\includegraphics[clip, width=0.5\columnwidth]{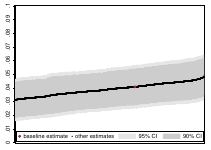}}
\subfigure[external inventors]
{\includegraphics[clip, width=0.5\columnwidth]{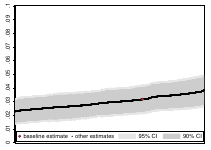}} \\
\footnotesize{\noindent
\emph{Notes:} Panels (a) and (b) illustrate the impacts of a top inventor inflow on the patent productivity of all local inventors and that of external inventors, respectively. Each specification curve is depicted using $12,288$ alternative specifications, as explained in footnote~\ref{fn_spec2}. The vertical axis is the value of $\widehat \phi^s$.}
\label{fig:spec_shift}
\medbreak
\end{figure}

Reassuringly, the estimates under this alternative identification strategy are not substantially different from our main estimates in Table~\ref{tab:tab5}. To assess the robustness of our findings obtained from the shift exogeneity assumption, we conduct a specification curve analysis as in Section~\ref{sec:sec4.2}. We employ different specifications of the IV regressions by considering various dimensions.\footnote{\label{fn_spec2}We consider (i) whether to use the ATR at the ninety-fifth or ninety-ninth percentile; (ii) whether to use $\ln(1 + Y_{dt})$ or drop commuting zones with $Y_{dt} = 0$;
(iii) whether to use the baseline or alternative detection of top inventor migrations;
(iv) whether to use the baseline or alternative definition of local inventors (see the analysis of local stayers in Section 5.3);
(v) whether to use ATRs or MTRs;
(vi) whether to include APTRs;
and (vii) whether to include each of the other controls (ATR50, CITR, ITC, RTC, manufacturing employment, and other employment variables such as ``finance and insurance,'' ``professional, scientific, and technical services,'' and ``management of companies and enterprises'' to capture other omitted variables,  e.g., access to venture capital). Since the usual caveat on weak instruments is applicable here, we adopt only specifications for which the null hypothesis of weak instruments is rejected.}
Figure~\ref{fig:spec_shift} plots the specification curve for $\widehat \phi^s$ with 90\% and 95\% confidence intervals. As seen from Panels (a) and (b), the productivity gains reported in Table~\ref{tab:shift} are fairly robust for $12,288$ alternative specifications, thus mitigating concerns about the sensitivity of our main results to different exogeneity assumptions and various choices of control variables.

\section{Mechanisms}
\label{sec:sec6}
The productivity gains estimated in the previous sections suggest that local inventors acquire knowledge from migrating top inventors, regardless of whether local inventors are internal or external. We now discuss the underlying mechanisms through which those productivity gains materialize. We first focus on patent citations that have been widely used as proxy for knowledge flows since Jaffe et al.~(1993). Specifically, we count how many times local inventors cite incoming top inventors and estimate the percentage change in the number of citations caused by top inventor inflows. Furthermore, Jaffe et al.~(1993) recognize the existence of other knowledge flows that cannot be captured by patent citations. We thus complement the foregoing analysis by using state-year variation in legal protection of trade secrets documented by Png (2017a, 2017b) as a quasi-natural experiment. We expect that knowledge flows from migrating top inventors to local external inventors would be greater in states where legal protection of trade secrets is weaker, so that there would be additional local productivity gains in those states. Our results presented below are consistent with Marshall's insight on knowledge spillovers since external inventors can not only learn patentable knowledge but also obtain other forms of knowledge from migrating top inventors as if those were in the air.

\subsection{Patent citations}
\label{sec:sec6.1}
To see the impact of top inventor inflows on local patent citations, we count how many times the patents of the top inventors who migrated into commuting zone $d$ in year $t$ were cited by the local inventors in commuting zone $d$ in year $t$ and denote it by $C_{dt}$. When constructing $C_{dt}$, we focus on the patents that had been applied over the last ten years. Replacing patent productivity $Y_{dt}$ in \eqref{eq:iv} with the number of citations $C_{dt}$, we consider the structural equation for citations as follows:
\vspace{-.1cm}
\begin{equation*}
\ln C_{dt} = \phi^{sc} M_{dt} + \xi^{sc} X_{dt} +\varepsilon^{sc}_{dt},
\vspace{-.1cm}
\end{equation*}
while retaining the same first-stage equation \eqref{eq:fs}. The coefficient $\phi^{sc}$ gauges the magnitude of knowledge flows from migrating top inventors to local inventors.

\begin{figure}[tp]
\caption{\mbox{Citations.}}
\subfigure[all local inventors]
{\includegraphics[clip, width=0.5\columnwidth]{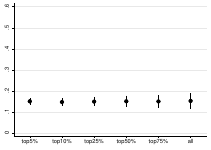}}
\subfigure[external inventors]
{\includegraphics[clip, width=0.5\columnwidth]{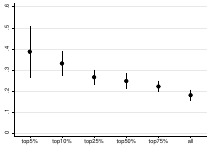}}
\footnotesize{\noindent
\emph{Notes:} Panels (a) and (b) illustrate the impact of a top inventor inflow on the number of all local inventors' citations to the incoming top inventor and that of external inventors' citations to the incoming top inventor (percentage change in decimal form), respectively. In each panel, we use $B_{dt}$ and $B_{dt}^\sigma$ as instruments.}
\label{fig:fig9}
\end{figure}

Panels (a) and (b) of Figure~\ref{fig:fig9} illustrate the estimated coefficients for the case of all local inventors and that of external inventors, respectively. In both panels, we consider the top 5\%, 10\%, 25\%, 50\%, and 75\% of local inventors according to their patent productivity and estimate the causal effect for each productivity group. In Panel (a), an additional top inventor inflow raises the number of local inventors' citations to the incoming top inventor by 10-20\% regardless of the productivity of local inventors. By contrast, in Panel (b), the external inventors, especially those with higher productivity, tend to exhibit a greater percentage change in the number of citations to the top inventors who moved in the same commuting zone. These results imply the existence of knowledge flows from the migrating top inventors to the local inventors, even when we focus on the external inventors who are not directly connected to the migrating top inventors.\footnote{It is perhaps puzzling that the impact is smaller in Panel (a). This result may stem from the possibility that internal inventors, who account for approximately 40\% of all local inventors, had already collaborated with or worked in the same organization as the incoming top inventors and thus had already cited them prior to their migration. In that case, we would expect a smaller percentage change in the number of internal inventors' citations after the arrival of the top inventors.}
This existence of external knowledge spillovers as evidenced by the citation flow from each incoming top inventor to each external inventor is related to but differs from Atkin et al.~(2022), who find that face-to-face meetings between workers in different establishments enhance between-establishment citations, as the latter aim to capture worker interactions by smartphone data while abstracting from who cites whom.

\subsection{Trade secrets}

Trade secrets were formerly defined and protected from misappropriation by common law in the United States. However, these definitions and protections have been codified into law with the enactment of federal legislation known as the Uniform Trade Secrets Act (UTSA). While most states had already adopted the UTSA, there had been substantial heterogeneity in the states' approaches to trade secrets due to the differences in the timing of the adoption of the UTSA and the strength of trade secrets protection during the common law era. We exploit the state-year variation in trade secrets protection to uncover productivity gains through knowledge flows that cannot be captured by patent citations.

Given the heterogeneity in legal protection of trade secrets, top inventors who migrate to a state with weaker protection would exchange knowledge more frequently with other inventors beyond organizational boundaries and co-inventor relationships, thereby bringing about additional productivity gains to local external inventors. In contrast, the change in legal protection of trade secrets would not influence knowledge sharing through organizations or co-inventor relationships, thus leaving the productivity gains of \mbox{local internal inventors unaffected.}

We examine those differential impacts of top inventor inflows on local patent productivity by using the state-level index of trade secrets in Png (2017a, 2017b).\footnote{Png (2017a) provides the index for the years 1979 to 1998, and Png (2017b) extends it to the years 1970 to 2010.}
This index captures both legal protection under common law and the enactment of the UTSA and ranges between 0 and 1, where a higher score implies stronger legal protection. Let $S_{\sigma(d)t}$ denote an indicator variable for whether the trade secrets index in state $\sigma(d)$ in year $t$ is below the median of the trade secrets index distribution.  If $S_{\sigma(d)t}=1$, the degree of trade secrets protection is low in commuting zone $d$ in year $t$, so that we expect higher patent productivity of external inventors due to a greater amount of knowledge brought about by top inventor inflows.

Let $\boldsymbol{\phi}^s=[\phi_{0}^s  \ \phi_{1}^s \  \phi_{2}^s]$ and $E_{dt}=[M_{dt} \ S_{\sigma(d)t} \  M_{dt}S_{\sigma(d)t} ]'$ denote a vector of coefficients and a vector of endogenous variables. The structural equation is then given by
\vspace{-.1cm}
\begin{equation}
\ln Y_{dt} = \boldsymbol{\phi}^s E_{dt} +\xi^s X_{dt} +\varepsilon_{dt}^s.
\label{eq:ts_sq}
\vspace{-.1cm}
\end{equation}
Our interest is in the coefficient $\phi_{2}^s$ on $M_{dt}S_{\sigma(d)t}$ when $ Y_{dt}$ is measured by the patent productivity of external inventors.  If $\phi_{2}^s>0$, the top inventor inflows lead to higher patent productivity of external inventors in commuting zones with weaker trade secrets protection, which suggests that the knowledge brought about by the top inventor inflows is more likely to spill over to external inventors in commuting zones with weaker trade secrets protection.

As before, we use the Bartik instruments for the top inventor inflows $M_{dt}$ to mitigate the endogeneity concern. To address the potential endogeneity of the trade secrets indicator $S_{\sigma(d)t}$, we follow Png (2017b) who argues that the enactment of the UTSA is related to the enactment of other state-level uniform laws such as the Uniform Determination of Death Act (UDDA), Uniform Federal Lien Registration Act (UFLRA), and Uniform Fraudulent Transfer Act (UFTA) because these laws were introduced to harmonize state laws. Since the three laws are unlikely to be associated with local patent productivity shocks, we use them as instruments for the trade secrets indicator. The first-stage equation that accompanies \eqref{eq:ts_sq} is thus given by
\vspace{-.1cm}
\begin{equation}
E_{dt} = \boldsymbol{\psi}^{f} Z_{dt} + \xi^{f} X_{dt}
+\varepsilon_{dt}^f,
\vspace{-.1cm}
\end{equation}
where $\boldsymbol{\psi}^{f}$ is a vector of coefficients and $Z_{dt}=[(B_{dt} , B^{\sigma}_{dt})\#\#(U_{\sigma(d)t}^{\rm DDA} , U_{\sigma(d)t}^{\rm FLRA} , U_{\sigma(d)t}^{\rm FTA} )]'$ is a vector of IVs, with $U_{\sigma(d)t}^\ell$ indicating whether uniform law $\ell = \{ {\rm DDA},  {\rm FLRA}, {\rm FTA}\}$ was in effect in state $\sigma(d)$ in year $t$, and $\#\#$ denotes an interaction-term operator.\footnote{The interaction-term operator $\#\#$ generates all possible combinations of elements for a given pair of sets. For example, let ${\mathfrak S}_1 = \{A,  B\}$ and ${\mathfrak S}_2 = \{C, D\}$, where each set ${\mathfrak S}_i$ has two elements. Then, $[{\mathfrak S}_1\#\#{\mathfrak S}_2]' = [\{A , B\}\#\#\{C, D\}]'=[A \ B \  C \  D \ AC \ AD \ BC \  BD]'$.}

\begin{figure}[tp]
\caption{Trade secrets.}
\subfigure[all local inventors]
{\includegraphics[clip, width=0.5\columnwidth]{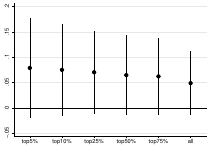}}
\subfigure[external inventors]
{\includegraphics[clip, width=0.5\columnwidth]{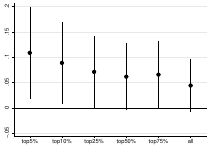}} \\
\footnotesize{\noindent
\emph{Notes:} Panels (a) and (b) illustrate the coefficients $\phi_2^s$ on the interaction term $M_{dt} S_{\sigma(d)t}$ in \eqref{eq:ts_sq} for all local inventors and for external inventors, respectively. In each panel, we use $B_{dt}$ and $B_{dt}^\sigma$ as instruments.}
\medbreak
\label{fig:fig10}
\vspace{-.3cm}
\end{figure}

Panels (a) and (b) in Figure~\ref{fig:fig10} illustrate the results for all local inventors and those for external inventors, respectively. In both panels, we consider the top 5\%, 10\%, 25\%, 50\%, and 75\% of local inventors according to their patent productivity and estimate the causal effect for each productivity group. Since the strength of trade secrets protection is unlikely to affect internal knowledge sharing within the same assignee and between co-inventors, it is not surprising that the overall impact in Panel (a) is insignificant, regardless of the productivity of local inventors. In contrast, in Panel (b) the impacts for the top 5\%, 10\%, and 25\% of inventors (top 50\% and 75\% of inventors) are significant at the 5\% (10\%) level. Hence, top inventor migration tends to enhance the patent productivity of external inventors in commuting zones with weaker trade secrets protection.

\section{A counterfactual experiment}
\label{sec:sec7}
We now conduct a counterfactual experiment. Using the baseline specification in Section~\ref{sec:sec4}, we consider what happens to the geographic distribution of patent productivity if all state individual income taxes are set to their average.\footnote{This experiment takes into account endogenous wage changes, driven by the hypothetical tax changes. To see this, one can view  \eqref{eq:inventors} and \eqref{eq:firms} as the relative labor supply and the relative labor demand, where the equilibrium relative wages are determined by their intersection. Since the changes in individual income taxes shift the relative supply curve, the counterfactual relative wages differ from the original ones.}
This experiment is useful for assessing to what extent state tax differences contribute to patent productivity differences across space.

Recalling that the changes in state taxes affect the choice probabilities $\widehat P_{odt}$ in \eqref{eq:migprob} and the Bartik instruments $B_{dt}$ in \eqref{eq:Bartik}, as well as $B_{dt}^\sigma$, the procedure of the counterfactual analysis can be summarized as follows. We first derive the counterfactual probabilities $\widetilde P_{odt}$ to construct the counterfactual Bartik instruments $\{\widetilde B_{dt}, \widetilde B_{dt}^\sigma \}$, which allow us to estimate the counterfactual top inventor flows $\widetilde M_{dt}$ via the first-stage regression \eqref{eq:fs}. We then define the counterfactual changes in the top inventor flows as $\widetilde \Delta M_{dt} = \left( \frac{\widetilde M_{dt} - \widehat M_{dt}}{\widehat M_{dt}} \right) M_{dt}$, where $M_{dt}$, $\widehat M_{dt}$, and $\widetilde M_{dt}$ are the actual, fitted, and counterfactual flows, respectively.\footnote{If the actual and fitted flows coincide, the definition reduces to $\widetilde \Delta M_{dt} =\widetilde M_{dt} - M_{dt}$. However, since the actual and fitted flows generally differ, we compute the percentage change in the top inventor flows $\left( \frac{\widetilde M_{dt} - \widehat M_{dt}}{\widehat M_{dt}} \right)$ based on the fitted and counterfactual flows in the tax-induced migration model and then multiply it by the actual flows $M_{dt}$.}
We finally apply $\widetilde \Delta M_{dt}$ to the structural equation \eqref{eq:iv} to construct the counterfactual changes in the log patent productivity $\widetilde \Delta \ln Y_{dt} = \widehat \phi^s  \widetilde \Delta M_{dt} + \widehat \xi^s \widetilde \Delta \ln  (1-{\rm ATR}_{\sigma(d)t})$, where $\widetilde \Delta \ln  (1-{\rm ATR}_{\sigma(d)t}) = \ln (1-\widetilde {\rm ATR}_{\sigma(d)t})- \ln (1-{\rm ATR}_{\sigma(d)t})$ captures the counterfactual tax changes.\footnote{When computing the counterfactual change, we replace $\widetilde \Delta \ln Y_{dt}$ with $\widetilde \Delta \ln(1 + Y_{dt} )$ as before to accommodate commuting zone $\times$ year observations with no patents.}
Thus, the overall impact of tax changes on $\widetilde \Delta \ln Y_{dt}$ can be decomposed into two: the direct effect from the tax changes, $\widehat \xi^s \widetilde \Delta \ln  (1-{\rm ATR}_{\sigma(d)t})$, and the indirect effect via the changes in top inventor flows induced by the tax changes, $\widehat \phi^s  \widetilde \Delta M_{dt}$. The indirect effect can be further decomposed into two: productivity gains due to internal knowledge sharing and those due to external knowledge spillovers.

\begin{figure}[tp]
\caption{Counterfactual experiment (setting state taxes to their average).}
\hskip 2.4cm
{\includegraphics[trim = 0 0 0 0, height=6cm, width=12cm, clip]{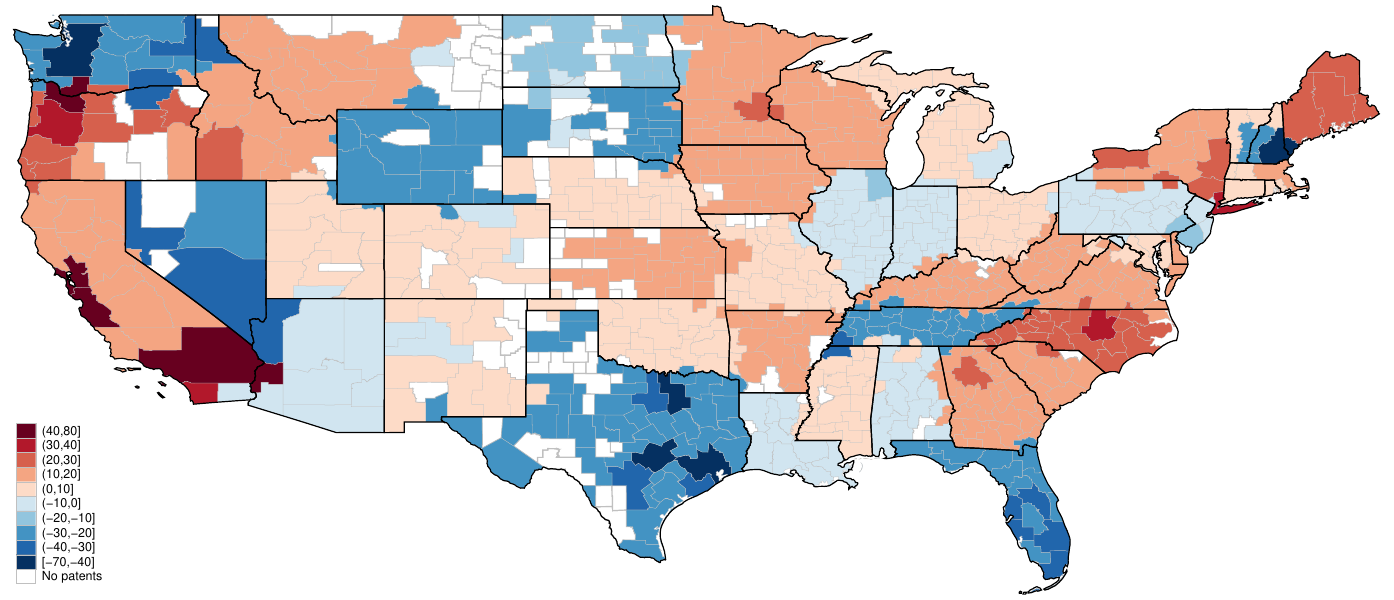}} \\
\footnotesize{\noindent
\emph{Notes:}
This figure illustrates the counterfactual percentage change in the number of patents at the commuting zone level when all state individual income taxes in 2009 are set to their average. We use $B_{dt}$ and $B_{dt}^\sigma$ as instruments.}
\label{fig:fig11}
\vspace{-.25cm}
\end{figure}

Figure~\ref{fig:fig11} illustrates the percentage change in local patent productivity when state taxes are set to their average. The overall impact tends to be positive in commuting zones in California, Oregon, North Carolina, and New York, where state taxes and initial patent productivity are high.\footnote{Notably, the counterfactual changes are heterogeneous even within states, although we equalize taxes between states. The reason is that the counterfactual choice probabilities $\widetilde P_{odt}$, which are obtained by setting state taxes equal in \eqref{eq:migprob}, include fixed effects at the commuting zone level.}
Table~\ref{tab:tab6} summarizes the top 10 commuting zones by patent productivity gains. For instance, if state taxes were equal, the number of patents in Santa Clara--Monterey--Santa Cruz (which is the commuting zone with the highest patent productivity in Table~\ref{tab:tab3}) would be larger by 72.3\%. In contrast, the overall impact tends to be negative in commuting zones in Texas, Washington, Florida, and New Hampshire, where state taxes are low and initial patent productivity is high. Table~\ref{tab:tab7} summarizes the bottom 10 commuting zones by patent productivity gains. For instance, if state taxes were equal, the number of patents in King--Pierce--Snohomish  (which is ranked as the tenth most productive commuting zone in Table~\ref{tab:tab3}) would be smaller by 64.8\%. These results suggest that the presence of state tax differences significantly distorts the spatial distribution of inventive activity.

\begin{table}[tp]
\caption{Top 10 commuting zones by patent productivity gains (\%).}
\label{tab:tab6}
{\footnotesize
\hskip 2.1cm \begin{tabular}{rrlcr}
\hline \hline
rank & cz number & counties & state & gains (\%) \\
\hline
1	&	37500	&	Santa Clara--Monterey--Santa Cruz	&	CA	&	72.291	\\
2	&	37800	&	Alameda--Contra Costa--San Francisco	&	CA	&	53.377	\\
3	&	38300	&	Los Angeles--Orange--San Bernardino	&	CA	&	47.115	\\
4	&	38801	&	Multnomah--Washington--Clackamas	&	OR	&	46.022	\\
5	&	38000	&	San Diego	&	CA	&	38.042	\\
6	&	1701	&	Wake--Durham--Orange	&	NC	&	35.253	\\
7	&	19400	&	Kings--Queens--New York	&	NY	&	35.080	\\
8	&	38901	&	Lane--Marion--Linn	&	OR	&	31.310	\\
9	&	35801	&	Ada--Canyon--Elmore	&	ID	&	29.754	\\
10	&	39203	&	Deschutes--Crook--Jefferson	&	OR	&	29.206	\\
\hline
\hline
\end{tabular}
}
\begin{tablenotes}[flushleft]
\footnotesize
\item
\hskip -.1cm \emph{Notes:} Patent productivity gains are defined as the percentage change in the number of patents when all state individual income taxes in 2009 are set to their average. We use $B_{dt}$ and $B_{dt}^\sigma$ as instruments.
\end{tablenotes}
\vspace{-.1cm}
\end{table}
\begin{table}[tp]
\caption{Bottom 10 commuting zones by patent productivity gains (\%).}
\label{tab:tab7}
{\footnotesize
\hskip 2.5cm \begin{tabular}{rrlcr}
\hline \hline
rank & cz number & counties & state & gains (\%) \\
\hline
1	&	39400	&	King--Pierce--Snohomish	&	WA	&	-64.777	\\
2	&	31201	&	Travis--Williamson--Hays	&	TX	&	-50.710	\\
3	&	32000	&	Harris--Fort Bend--Galveston	&	TX	&	-49.002	\\
4	&	33100	&	Dallas--Denton--Collin	&	TX	&	-46.491	\\
5	&	20600	&	Hillsborough--Rockingham--York	&	NH	&	-41.326	\\
6	&	7100	&	Palm Beach--St. Lucie--Martin	&	FL	&	-38.194	\\
7	&	7400	&	Orange--Seminole--Lake	&	FL	&	-35.219	\\
8	&	5202	&	Shelby--DeSoto--Tipton	&	TN	&	-35.020	\\
9	&	6900	&	Sarasota--Manatee--Charlotte	&	FL	&	-34.585	\\
10	&	7000	&	Dade--Broward--Monroe	&	FL	&	-34.561	\\
\hline
\hline
\end{tabular}
}
\begin{tablenotes}[flushleft]
\footnotesize
\item \hskip -.1cm \emph{Notes:} Patent productivity gains are defined as the percentage change in the number of patents when all state individual income taxes in 2009 are set to their average. We use $B_{dt}$ and $B_{dt}^\sigma$ as instruments.
\end{tablenotes}
\end{table}

To see which states are most affected by the presence of tax differences, we first define, for each commuting zone $d$, the counterfactual change in the number of patents $\widetilde \Delta Y_{dt} = \left( \frac{\widetilde Y_{dt} - \widehat Y_{dt}}{\widehat  Y_{dt}} \right) Y_{dt}$ in the same way as $\widetilde \Delta M_{dt}$, where $Y_{dt}$, $\widehat Y_{dt}$, and $\widetilde Y_{dt}$ are the actual, fitted, and counterfactual numbers of patents in $d$, respectively. We then aggregate $\widetilde \Delta Y_{dt}$ within each state $\sigma$ to obtain the counterfactual changes in the number of patents $\widetilde \Delta Y_{\sigma t}=\sum_{d \in \sigma} \widetilde \Delta Y_{dt}$. Denoting by $Y_{\sigma t} = \sum_{d \in \sigma} Y_{dt}$ the actual number of patents at the state level, we finally compute the percentage change in the number of patents at the state level $\frac{\widetilde \Delta Y_{\sigma t}}{Y_{\sigma t}}$.\footnote{Observe that $\frac{\widetilde \Delta Y_{\sigma t}}{Y_{\sigma t}}= \frac{\sum_{d \in \sigma} \widetilde \Delta Y_{dt}}{\sum_{d \in \sigma} Y_{dt}}=\sum_{d \in \sigma} \left( \frac{\widetilde Y_{dt} - \widehat Y_{dt}}{\widehat  Y_{dt}} \right) \frac{Y_{dt}}{\sum_{d \in \sigma} Y_{dt}}$. The latter is the weighted average of the percentage change $\left( \frac{\widetilde Y_{dt} - \widehat Y_{dt}}{\widehat  Y_{dt}} \right)$ with weight being the share of actual number of patents $\frac{Y_{dt}}{\sum_{d \in \sigma} Y_{dt}}$. If the actual and fitted numbers of patents coincide, the percentage change in the number of patents at the state level reduces to $\frac{\widetilde \Delta Y_{\sigma t}}{Y_{\sigma t}} = \frac{ \sum_{d \in \sigma} \widetilde Y_{dt} -  \sum_{d \in \sigma} Y_{dt} }{\sum_{d \in \sigma} Y_{dt}} $.}

\begin{figure}[t]
\caption{\mbox{Counterfactual experiment (setting state taxes to their average).}}
\vskip -0.4cm
\hskip 2.5cm
\includegraphics[clip, width=.7\columnwidth]{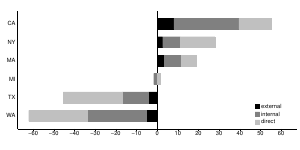} \\
\vskip -0.9cm
\footnotesize{\noindent
\emph{Notes:}
This figure illustrates the counterfactual percentage change in the number of patents for the states of California, New York, Massachusetts, Michigan, Texas, and Washington when all state individual income taxes in 2009 are set to their average. The overall change for each state is decomposed into the direct effect via the change in state taxes and the indirect effect via the tax-induced top inventor migration. The latter consists of the internal knowledge sharing effect and the external knowledge spillover effect. We use $B_{dt}$ and $B_{dt}^\sigma$ as instruments.}
\label{fig:fig12}
\vspace{-.2cm}
\end{figure}

Figure~\ref{fig:fig12} illustrates the percentage change in the number of patents for selected states when state taxes are set to their average. For instance, if state taxes were equal, the number of patents in California (where state taxes and patent productivity are high) would be greater by $55.1\%$, which can be decomposed into the direct effect via the reduction in California state taxes ($15.6\%$) and the indirect effect via the tax-induced top inventor migration ($39.5\%$). The indirect effect can be further decomposed into the internal sharing effect ($31.4\%$) and the external spillover effect ($8.1\%$). In contrast,  the number of patents in Texas (where state taxes are low and patent productivity is high) would be smaller by $45.3\%$, which can be decomposed into the direct effect via the rise in Texas state taxes ($28.5\%$) and the indirect effect via the tax-induced migration  ($16.8\%$). The indirect effect can be further decomposed into the internal sharing effect ($12.7\%$) and the external spillover effect ($4.1\%$).

These results suggest that the indirect effect via the tax-induced migration of top inventors can be substantial. To see the relative importance of the direct and indirect effects on patent productivity at the national level, we aggregate those changes in the number of patents across all commuting zones in all states that we consider in the paper. We find that the share of the indirect effect is $0.725$, while that of the direct effect is $0.275$. Our results thus complement Akcigit et al. (2022) who assess the direct impact of state taxes on innovation.

\section{Concluding remarks}
In this paper, we have uncovered the idea-generating process described by Marshall (1890) using Bartik (1991) instruments. We have identified a significant causal effect of a top inventor inflow on the patent productivity of all local inventors. Even when we focus on local external inventors who are not directly connected to incoming top inventors through organizations or co-inventor relationships, the effect remains significant and is approximately 4\%, thus implying that the mysteries of the trade are in the air.

We have disentangled productivity gains due to external knowledge spillovers from those due to internal knowledge sharing. Thus, our findings are consistent with the partially nonexcludable good nature of knowledge, whose implications have been explored theoretically in the technology and growth literature. Since the existence of the productivity gains from external knowledge spillovers leads to market failures and constitutes a rationale for spatial agglomeration of inventive activity, our analysis would be useful for innovation policies that consider both the benefits and costs of entrepreneurial clusters.

Our counterfactual experiment suggests that the presence of tax differences across states distort the spatial distribution of inventive activity up to $-64.8$\% to $72.3$\%, with considerable spatial heterogeneity. The decomposition of those gains and losses reveals that not only the direct gains from tax changes but also the indirect gains via the top inventor migration driven by tax changes are important.

While this paper has considered the tax-induced domestic migration of top inventors, our model-based Bartik instruments can be used in any setting where origin-destination flows are affected by changes in location-specific policies. Thus, our framework would be applicable to various settings where the movement of goods, people, and ideas across space is influenced by policy differences between locations.

{

}

\newpage
\clearpage

\pagenumbering{Roman}
\setcounter{page}{1}
\appendix

\section*{Online Appendix}

\renewcommand{\thesection}{Appendix \Alph{section}}
\renewcommand{\thesubsection}{A.\arabic{subsection}}
\renewcommand{\thefigure}{A\arabic{figure}}
\setcounter{figure}{0}
\renewcommand{\thetable}{A\arabic{table}}
\setcounter{table}{0}

\titleformat*{\section}{\large\bfseries}
\titleformat*{\subsection}{\normalsize\bfseries}
\vspace{-.25cm}
\section{Data appendix}
\label{sec:secA}
\vspace{-.25cm}
\subsection{Data sources and construction}
\label{sec:A1}
\vspace{-.25cm}
\paragraph{Patent data.}
The main data come from USPTO PatentsView (\url{https://patentsview.org/}). It includes data on patents, inventors, inventors' addresses, assignees, and patent citations and provides data files regarding the disambiguation of inventor and assignee names (\url{https://patentsview.org/disambiguation/}). Additional procedures are used to allocate inventors' addresses to commuting zones. We first use the latitude and longitude of each inventor's address (which are taken from USPTO PatentsView) to identify his/her county of residence. We then relate it to the commuting zone in which the inventor resides based on the correspondence table between counties and commuting zones in 1990 on the IPUMS USA website (\url{https://usa.ipums.org/usa/volii/1990lma.shtml}).

The disambiguation algorithm adopted in PatentsView is known to be highly accurate, as it copes with the two problems involving false positives and negatives pointed out by Trajtenberg et al.~(2006) and Monath et al.~(2021). One is multiple names for the same entity (assignee, inventor, or location), e.g., the spelling of an assignee's name may differ from one patent to another. The other is multiple different entities with the same name, e.g., different inventors might share the exact same name (also known as the ``John Smith'' problem).

However, the disambiguation process is not error-free (see Toole et al., 2021). To address the former problem, we check the most typical assignee names: International Business Machines Corporation and IBM Corporation. In our dataset, the number of applications for International Business Machines Corporation is $207,139$, whereas that for IBM Corporation is $1,074$, which implies that $\frac{207,139}{207,139 + 1,074} \times 100 \approx 99.484$\% of the IBM-related applications are classified into the same assignee. The latter problem implies that different inventors with the same name might be mistakenly recognized as the same inventor, which could generate seemingly frequent moves. To cope with possible overdetection of moves, we focus on the top inventors who moved fewer than eight times. This still leaves us with $59,770$ out of $60,294$ unique top inventors---more than $\frac{59,770}{60,294} \times 100 \approx 99.131$\% of the inventors who applied for patents between two consecutive years \emph{and} who qualified as top inventors at least once.

\paragraph{State taxes and tax credits.}

Data on U.S. state taxes and tax credits for 1976-2019 are obtained from openICPSR (\url{https://www.openicpsr.org/openicpsr/project/113057/version/V1/view}). Summary statistics are presented in the Online Appendix for Moretti and Wilson (2017).

\paragraph{Employment data.}

The employment data are taken from the County Business Patterns (CBP) database (\url{http://fpeckert.me/cbp/}). Eckert et al.~(2021) provide a detailed description of the data. Since the original employment data are at the county level, we aggregate them at the commuting zone level. We use the 2012 NAICS codes 31-33 to obtain the number of employees for manufacturing, code 52 for finance and insurance, code 54 for professional, scientific, and technical services, and code 55 for management of companies and enterprises.

\paragraph{Trade secrets index and state-level uniform laws data.}

The trade secrets index is compiled by Png (2017a, 2017b). Each state has six binary scores regarding the strength of legal protection of trade secrets under the common law and the Uniform Trade Secrets Act (UTSA). The trade secrets index used in our analysis is the sum of the six scores divided by six, which takes a value between zero and one. Data on state-level uniform laws such as the Uniform Determination of Death Act (UDDA), Uniform Federal Lien Registration Act (UFLRA), and Uniform Fraudulent Transfer Act (UFTA) are provided by Png (2017b).

\paragraph{Data for state tax competition analysis.}

Data on the socio-politico-economic characteristics used in the state tax competition analysis are taken from multiple sources. The data on population for various age and race groups are from the ``U.S. Intercensal County Population Data by Age, Sex, Race, and Hispanic Origin'' web page (\url{https://www.nber.org/research/data/us-intercensal-county-population-data-age-sex-race-and-hispanic-origin}) \linebreak
operated by NBER. The key economic and state finance data are from the ``State Economic and Government Finance Data'' web page (\url{https://doi.org/10.7910/DVN/CJBTGD}) provided by Klarner (2015). The political party affiliation of each state governor is from the National Governors Association web page (\url{https://www.nga.org/governors/}).

\newpage

\subsection*{A.2 Other summary statistics}
\vskip -.5cm
\begin{table}[hbtp]
\caption{Summary statistics at the commuting zone level (other variables).}
\label{tab:tabA1}
{\footnotesize
\hskip 2.1cm \begin{tabular}{lrrrr}
\hline\hline
 & \multicolumn{1}{c}{mean} & \multicolumn{1}{c}{sd} & \multicolumn{1}{c}{min} & \multicolumn{1}{c}{max} \\
\hline
ATR & 0.238  & 0.030  & 0.164  & 0.330  \\
ATR99 & 0.315  &  0.032 & 0.244 & 0.410 \\
ATR50 & 0.108  &  0.027 & 0.033 & 0.169 \\
CITR & 0.064  & 0.027  & 0.000  & 0.138  \\
ITC & 0.009  & 0.023  & 0.000  & 0.100  \\
RTC & 0.022  & 0.038  & 0.000  & 0.250  \\
TSI	&	0.340	&	0.235	&	0.000	&	0.767	\\
UDDA       &         0.490     &   0.500 &       0.000   &     1.000 \\
UFLRA       &        0.596     &   0.491    &    0.000    &    1.000 \\
UFTA         &       0.522   &     0.500     &   0.000   &     1.000 \\
MFG &	22,398.524	&	61,797.154	&	0.000	&	1,152,493.572	\\
FIN     &    7,304.904  &  27,786.213   &     0.000 &  541,668.019 \\
PRO    &     7,342.944 &   31,249.756    &    0.000 &  616,664.500 \\
MNG   &      3,463.924  &  13,608.328     &   0.000  & 168,097.910 \\
CITES ALL	&	21.966	&	347.065	&	0.000	&	17,344.000	\\
CITES EXT	&	2.098	&	37.218	&	0.000	&	2,215.000	\\
\hline
Number of observations& \multicolumn{4}{r}{23,628}  \\
Number of commuting zones  & \multicolumn{4}{r}{716} \\
Number of years &  \multicolumn{4}{r}{33}  \\
\hline\hline
\end{tabular}}
\begin{tablenotes}[flushleft]
\footnotesize
\item \hskip -.1cm \emph{Notes:} Summary statistics are based on the data described in Section~\ref{sec:sec2} for the years 1977 to 2009. ATR (ATR99, ATR50), CITR, ITC, RTC, TSI, UDDA, UFLRA, and UFTA stand for the individual income average tax rate at the ninety-fifth (ninety-ninth,  fiftieth) percentile of the U.S. income distribution, corporate income tax rate, investment tax credits, R\&D tax credits, trade secrets index, Uniform Determination of Death Act, Uniform Federal Lien Registration Act, and Uniform Fraudulent Transfer Act, respectively. MFG, FIN, PRO, and MNG denote the employment in ``manufacturing,'' ``finance and insurance,'' ``professional, scientific, and technical services,'' and ``management of companies and enterprises.'' CITES ALL and CITES EXT are the number of citations by all local inventors and the number of citations by external inventors.
\end{tablenotes}
\vspace{-.1cm}
\end{table}

\begin{table}[hbtp]
\caption{Summary statistics at the state level (other variables).}
\label{tab:tabA2}
{\footnotesize
\hskip 4cm
\begin{tabular}{lrrrr}
\hline\hline
 & \multicolumn{1}{c}{mean} & \multicolumn{1}{c}{sd} & \multicolumn{1}{c}{min} & \multicolumn{1}{c}{max} \\
\hline
ATR	&	0.240	&	0.030	&	0.164	&	0.330	\\
ATR99 & 0.317  &      0.032  &      0.244  &      0.410 \\
ATR50 & 0.108   &     0.026    &    0.033    &    0.169 \\
CITR	&	0.067	&	0.028	&	0.000	&	0.138	\\
ITC	&	0.009	&	0.022	&	0.000	&	0.100	\\
RTC	&	0.024	&	0.044	&	0.000	&	0.250	\\
TSI	&	0.339	&	0.227	&	0.000	&	0.767	\\
UDDA          &      0.522     &   0.500     &   0.000     &   1.000 \\
UFLRA        &       0.580  &      0.494     &   0.000      &  1.000 \\
UFTA           &     0.522       & 0.500&        0.000      &  1.000 \\
\hline
Number of observations& \multicolumn{4}{r}{1,584}  \\
Number of states  & \multicolumn{4}{r}{48} \\
Number of years &  \multicolumn{4}{r}{33}  \\
\hline\hline
\end{tabular}}
\begin{tablenotes}[flushleft]
\footnotesize
\item \hskip -.1cm
\emph{Notes:} Summary statistics are based on the data described in Section~\ref{sec:sec2} for the years 1977 to 2009. ATR (ATR99, ATR50), CITR, ITC, RTC, TSI, UDDA, UFLRA, and UFTA stand for the individual income average tax rate at the ninety-fifth (ninety-ninth, fiftieth) percentile of the U.S. income distribution, corporate income tax rate, investment tax credits, R\&D tax credits, trade secrets index, Uniform Determination of Death Act, Uniform Federal Lien Registration Act, and Uniform Fraudulent Transfer Act, respectively.
\end{tablenotes}
\vspace{-3cm}
\end{table}

\renewcommand{\thesection}{Appendix \Alph{section}}
\renewcommand{\thesubsection}{B.\arabic{subsection}}
\renewcommand{\thefigure}{B\arabic{figure}}
\setcounter{figure}{0}
\renewcommand{\thetable}{B\arabic{table}}
\setcounter{table}{0}

\clearpage
\newpage

\section{Derivation}
\subsection{Derivation of equation \eqref{eq:eqm}}
\label{sec:secB1}

To derive \eqref{eq:eqm}, we first solve \eqref{eq:firms} for $\ln w_{dt} - \ln w_{ot}$ as follows
\begin{equation*}
\ln w_{dt} - \ln w_{ot}  = \beta [ \ln (1-\tau'_{\sigma(d)t})  -  \ln (1-\tau'_{\sigma(o)t})  ]
+ [ Z'_d -Z'_o ] - C'_{od} - \ln (P'_{odt}/P'_{oot}).
\end{equation*}
Plugging this expression into  \eqref{eq:inventors} and setting $\ln (P'_{odt}/P'_{oot}) = \ln (P_{odt}/P_{oot})$, we obtain
\begin{eqnarray*}
\ln (P_{odt}/P_{oot}) &=&  \alpha [ \ln (1-\tau_{\sigma(d)t})  -  \ln (1-\tau_{\sigma(o)t})  ] \\
&& +\alpha \{
\beta [ \ln (1-\tau'_{\sigma(d)t})  -  \ln (1-\tau'_{\sigma(o)t})  ] 
+ [ Z'_d -Z'_o ] - C'_{od} - \ln (P_{odt}/P_{oot}) \}
\\
&& + [ Z_d -Z_o ] - C_{od},
\end{eqnarray*}
which yields
\begin{eqnarray*}
(1+\alpha) \ln (P_{odt}/P_{oot}) &=&  \alpha [ \ln (1-\tau_{\sigma(d)t})  -  \ln (1-\tau_{\sigma(o)t})  ] \\
&& +\alpha \{ \beta [ \ln (1-\tau'_{\sigma(d)t})  -  \ln (1-\tau'_{\sigma(o)t})  ] 
+ [ Z'_d -Z'_o ] - C'_{od} \}
\\
&& + [ Z_d -Z_o ] - C_{od}.
\end{eqnarray*}
We thus have
\begin{eqnarray*}
\textstyle \ln (P_{odt}/P_{oot}) &=&  \textstyle \frac{\alpha}{1+\alpha} [ \ln (1-\tau_{\sigma(d)t})  -  \ln (1-\tau_{\sigma(o)t})  ]
+  \frac{\alpha \beta}{1+\alpha}  [ \ln (1-\tau'_{\sigma(d)t})  -  \ln (1-\tau'_{\sigma(o)t})  ]  \notag \\
&& \textstyle + \frac{1}{1+\alpha} [ Z_d -Z_o ]  + \frac{\alpha}{1+\alpha} [  Z'_d -  Z'_o ]
 -  \frac{1}{1+\alpha} [ C_{od} + \alpha C'_{od}   ]   \\
 &=& \textstyle \frac{\alpha}{1+\alpha} [ \ln (1-\tau_{\sigma(d)t})  -  \ln (1-\tau_{\sigma(o)t})  ]
+  \frac{\alpha \beta}{1+\alpha}  [ \ln (1-\tau'_{\sigma(d)t})  -  \ln (1-\tau'_{\sigma(o)t})  ]  \notag \\
&& \textstyle + \frac{1}{1+\alpha} [ Z_d +\alpha Z'_d ]  - \frac{1}{1+\alpha} [  Z_o  + \alpha  Z'_o ]
 -  \frac{1}{1+\alpha} [ C_{od} + \alpha C'_{od} ].
\end{eqnarray*}
Setting $\eta =  \frac{\alpha}{1+\alpha} $, $\eta'= \frac{\alpha \beta}{1+\alpha}$,
$\gamma_d= \frac{1}{1+\alpha} [ Z_d +\alpha Z'_d ]$, $\gamma_o= - \frac{1}{1+\alpha} [  Z_o  + \alpha  Z'_o ]$, and $\gamma_{od}= -\frac{1}{1+\alpha} [ C_{od} + \alpha C'_{od} ]$, the foregoing equation can be rewritten as
\begin{eqnarray}
\ln (P_{odt}/P_{oot})
&=&  \eta [ \ln (1-\tau_{\sigma(d)t})  -  \ln (1-\tau_{\sigma(o)t}) ] +  \eta' [ \ln (1-\tau'_{\sigma(d)t})  -  \ln (1-\tau'_{\sigma(o)t})  ]  \notag \\
&&  + \gamma_d  + \gamma_o +\gamma_{od}.
\label{eq:logodds_pre}
\end{eqnarray}
Adding an error term $u_{odt}$ to the right-hand side of \eqref{eq:logodds_pre}, we obtain the expression in \eqref{eq:eqm}.

\newpage
\subsection{Derivation of equation \eqref{eq:migprob}}
\label{sec:secB2}
We derive the predicted probability $\widehat{P}_{odt}$ in \eqref{eq:migprob} that top inventors migrate from $o$ to $d$ in year $t$ by the following three steps. First, equation \eqref{eq:logodds_pre}
implies that for any pair of commuting zones $c$ and $d$, ${P}_{oct}$ and ${P}_{odt}$ must satisfy
\begin{equation*}
\frac{ P_{oct}}{P_{odt}} = \frac{
 \exp \{  \eta  \ln (1-\tau_{\sigma(c)t}) +   \eta'  \ln (1-\tau'_{\sigma(c)t})  +  \gamma_c
 + \gamma_{oc} \}  }{ \exp \{   \eta  \ln (1-\tau_{\sigma(d)t})  +    \eta'  \ln (1-\tau'_{\sigma(d)t})    +  \gamma_d 
 + \gamma_{od} \} }.
\end{equation*}
Second, let ${\cal C}$ denote the set of \emph{all} commuting zones including origin commuting zone $o$ and destination commuting zone $d$. Since $\sum_{c \in  {\cal C}} P_{oct} = P_{oot}+P_{odt}+ \sum_{c \in  {\cal C}, c\neq \{o,d\}} P_{oct}  = 1$ holds, we have
\vspace{-.1cm}
\begin{equation*}
\frac{\sum_{c \in  {\cal C}} P_{oct}}{P_{odt}} = \frac{1}{P_{odt}} = \frac{\sum_{c \in {\cal C}}
 \exp \{  \eta  \ln (1-\tau_{\sigma(c)t}) +   \eta'  \ln (1-\tau'_{\sigma(c)t})  +  \gamma_c
 + \gamma_{oc} \}  }{ \exp \{   \eta  \ln (1-\tau_{\sigma(d)t})  +    \eta'  \ln (1-\tau'_{\sigma(d)t})    +  \gamma_d  + \gamma_{od} \} },
 \vspace{-.1cm}
\end{equation*}
so that
\vspace{-.1cm}
\begin{equation*}
P_{odt} = \frac{ \exp \{   \eta  \ln (1-\tau_{\sigma(d)t})  +    \eta'  \ln (1-\tau'_{\sigma(d)t})    +  \gamma_d  + \gamma_{od} \} }{\sum_{c \in {\cal C}}
 \exp \{  \eta  \ln (1-\tau_{\sigma(c)t}) +   \eta'  \ln (1-\tau'_{\sigma(c)t})  +  \gamma_c
 + \gamma_{oc} \}}.
 \vspace{-.1cm}
\end{equation*}
Finally, replacing the parameters with the estimates,
\mbox{$\widehat \eta$, $\widehat \eta'$, and $\{\widehat \gamma_c, \widehat \gamma_{oc} \}_{c \in \cal C}$, from \eqref{eq:eqm} yields \eqref{eq:migprob}.}

\renewcommand{\thesection}{Appendix \Alph{section}}
\renewcommand{\thesubsection}{C.\arabic{subsection}}
\renewcommand{\thefigure}{C\arabic{figure}}
\setcounter{figure}{0}
\renewcommand{\thetable}{C\arabic{table}}
\setcounter{table}{0}

\section{State tax competition}
\label{app:taxc}

We examine the possibility of strategic interactions among state governments by estimating a reaction function, where the income tax in one state responds to the income taxes in other states as follows (see, e.g., Brueckner, 2003):
\vspace{-.1cm}
\begin{equation}\label{eq:tax_comp}
\textstyle \tau_{\sigma t} = \rho\sum_{\sigma'\neq \sigma}\omega_{\sigma \sigma'}\tau_{\sigma't}
+\boldsymbol{\beta} \mathbf{X}_{\sigma t-1} + \chi_{\sigma} + \chi_{t} + \varepsilon_{\sigma t}.
\vspace{-.1cm}
\end{equation}
The dependent variable $\tau_{\sigma t}$ is the tax rate in state $\sigma$ in year $t$, and the term $\sum_{\sigma'\neq \sigma}\omega_{\sigma \sigma'}\tau_{\sigma' t}$ on the right-hand side is the weighted sum of the tax rates in the neighboring states with the weight being $\omega_{\sigma \sigma'}$. $\mathbf{X}_{\sigma t-1}$ denotes a vector of socio-politico-economic characteristics for state $\sigma $ in year $t-1$ (see Table \ref{tax_comp_vars}).\footnote{When the state governments choose a tax rate, they observe the information on the socio-politico-economic conditions in the previous year. We thus use the lagged variables $\mathbf{X}_{\sigma t-1}$.}
$\chi_{\sigma}$ and $\chi_{t}$ stand for state and year fixed effects, respectively, and $\varepsilon_{\sigma t}$ is the error term.

\begin{table}[htp]
    \caption{Summary statistics (state tax competition).}
    \label{tax_comp_vars}
    \footnotesize{
    \hskip 1.25cm
 \begin{tabular}{lrrrrr} 
    \hline\hline
        & \multicolumn{1}{c}{mean} & \multicolumn{1}{c}{sd} & \multicolumn{1}{c}{min} & \multicolumn{1}{c}{max} & \multicolumn{1}{c}{obs} \\ \hline
        log patent productivity & 6.285 & 1.422 & 2.881 & 10.201 & 1,617 \\ 
        log population & 15.011 & 1.009 & 12.951 & 17.430 & 1,617 \\ 
        share of black or African American population & 0.114 & 0.121 & 0.002 & 0.705 & 1,617 \\ 
        share of population younger than 20 & 0.292 & 0.028 & 0.205 & 0.414 & 1,617 \\ 
        share of population older than 64 & 0.124 & 0.018 & 0.075 & 0.184 & 1,617 \\ 
        unemployment rate & 5.919 & 2.046 & 2.342 & 17.350 & 1,617 \\ 
        log total income & 18.015 & 1.185 & 15.010 & 21.186 & 1,617 \\ 
        log gross state product & 18.191 & 1.170 & 15.106 & 21.403 & 1,617 \\ 
        democrat governor (dummy variable) & 0.511 & 0.500 & 0.000 & 1.000 & 1,617 \\ 
	log tax revenue & 15.291     &   1.159  & 12.318 & 18.581 & 1,584 \\ 
        log government debt & 15.302    &    1.253 & 10.872 & 18.819 & 1,584 \\ 
        log government revenue & 16.110    &    1.134 & 13.301 & 19.511 & 1,584 \\ 
        \hline\hline \end{tabular}
        }
\begin{tablenotes}[flushleft]
\footnotesize
\item \hskip -.1cm
\emph{Notes:}
Summary statistics are based on the data described in Appendix~\ref{sec:A1} for the years 1977 to 2009.
\end{tablenotes}
\vspace{-.2cm}
\end{table}

Our null hypothesis is that $\rho=0$. If $\rho \neq 0$ were to hold, there would be strategic tax competition between state governments, which would induce interstate correlation between taxes and productivity. In that case, the share exogeneity would be violated.

When estimating the value of $\rho$ in \eqref{eq:tax_comp}, we address the potential endogeneity arising from the simultaneous determination of state taxes, which leads to the main regressor $\sum_{\sigma'\neq \sigma}\omega_{\sigma \sigma'}\tau_{\sigma't}$ being correlated with the error term $\varepsilon_{\sigma t}$. Specifically, we follow Kelejian and Prucha (1998) and use the weighted sum of neighboring states' socio-politico-economic characteristics as an instrument. Let $\mathbf{X}_{\sigma t-1}= [x_{1, \sigma t-1},\cdots,x_{K,\sigma t-1} ]'$, where $x_{k,\sigma t-1}$ is the $k$-th characteristic. We first generate the weighted sum $\sum_{\sigma'\neq \sigma}\omega_{\sigma \sigma'} x_{k,\sigma' t-1}$ for each characteristic \{$x_{k,\sigma t-1}\}_{k=1}^K$. These $K$ instruments are then used to estimate the predicted value of the weighted sum of neighboring states' tax rates,  $\sum_{\sigma'\neq \sigma}\omega_{\sigma \sigma'}\tau_{\sigma' t}$, in the first-stage regression.

We consider several different weights and examine the sensitivity of the estimates. In the baseline cases, we use the following two types of weights. One is the first-order contiguity weight: $\omega_{\sigma \sigma'}=1$ if state $\sigma'$ is contiguous with state $\sigma$ and $\omega_{\sigma \sigma'}=0$ otherwise. The other is constructed such that it is proportional to top inventor flows from state $\sigma'$ to state $\sigma$.\footnote{\label{fn46}In the specification curve analysis below, we consider three other weights: (i) the second-order contiguity weight, i.e., $\omega_{\sigma  \sigma'}=1$ if state $\sigma'$ is contiguous with state $\sigma$ or is contiguous with the states that are contiguous with state $\sigma$; (ii) the inverse-distance weight, i.e., $\omega_{\sigma  \sigma'}$ is inversely related to the geographical distance between $\sigma$ and $\sigma'$; and (iii) the inverse-distance weight with a cutoff distance, i.e., the interstate effect is assumed to be zero beyond 1000 miles. When estimating \eqref{eq:tax_comp}, all weights are normalized such that $\sum_{ \sigma' \neq  \sigma}\omega_{ \sigma  \sigma'}=1$ for any $\sigma$.}

\begin{table}[htp]
\caption{State tax competition.}
\label{tab:tabD2}
    {\footnotesize
        \hskip 1.75cm
  \begin{tabular}{lrrrr}
\hline \hline				
	&	\multicolumn{1}{c}{(1)}	&	\multicolumn{1}{c}{(2)}	&	\multicolumn{1}{c}{(3)}	&	\multicolumn{1}{c}{(4)} \\  \hline
	$\sum_{\sigma'\neq \sigma}\omega_{\sigma \sigma'}\tau_{\sigma't}$	&	 -0.163    &       0.027    &      -1.364    &      -1.071   	\\
	&	(0.314)      &   (0.260)  &       (0.822)   &      (0.797)   	\\
log patent productivity	&		&	0.005	&		&	0.005	\\
	&		&	(0.003)	&		&	(0.003)	\\
log population	&	0.016    &       0.016     &      0.017    &       0.013	\\
	&	(0.038)   &      (0.033)   &      (0.029)      &   (0.029)  	\\
share of black or African American population	&	0.193 &         0.236 &         0.111        &   0.165   	\\
	&	 (0.107)      &   (0.117)      &   (0.120)    &     (0.124)   	\\
share of population younger than 20 &	-0.175   &       -0.160   &       -0.192   &       -0.167  	\\
	&	(0.179)    &     (0.168)      &   (0.165)       &  (0.160)   	\\
share of population older than 64	&	 0.024    &       0.029     &      0.051  &         0.088 	\\
	&	(0.287)   &      (0.249)  &       (0.222)    &     (0.212)   	\\
unemployment rate	&	 -0.000     &     -0.000 &         -0.000     &     -0.000   	\\
	&	(0.001)	&	(0.001)	&	(0.001)	&	(0.000)	\\
log total income	&	0.003      &    -0.005    &      -0.009   &       -0.013   	\\
	&	(0.032)      &   (0.031)    &     (0.028) &        (0.029)   	\\
log gross state product	&	0.029     &     0.028   &       0.036    &     0.036	\\
	&	 (0.015)      &   (0.016)    &     (0.014) &        (0.014)   	\\
democrat governor	&	0.001	&	0.002	&	0.001	&	0.002	\\
	&	(0.001)    &     (0.001)    &     (0.001)     &    (0.001) 	\\
log tax revenue	&	-0.024 &         -0.027 &        -0.023 &        -0.025 	\\
	&	(0.012)   &      (0.011)    &     (0.011)    &     (0.010) 	\\
log government debt	&	0.000  &        -0.000     &     -0.000   &       -0.000	\\
	&	(0.002)	&	(0.002)	&	(0.002)	&	(0.002)	\\
log government revenue	&	-0.002      &     0.001       &   -0.002      &    -0.000  	\\
	&	(0.004)     &    (0.003)     &    (0.004)    &     (0.003) 	\\
	 \\
Effective $F$ statistic & 15.924    &   18.517 &      50.742   &    50.307 \\
$\tau=5\%$ & 26.392    &   26.689 &      25.159  &     25.143 \\
$\tau=10\%$	&	15.173   &    15.368    &   14.444   &    14.355	\\
$\tau=20\%$	&	9.191    &    9.320    &    8.745     &   8.639	\\
$\tau=30\%$	&	7.055   &     7.156      &  6.715  &      6.612	\\
 \\
Observations	&	1,584       &    1,584     &      1,584    &       1,584	\\
        \hline\hline \end{tabular}
        }
\begin{tablenotes}[flushleft]
\footnotesize
\item \hskip -.1cm
\emph{Notes:}
Columns 1-2 (Columns 3-4) use the first-order contiguity (top-inventor-inflow) weights. Columns 2
and 4 include the lagged own-state patent productivity $\ln Y_{\sigma t-1}$.
\end{tablenotes}
\end{table}

Table~\ref{tab:tabD2} reports the regression results for the sample period 1977-2009.\footnote{We proxy $\tau_{\sigma t}$ by the ATR at the ninety-fifth percentile in state $\sigma$ in year $t$. We drop Washington D.C. from the sample because some state characteristics are unavailable.}
Columns 1-2 (Columns 3-4) use the first-order contiguity (top-inventor-inflow) weights, where Columns 2 and 4 include the lagged own-state patent productivity. In all cases, the estimated values of $\rho$ are not significantly different from zero at the conventional 5\% level. These results imply that there is neither strategic tax competition between states nor a direct tax response to previous patent productivity within each state. We further apply a test by Montiel Olea and Pflueger (2013) to each specification in Table~\ref{tab:tabD2}. The effective $F$-statistics indicate that in all cases we can reject the null hypothesis of a weak instrument at the conventional level.

We check the robustness of the results using the specification curve analysis as in Simonsohn et al.~(2020). We consider different specifications of equation (\ref{eq:tax_comp}) by focusing on three dimensions. First, we include three other weights discussed in footnote~\ref{fn46} in the specification curve analysis. Second, we estimate the model for every possible combination of the variables listed in Table \ref{tax_comp_vars}. Last, since the model is overidentified, we use different combinations of neighboring states' characteristics as instruments, which allows us to explore the sensitivity of the estimates.\footnote{Since the usual caveat on weak instruments is applicable here, we adopt only specifications for which the null hypothesis of weak instruments is rejected.}

\begin{figure}[tp]
\caption{Specification curve (state tax competition).}
\label{speccurve}
\hskip 4.1cm
\includegraphics[clip, width=0.5\columnwidth]{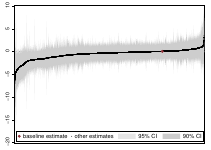}  \\
\footnotesize{\noindent
\emph{Notes:} The specification curve is depicted using $30, 843$ alternative specifications, as explained in \ref{app:taxc}. The vertical axis is the value of $\widehat \rho$.}
\end{figure}

Figure \ref{speccurve} plots the specification curve for $\widehat{\rho}$ using $30,843$ alternative specifications. The figure shows that the great majority of the estimated coefficients are not significantly different from zero at the 5\% level. Hence, we do not find strong evidence of strategic tax competition between states during the sample period.

\renewcommand{\thesection}{Appendix \Alph{section}}
\renewcommand{\thesubsection}{D.\arabic{subsection}}
\renewcommand{\thefigure}{D\arabic{figure}}
\setcounter{figure}{0}
\renewcommand{\thetable}{D\arabic{table}}
\setcounter{table}{0}

\section{Share exogeneity}

\subsection{Two main assumptions to ensure share exogeneity}
\label{app:exo1}

We elaborate on the two main assumptions to ensure the share exogeneity. Recall that we let $X_{dt} = \{\tau_{\sigma(d)t}, \delta_d, \delta_t  \}$ in the main analysis, where $\tau_{\sigma(d)t}$ is the tax rate in state $\sigma (d)$ in year $t$, $\delta_d$ is the destination commuting zone fixed effect, and $\delta_t$ is the year fixed effect. We can then rewrite \eqref{eq:iv} and \eqref{eq:fs} as
\begin{eqnarray}
\ln Y_{dt} &=& \phi^s M_{dt} +\xi^s \tau_{\sigma(d)t} + \delta_d+  \delta_t + \varepsilon_{dt}^s 
\label{eq:iv_rewrite} \\
M_{dt} &=& \textstyle \psi^f B_{dt} +\xi^f \tau_{\sigma(d)t} + \delta_d+  \delta_t +\varepsilon_{dt}^f,
\label{eq:fs_rewrite}
\end{eqnarray}
where $\varepsilon_{dt}^s$ and $\varepsilon_{dt}^f$ are error terms.

We impose the following two assumptions to ensure the share exogeneity, $E(\varepsilon_{dt}^s \widehat P_{odt}| X_{dt})=0$, for the IV regression \eqref{eq:iv_rewrite} and \eqref{eq:fs_rewrite}.
First, $\varepsilon_{dt}^s$ is mean zero conditional on $X_{dt}$, i.e.,
\begin{equation}
{E}(\varepsilon_{dt}^s|\tau_{\sigma(d)t}, \delta_d, \delta_t  ) =0.
\label{eq:CMA}
\end{equation}
This relies on $E(\varepsilon_{dt}^s)=0$ and $\varepsilon_{dt}^s \perp \{ \tau_{\sigma(d)t}, \delta_d, \delta_t  \}$, both of which are assumed to hold.\footnote{${\mathfrak A}\perp {\mathfrak B}$ denotes the independence of ${\mathfrak A}$ and ${\mathfrak B}$.}
Second, $\varepsilon_{dt}^s$ and \textit{other state taxes} $\{\tau_{\sigma(c)t}\}_{c \notin \sigma (d)}$ are independent conditional on $X_{dt}$, i.e.,\footnote{${\mathfrak A}\perp {\mathfrak B}| {\mathfrak C}$ denotes the independence of ${\mathfrak A}$ and ${\mathfrak B}$ conditional on ${\mathfrak C}$.}
\begin{equation}
\varepsilon_{dt}^s \perp \tau_{\sigma(c)t}  | \{\tau_{\sigma(d)t}, \delta_d, \delta_t  \}
\quad
\mbox{for \  $c \notin \sigma (d)$ \  and \  $c ,d  \in {\cal C}$.}
\label{eq:CIA}
\end{equation}
These two assumptions imply the exclusion restriction (see Section~\ref{sec:4.3} for details), so that once we account for the local tax rate, the local commuting zone fixed effect, and the year fixed effect, the local patent productivity shock should be unaffected by the tax policies of other states. This would be satisfied in a situation where local patent productivity primarily responds to its own local factors and conditions.

Under these assumptions, we can show that the share exogeneity, $E(\varepsilon_{dt}^s \widehat P_{odt}| \tau_{\sigma(d)t}, \delta_d, \delta_t) =0$, holds as follows. By the property of conditional expectations, we have
\begin{equation*}
{E}(\varepsilon_{dt}^s \widehat P_{odt}|\tau_{\sigma(d)t}, \delta_d, \delta_t)=
{E}
[
{E}(\varepsilon_{dt}^s \widehat P_{odt} | \{\tau_{\sigma(c) t} \}_{\forall c\in \cal C}, \delta_d, \delta_t )
| \tau_{\sigma(d)t},\delta_d, \delta_t
].
\end{equation*}
The right-hand side of the above equation becomes:
\begin{eqnarray*}
&& E [
E (\varepsilon_{dt}^s \widehat P_{odt} |
\{\tau_{\sigma(c) t} \}_{\forall c\in \cal C},
\delta_d, \delta_t )
| \tau_{\sigma(d)t}, \delta_d, \delta_t ] \\
&=& E [
\widehat P_{odt} 
E (\varepsilon_{dt}^s |
\{\tau_{\sigma(c) t} \}_{\forall c\in \cal C},
\delta_d, \delta_t )
| \tau_{\sigma(d)t}, \delta_d, \delta_t ] \\
&=& E [
\widehat P_{odt} 
E (\varepsilon_{dt}^s | \tau_{\sigma(d)t}, \delta_d, \delta_t,  \{\tau_{\sigma(c)t} \}_{c \notin \sigma (d)}  )
| \tau_{\sigma(d)t}, \delta_d, \delta_t ] \\
&=& E [
\widehat P_{odt} 
E (\varepsilon_{dt}^s | \tau_{\sigma(d)t}, \delta_d, \delta_t  )
| \tau_{\sigma(d)t}, \delta_d, \delta_t ] \\
&=& E [
\widehat P_{odt} \cdot 0
| \tau_{\sigma(d)t}, \delta_d, \delta_t ] 
= 0.
\end{eqnarray*}
The first equality holds because $\widehat{P}_{odt}$ is a function of $\{\tau_{\sigma(c) t} \}_{\forall c\in \cal C}$.\footnote{One may worry that $\widehat{P}_{odt}$ in \eqref{eq:migprob} depends not only on $\{\tau_{\sigma(c) t} \}_{\forall c\in \cal C}$ but also on the set of fixed effects $\{\widehat \gamma_{c}, \widehat \gamma_{oc}\}_{\forall c\in \cal C}$ obtained from the log odds regression in  \eqref{eq:eqm}. We show in Appendix~\ref{app:exo3} that, even in that case, a similar procedure can be used to establish the share exogeneity by imposing an additional assumption.}
The second equality holds since $\{\tau_{\sigma(c) t} \}_{\forall c\in \cal C}=\{ \tau_{\sigma(d)t}, \{\tau_{\sigma(c)t} \}_{c \notin \sigma (d)} \}$. The third equality is due to the conditional independence assumption \eqref{eq:CIA}. \mbox{The last equality comes from the conditional mean assumption~\eqref{eq:CMA}.}

\subsection{Alternative assumptions to ensure share exogeneity}
\label{app:exo_temp}
\vspace{-.1cm}
As discussed in Section~\ref{sec:4.3}, the assumption in \eqref{eq:CMA} may not hold due to a possible correlation between $\varepsilon_{dt}^s$ and $\tau_{\sigma(d)t}$ through unobserved state-specific time-varying factors. To alleviate potential concerns that state taxes may respond to local economic conditions or be correlated with local economic policies affecting innovation, we follow Akcigit et al.~(2022) and employ alternative specifications with state $\times$ year fixed effects. Specifically, we replace
$X_{dt} = \{\tau_{\sigma(d)t}, \delta_d, \delta_t  \}$ with $X_{dt}' = \{\delta_{\sigma(d)t}, \delta_d, \delta_t  \}$ and consider
\begin{eqnarray}
\ln Y_{dt} &=& \phi^s M_{dt} + \delta_{\sigma(d)t} + \delta_d+  \delta_t + \zeta_{dt}^s 
\label{eq:iv_rewrite2} \\
M_{dt} &=& \textstyle \psi^f B_{dt} + \delta_{\sigma(d)t} + \delta_d+  \delta_t +\zeta_{dt}^f,
\label{eq:fs_rewrite2}
\end{eqnarray}
where $\zeta_{dt}^s$ and $\zeta_{dt}^f$ are error terms. We then impose the following two assumptions to obtain the share exogeneity, $E(\zeta_{dt}^s \widehat P_{odt}| X_{dt}')=0$, for the IV regression \eqref{eq:iv_rewrite2} and \eqref{eq:fs_rewrite2}. First, $\zeta_{dt}^s$ is mean zero conditional on $X_{dt}'$, i.e.,
\vspace{-.25cm}
\begin{equation*}
{E}(\zeta_{dt}^s|\delta_{\sigma(d)t}, \delta_d, \delta_t  )  =0.
\vspace{-.1cm}
\end{equation*}
Second, $\zeta_{dt}^s$ and \textit{other state taxes} $\{\tau_{\sigma(c)t}\}_{c \notin \sigma (d)}$ are independent conditional on $X_{dt}'$, i.e.,
\begin{equation*}
\zeta_{dt}^s \perp \tau_{\sigma(c)t}  | \{\delta_{\sigma(d)t}, \delta_d, \delta_t  \}
\quad
\mbox{for \  $c \notin \sigma (d)$ \  and \  $c ,d  \in \cal{C}$.}
\end{equation*}
Under these assumptions, we can show that the share exogeneity, $E(\zeta_{dt}^s \widehat P_{odt}| \delta_{\sigma(d)t}, \delta_d, \delta_t)=0$, holds in the same way as above.

\subsection{An additional assumption to ensure share exogeneity}
\label{app:exo3}
\vspace{-.1cm}
One may worry that $\widehat{P}_{odt}$ in \eqref{eq:migprob} depends not only on $\{\tau_{\sigma(c) t} \}_{\forall c\in \cal C}$ but also on the set of fixed effects $\{\widehat \gamma_{c}, \widehat \gamma_{oc}\}_{\forall c\in \cal C}$ obtained from the log odds regression in \eqref{eq:eqm}. Recall that when estimating \eqref{eq:eqm} we do not simultaneously use $\{\widehat \gamma_{c}, \widehat \gamma_{oc}\}_{\forall c\in \cal C}$ (see footnote~\ref{fn12}). Since we consider $\{ \widehat \gamma_{oc}\}_{\forall c\in \cal C}$ in the baseline specification, we focus on $\{ \widehat \gamma_{oc}\}_{\forall c\in \cal C}$ below.\footnote{The following discussion can be readily modified to accommodate the case with $\{\widehat \gamma_{c}\}_{\forall c\in \cal C}$.}
We now show that, even if we take into account these fixed effects, a similar procedure as in Appendix~\ref{app:exo1} can be used to establish the share exogeneity, which is the exclusion restriction for the exogenous-share approach to shift-share designs as stated in Goldsmith-Pinkham et al.~(2020), by imposing an additional assumption.

We assume that the time-varing error term $\varepsilon_{dt}^s$ is independent of the time-invariant fixed effects $\gamma_{oc}$ that are specific to each pair of commuting zones conditional on $\{\{\tau_{\sigma(c)t}\}_{\forall c \in \mathcal{C}}, \delta_{d}, \delta_{t} \}$:
\vspace{-.6cm}
\begin{equation}
\label{eq:3}
\varepsilon_{dt}^s \perp \gamma_{oc} \ | \{\{\tau_{\sigma(c)t}\}_{\forall c \in \mathcal{C}}, \delta_{d}, \delta_{t} \}
\quad
\mbox{for $o, c ,d  \in \mathcal{C}$}.
\vspace{-.1cm}
\end{equation}
This would be satisfied in a situation where patent productivity shocks are unrelated to factors affecting inventors' migration costs or firms' relocation costs between origin and destination commuting zones, after controlling for time invariant factors specific to the destination commuting zone and for contemporaneous factors common to all commuting zones.

Using this additional assumption, as well as the conditional mean assumption \eqref{eq:CMA} and the conditional independence assumption \eqref{eq:CIA} in Appendix~\ref{app:exo1}, we can show that the share exogeneity,
$E(\varepsilon_{dt}^s \widehat P_{odt}| \tau_{\sigma(d)t}, \delta_{d}, \delta_{t})
=0$, holds, thus implying the exclusion restriction as discussed in Section~\ref{sec:4.3}, as follows.  By the property of conditional expectations, we have
\vspace{-.1cm}
\begin{equation*}
{E}(\varepsilon_{dt}^s \widehat P_{odt}|\tau_{\sigma(d)t}, \delta_{d}, \delta_{t})=
{E}
[
{E}(\varepsilon_{dt}^s \widehat P_{odt} |
\{ \gamma_{oc} \}_{\forall o,c\in \mathcal{C}},
\{\tau_{\sigma(c)t}\}_{\forall c \in \mathcal{C}}, \delta_{d}, \delta_{t} )
| \tau_{\sigma(d)t},\delta_{d}, \delta_{t}
].
\vspace{-.1cm}
\end{equation*}
The right-hand side of the above equation becomes:
\vspace{-.1cm}
\begin{eqnarray*}
&&{E}
[
{E}(\varepsilon_{dt}^s \widehat P_{odt} |
\{ \gamma_{oc} \}_{\forall o,c\in \mathcal{C}},
\{\tau_{\sigma(c)t}\}_{\forall c \in \mathcal{C}}, \delta_{d}, \delta_{t} )
| \tau_{\sigma(d)t},\delta_{d}, \delta_{t}
] \\
&=& {E}
[ \widehat P_{odt}
{E}(\varepsilon_{dt}^s  |
\{ \gamma_{oc} \}_{\forall o,c\in \mathcal{C}},
\{\tau_{\sigma(c)t}\}_{\forall c \in \mathcal{C}}, \delta_{d}, \delta_{t} )
| \tau_{\sigma(d)t},\delta_{d}, \delta_{t}
] \\
&=& {E}
[ \widehat P_{odt}
{E}(\varepsilon_{dt}^s  |
\{\tau_{\sigma(c)t}\}_{\forall c \in \mathcal{C}}, \delta_{d}, \delta_{t} )
| \tau_{\sigma(d)t},\delta_{d}, \delta_{t}
] \\
&=& E [
\widehat P_{odt} 
E (\varepsilon_{dt}^s | \tau_{\sigma(d)t},
\{\tau_{\sigma(c)t} \}_{c \in \mathcal{C}, c \notin \sigma(d)}, \delta_{d}, \delta_{t},   )
| \tau_{\sigma(d)t}, \delta_{d}, \delta_{t} ] \\
&=& E [
\widehat P_{odt} 
E (\varepsilon_{dt}^s | \tau_{\sigma(d)t}, \delta_{d}, \delta_{t}  )
| \tau_{\sigma(d)t}, \delta_{d}, \delta_{t} ] \\
&=& E [
\widehat P_{odt} \cdot 0
| \tau_{\sigma(d)t}, \delta_{d}, \delta_{t} ] 
= 0.
\vspace{-.1cm}
\end{eqnarray*}
The sequence of equalities can be explained  as follows: The first equality holds because $\widehat P_{odt}$ is a function of $\{\gamma_{oc} \}_{\forall o,c\in \mathcal{C}}$ and $\{\tau_{\sigma(c)t}\}_{\forall c \in \mathcal{C}}$. The second equality is derived from assumption \eqref{eq:3}. The last two equalities follow from assumptions \eqref{eq:CIA} and \eqref{eq:CMA}.

\renewcommand{\thesection}{Appendix \Alph{section}}
\renewcommand{\thesubsection}{E.\arabic{subsection}}
\renewcommand{\thefigure}{E\arabic{figure}}
\setcounter{figure}{0}
\renewcommand{\thetable}{E\arabic{table}}
\setcounter{table}{0}

\vspace{-.15cm}
\section{Decomposition of the Bartik estimator}
\label{app:rotem}
\vspace{-.1cm}
To assess the validity of the Bartik estimator, we follow Goldsmith-Pinkham~et~al.~(2020) and decompose it into the weighted sum of just-identified instrumental variable estimators, $\widehat{\phi}^{s} = \sum_{o \in {\cal C}}\widehat{\omega}_{o}\widehat{\phi}^{s}_{o}$, where $\widehat{\omega}_{o}$ and $\widehat{\phi}^{s}_{o}$ are the Rotemberg weight and the origin-specific productivity effect, respectively.

\begin{table}[hbtp]
\caption{Summary of the decomposition of the Bartik estimator.}
\label{tab:tabF1}
\vspace{-.1cm}
\hskip -.35cm
{\footnotesize
\begin{tabular}{lrrrrr}
\hline
\hline
\multicolumn{6}{l}{Panel A: Negative and positive weights}\\
& \multicolumn{1}{c}{sum} & \multicolumn{1}{c}{mean} & \multicolumn{1}{c}{share} & & \\
\cline{2-4}
Negative & -0.0002  & -0.0000  & 0.0002  &  &  \\
Positive & 1.0002  & 0.0099  & 0.9998  &  &  \\
\hline
\multicolumn{6}{l}{Panel B: Correlations}  \\
 & \multicolumn{1}{c}{$\widehat{\omega}_{o}$} & \multicolumn{1}{c}{$\overline {I}_{o}$} & \multicolumn{1}{c}{$\widehat{\phi}_{o}^s$} & \multicolumn{1}{c}{$\widehat{F}_{o}$ \  } & \multicolumn{1}{c}{$\operatorname{var}(\widehat{P}_{o})$ \ \ \ \ \ } \\
\cline{2-6}
$\widehat{\omega}_{o}$ & 1.0000  &  &  &  &  \\
$\overline{I}_{o}$ & 0.9204  & 1.0000  &  &  &  \\
$\widehat{\phi}_{o}^s$ & -0.0478  & -0.0365  & 1.0000  &  &  \\
$\widehat{F}_{o}$ & 0.0372  & -0.0603  & -0.0268  & 1.0000 \ \ \ \ \  &  \\
$\operatorname{var}(\widehat{P}_{o})$ & 0.7865  & 0.7801  & -0.0417  & -0.0845  \ \ \ \ \  & 1.0000 \ \ \ \ \ \  \\
\hline
\multicolumn{6}{l}{Panel C: Top five Rotemberg weight commuting zones}  \\
 & \multicolumn{1}{c}{$\widehat{\omega}_{o}$} & \multicolumn{1}{c}{$\overline{I}_{o}$} & \multicolumn{1}{c}{$\widehat{\phi}_{o}^s$} & \multicolumn{1}{c}{95\% CI} &  \\
\cline{2-5}
19600 \ Bergen-Essex-Middlesex (NJ) & 0.0838  & 186  & 0.0507  & (0.033, 0.097) &  \\
24300 \ Cook-DuPage-Lake (IL) & 0.0535  & 113  & 0.0617  & (0.037, 0.170) &  \\
19400 \ Kings-Queens-New York (NY) & 0.0448  & 63  & 0.0402  & (0.011, 0.065) &  \\
19700 \ Philadelphia-Montgomery-Delaware (PA) & 0.0414  & 68  & 0.0512  & (0.033, 0.170) &  \\
16300 \ Allegheny-Westmoreland-Washington (PA) & 0.0396  & 48  & 0.0358  & (0.018, 0.059) &    \\
\hline
\multicolumn{6}{l}{Panel D: Estimates of $\widehat{\phi}_{o}^s$ for positive and negative weights}  \\
 &\multicolumn{2}{c}{$\widehat{\omega}$-weightd sum} & \multicolumn{2}{c}{share of  overall $\widehat{\phi}^s$} & mean   \\
\cline{2-6}
Negative &\multicolumn{2}{c}{0.0001}   &\multicolumn{2}{c}{0.0023}  & -0.5575 \\
Positive &\multicolumn{2}{c}{0.0417}  & \multicolumn{2}{c}{0.9977}  & 0.0478    \\
\hline
\hline
\multicolumn{6}{p{16.9cm}}{
\textit{Notes:} Panel A presents the sum, mean, and share of the positive and negative Rotemberg weights. Panel B reports correlations between the Rotemberg weights ($\widehat{\omega}_{o}$), the number of top inventors ($\overline{I}_{o} = (1/|{\cal S}|) \sum_{t \in {\cal S}} {I}_{ot}$), the just-identified coefficient estimates ($\widehat{\phi}_{o}^s$), the first stage $F$-statistic of the share ($\widehat{F}_{o}$), and the variance of the shares across destinations and years (${\rm var}(\widehat{P}_{o})$). Panel C reports the origin commuting zones with the top five highest Rotemberg weights. The state of the representative county of each commuting zone is in parentheses, where the representative county is the one with the largest number of inventors. The 95\% confidence interval is the weak instrument robust confidence interval as in Chernozhukhov and Hansen (2008) over a range from 0 to 0.5. In Panel D ``$\widehat{\omega}$-weighted sum'' reports $\sum_{o|\widehat{\omega}_{o}<0}\widehat{\omega}_{o} \widehat{\phi}^{s}_{o}$ for negative and $\sum_{o|\widehat{\omega}_{o} > 0}\widehat{\omega}_{o} \widehat{\phi}^{s}_{o}$ for positive cases, and ``share of overall $\widehat{\phi}^s$'' reports $({1}/{\widehat{\phi}^{s}})\sum_{o|\widehat{\omega}_{o}<0}\widehat{\omega}_{o}\widehat{\phi}^{s}_{o}$ for negative and $({1}/{\widehat{\phi}^{s}})\sum_{o|\widehat{\omega}_{o} > 0} \widehat{\omega}_{o} \widehat{\phi}^{s}_{o}$ for positive cases.}
\end{tabular}}
\vspace{-.3cm}
\end{table}

Table \ref{tab:tabF1} presents the summary statistics.\footnote{Since the decomposition is applicable to a single estimator, we focus on $B_{dt}=\sum_{o\neq d}\widehat{P}_{odt}I_{ot}$.}
Panel~A shows that the Rotemberg weights are positive in almost all cases. Panel~B shows that the weights are highly correlated with the variances of the shares $\operatorname{var}(\widehat{P}_{o})$, where the variances are taken across destination commuting zones $d$ and years $t$. Panel~C reports origin commuting zones with the top five highest Rotemberg weights. These commuting zones account for $26.3$\% of the total share.

Panel D provides the heterogeneous effects interpretation of the Bartik instrument. The Bartik estimator can be rewritten as $\widehat{\phi}^{s} \approx \sum_{d} {\phi}^{s}_{d}\sum_{o} {\omega}_{o} {v}_{od}$, where ${\phi}^{s}_{d}$ is the destination-specific productivity effect and ${v}_{od} \ge 0$ is defined in the same way as in Proposition 4 in Goldsmith-Pinkham et al. (2020). Since negative Rotemberg weights for some origin commuting zones $o$ may make $\sum_{o} {\omega}_{o} {v}_{od}$ negative, the Bartik estimator $\widehat{\phi}^{s}$ may become a \textit{nonconvex} combination. However, since Panel D shows that the positive part $\sum_{o|\widehat{\omega}_{o} > 0}\widehat{\omega}_{o} \widehat{\phi}^{s}_{o}$ is much larger than the negative part  $\sum_{o|\widehat{\omega}_{o}<0}\widehat{\omega}_{o}\widehat{\phi}^{s}_{o}$, the negative Rotemberg weights are unlikely to be a problem for the LATE (local average treatment effect)-like interpretation of the productivity effect.

\begin{table}[t]
\caption{Destinations to which top inventors migrated from the highest-weight origins.}
\label{tab:tabF2}
\vspace{-.1cm}
\hskip -.35cm
{\footnotesize
\begin{tabular}{lcc}
\hline
\hline
 & \multicolumn{1}{c}{\begin{tabular}{c}
 predicted \\ migration \\ probability
 \end{tabular}}
 & \multicolumn{1}{c}{\begin{tabular}{c}
state of the \\ representative \\ county
\end{tabular}} \\
\hline
(A) 19600 \ Bergen-Essex-Middlesex (the highest weight) &  & NJ \\
\quad \quad 38000 \ San Diego & 0.0078 & CA \\
\quad \quad 37500 \ Santa Clara-Monterey-Santa Cruz & 0.0073 & CA \\
\quad \quad 39400 \ King-Pierce-Snohomish & 0.0071 & WA \\
\quad \quad 35801 \ Ada-Canyon-Elmore & 0.0070 & ID \\
\quad \quad 37800 \ Alameda-Contra Costa-San Francisco & 0.0069 & CA \\[2mm]
(B) 24300 \ Cook-DuPage-Lake (the second highest weight)&  & IL \\
\quad \quad \hphantom{1}7100 \ Palm Beach-St. Lucie-Martin & 0.0113 & FL \\
\quad \quad \hphantom{1}9100 \ Fulton-DeKalb-Cobb & 0.0091 & GA \\
\quad \quad 11304 \ Fairfax-Montgomery-Prince George's & 0.0081 & MD \\
\quad \quad 37000 \ Stanislaus-Merced-Tuolumne & 0.0079 & CA \\
\quad \quad 37500 \ Santa Clara-Monterey-Santa Cruz & 0.0079 & CA \\[2mm]
(C) 19400 \ Kings-Queens-New York (the third highest weight) &  & NY \\
\quad \quad 37500 \ Santa Clara-Monterey-Santa Cruz & 0.0118 & CA \\
\quad \quad 18600 \ Albany-Saratoga-Rensselaer & 0.0114 & NY \\
\quad \quad 37800 \ Alameda-Contra Costa-San Francisco & 0.0111 & CA \\
\quad \quad 38801 \ Multnomah-Washington-Clackamas & 0.0100 & OR \\
\quad \quad 39400 \ King-Pierce-Snohomish & 0.0097 & WA \\[2mm]
(D) 19700 \ Philadelphia-Montgomery-Delaware (the fourth highest weight) &  & PA \\
\quad \quad 39400 \ King-Pierce-Snohomish & 0.0151 & WA \\
\quad \quad 28900 \ Denver-Jefferson-Arapahoe & 0.0143 & CO \\
\quad \quad 38000 \ San Diego & 0.0131 & CA \\
\quad \quad 38801 \ Multnomah-Washington-Clackamas & 0.0116 & OR \\
\quad \quad 37800 \ Alameda-Contra Costa-San Francisco & 0.0113 & CA \\[2mm]
(E) 16300 \ Allegheny-Westmoreland-Washington (the fifth highest weight) &  & PA \\
\quad \quad 37500 \ Santa Clara-Monterey-Santa Cruz & 0.0217 & CA \\
\quad \quad 37800 \ Alameda-Contra Costa-San Francisco & 0.0211 & CA \\
\quad \quad 38300 \ Los Angeles-Orange-San Bernardino & 0.0195 & CA \\
\quad \quad 21501 \ Hennepin-Ramsey-Dakota & 0.0180 & MN \\
\quad \quad \hphantom{1}7000 \ Dade-Broward-Monroe & 0.0175 & FL \\
\hline
\hline
\multicolumn{3}{p{17.1cm}}{
\textit{Notes:}
The top five destination commuting zones are defined by the predicted migration probability of the top inventors from each origin commuting zone given in the first column. The second column reports the state of the representative county of each commuting zone, where the representative county is the one with the largest number of inventors.}
\end{tabular}}
\vspace{-.5cm}
\end{table}

We further assess the validity of the identification assumption. For each of the top five \emph{origin} commuting zones in Panel C of Table \ref{tab:tabF1}, Table \ref{tab:tabF2} lists the top five \emph{destination} commuting zones by the predicted top inventor migration probability. The result that origin and destination states differ in almost all cases is line with the assumption that the main source of identifying variation comes from interstate top inventor migrations induced by individual income tax differences across states.

We finally examine the relationship between the shares $\widehat{P}_{odt}$ associated with top five origin commuting zones on the one hand and the location-specific characteristics that may be correlated with the outcome $Y_{dt}$ on the other hand as suggested by Goldsmith-Pinkham~et~al.~(2020). For the shares $\widehat{P}_{odt}$ to satisfy the share exogeneity assumption in Section~\ref{sec:4.3}, they should not be correlated with destination commuting zone characteristics. In our analysis, we use the log of lagged employment in four sectors---``manufacturing,'' ``finance and insurance,'' ``professional, scientific, and technical services,'' and ``management of companies and enterprises''---in commuting zone $d$ as such characteristics.

Table~\ref{tab:tabF3} presents the results of this analysis, where we take the time difference of the variables to control for commuting zone fixed effects. The first panel reports the correlation between the shares related to the highest-weight origins and the log of lagged sectoral employment, which shows that the correlations are low in all cases. The second panel reports the coefficients from regressing the shares related to the highest-weight origins on the lagged sectoral employment while controlling for year fixed effects.
The coefficients are not statistically significant at the conventional 5\% level.\footnote{The absence of significance at the 5\% level does not necessarily imply that the true coefficient is zero. Indeed, in one commuting zone, the null hypothesis is marginally rejected at the 10\% level for some sectors. However, since our analysis involves multiple regressions---each estimating the effect of lagged sectoral employment on the shares related to the highest-weight origin, the rejection of the null hypothesis, if any, might be an artifact of multiple hypothesis testing rather than a genuine effect. To address spurious rejections arising from multiple hypothesis testing, we employ two approaches. First, we report p-values adjusted to control the family-wise error rate (FWER) using the Romano-Wolf multiple hypothesis correction method. Second, we compute the sharpened false discovery rate (FDR) q-values using the code provided by Anderson (2008).}
Hence, the results in Table~\ref{tab:tabF3} suggest that there is no compelling evidence of a significant correlation between the shares related to the highest-weight origins and the log of lagged employment across sectors and commuting zones.

\clearpage
\newpage

\begin{table}[hbtp]
\caption{Relationship between the shares and the lagged employment.}
\label{tab:tabF3}
\hskip -.475cm
{\scriptsize
\begin{tabular}{p{4cm}ccccc}
\hline
\hline
& \multicolumn{5}{c}{share: the predicted probability of top inventors from}\\
 & 19600 & 24300 & 19400 & 19700 & 16300 \\
&\multicolumn{1}{c}{\begin{tabular}{c}
Bergen- \\ Essex- \\  \ \ Middlesex \ \
\end{tabular}} &\multicolumn{1}{c}{\begin{tabular}{c}
Cook- \\ \ \ \ \ DuPage- \  \ \ \ \\Lake
\end{tabular}} &\multicolumn{1}{c}{\begin{tabular}{c}
Kings- \\Queens- \\ \ \  New York \ \ 
\end{tabular}} &\multicolumn{1}{c}{\begin{tabular}{c}
Philadelphia- \\ Montgomery- \\Delaware
\end{tabular}}&\multicolumn{1}{c}{\begin{tabular}{c}
Allegheny- \\ Westmoreland- \\ Washington
\end{tabular}}\\
\hline
    Correlation coefficients & & & & & \\
    \hspace{.5em}  Manufacturing & -0.007549 & 0.003165 & -0.005326 & -0.002585 & -0.006714 \\
     \\
    \hspace{.5em}  Finance and insurance & -0.002242 & -0.005738 & -0.000737 & 0.000201 & -0.008787 \\
     \\
    \hspace{.5em}  Professional, scientific,  & -0.002235 & -0.002048 & -0.001205 & -0.001759 & -0.000779 \\
    \hspace{.5em}  \ \ and technical services &&&&&\\
    \hspace{.5em}  Management of companies  & 0.002552 & -0.001077 & -0.000365 & 0.000917 & -0.001256 \\
    \hspace{.5em}  \ \ and enterprises &&&&&\\ 
    \\
   Regression coefficients & & & & & \\
   \hspace{.5em} Manufacturing & -0.000007 & 0.000012 & -0.000011 & 0.000005 & -0.000017 \\
                                   & (0.000011) & (0.000014) & (0.000012) & (0.000010) & (0.000010) \\
                                   & [0.542358] & [0.401073] & [0.368871] & [0.648052] & [0.094328] \\
                                   & \{0.989000\} & \{0.615400\} & \{0.989000\} & \{1.000000\} & \{0.347700\} \\
                                   & $<$1.000000$>$ & $<$1.000000$>$ & $<$1.000000$>$ & $<$1.000000$>$ & $<$1.000000$>$ \\[2mm]
    \hspace{.5em} Finance and insurance & -0.000013 & -0.000043 & -0.000010 & -0.000001 & -0.000078 \\
                                   & (0.000018) & (0.000031) & (0.000021) & (0.000022) & (0.000042) \\
                                   & [0.460584] & [0.174263] & [0.647973] & [0.963237] & [0.060422] \\
                                   & \{0.965000\} & \{0.615400\} & \{0.989000\} & \{1.000000\} & \{0.200800\} \\
                                   & $<$1.000000$>$ & $<$1.000000$>$ & $<$1.000000$>$ & $<$1.000000$>$ & $<$1.000000$>$ \\[2mm]
    \hspace{.5em} Professional, scientific,  & 0.000000 & -0.000009 & 0.000001 & -0.000001 & 0.000005 \\ 
     \hspace{.5em}   \ \  and  technical services                                  & (0.000007) & (0.000009) & (0.000007) & (0.000007) & (0.000010) \\
                                   & [0.950435] & [0.281902] & [0.882331] & [0.838880] & [0.574786] \\
                                   & \{1.000000\} & \{0.615400\} & \{1.000000\} & \{1.000000\} & \{0.200800\} \\
                                   & $<$1.000000$>$ & $<$1.000000$>$ & $<$1.000000$>$ & $<$1.000000$>$ & $<$1.000000$>$ \\[2mm]
     \hspace{.5em} Management of companies  & 0.000002 & -0.000002 & -0.000000 & 0.000001 & -0.000002 \\
     \hspace{.5em} \ \                     and enterprises           & (0.000002) & (0.000003) & (0.000001) & (0.000002) & (0.000002) \\
                                   & [0.178380] & [0.544422] & [0.872453] & [0.386848] & [0.344007] \\
                                   & \{0.615400\} & \{0.989000\} & \{1.000000\} & \{1.000000\} & \{0.200800\} \\
                                   & $<$1.000000$>$ & $<$1.000000$>$ & $<$1.000000$>$ & $<$1.000000$>$ & $<$1.000000$>$ \\
\hline
\hline
\multicolumn{6}{p{17.2cm}}{
\textit{Notes:}
The first panel reports the correlation coefficient between the shares related to the highest-weight origin and the log of lagged sectoral employment. The second panel reports the coefficients obtained from regressing the shares related to the highest-weight origin on the lagged sectoral employment. The parentheses contain the conventional standard errors, while the square brackets show the p-values. The braces present the family-wise error rate (FWER) adjusted p-values, computed by using the Romano-Wolf multiple hypothesis correction method, and the angle brackets present the sharpened false discovery rate (FDR) q-values, calculated by using the code provided by Anderson (2008).}
\end{tabular}
}
\vspace{-.75cm}
\end{table}

\renewcommand{\thesection}{Appendix \Alph{section}}
\renewcommand{\thesubsection}{F.\arabic{subsection}}
\renewcommand{\thefigure}{F\arabic{figure}}
\setcounter{figure}{0}
\renewcommand{\thetable}{F\arabic{table}}
\setcounter{table}{0}

\section{Robustness to possible violations of the parallel trends assumption}
\label{sec:G}

We examine the robustness to possible violations of the parallel trends assumption for the event study analysis in Section~\ref{sec:sec5}. Following Rambachan and Roth (2023) we specify, for each event study regression, a set $\Upsilon^{\rm SD} (M)=\{\upsilon :  |(\upsilon_{t+1}-\upsilon_{t}) - (\upsilon_{t}-\upsilon_{t-1})| \le M \}$ to bound the degree to which the slope of differential trend $\upsilon$ can vary between consecutive periods. We use the default setting in the R package \texttt{HonestDiD} provided by Rambachan and Roth (2023), namely that the value of $M$ ranges from 0, which corresponds to a linear trend, to half a standard deviation of the parameter of interest.

\begin{figure}[htp]
\caption{Robustness to possible violations of parallel trends assumption.}
\subfigure[all local inventors]
{\includegraphics[clip, width=0.5\columnwidth]{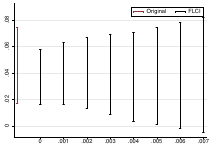}}
\subfigure[external inventors]
{\includegraphics[clip, width=0.5\columnwidth]{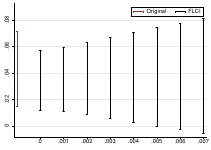}} \\
\footnotesize{\noindent
\emph{Notes:} Panels (a) and (b) illustrate the sensitivity analysis for the event study model in Section~\ref{sec:sec5}, where we consider all local inventors and external inventors, respectively. In each panel, the leftmost bar is the 95\% confidence interval of the estimate $\widehat{\mu}_{1}^{es}$ and the other bars are fixed length confidence intervals (FLCIs) proposed by Rambachan and Roth (2023).}
\label{fig:figG1}
\end{figure}

Figure~\ref{fig:figG1} illustrates the robustness test result, where we evaluate $\Upsilon^{\rm SD}(M)$ at $t=0$ by normalizing $\upsilon_0=0$ to conduct a sensitivity analysis for the first-period effect $\widehat \mu_{1}^{es}$. In Panel (a) (Panel (b)), the ``break-down'' value of $M$, at which the null hypothesis that the first-period effect is zero can no longer be rejected, is $0.006$ ($0.006$). Since the value is $40.00$\% ($42.90$\%) of the standard error of the estimated effect $\widehat{\mu}_{1}^{es}$, the parallel trends assumption holds for a reasonable deviation from a linear trend.\footnote{We examine the parallel trends assumption for the IV event study regression using $B_{dt}$ and $B^{\sigma}_{dt}$ as instruments, which corresponds to the IV ES1 case presented in Figure 5. The standard error of the first-period effect $\widehat{\mu}_{1}^{es}$ for all local inventors (for external inventors) is $0.015$ ($0.014$).}

We further perform a permutation-based placebo analysis to see how likely the ``break-down'' value of $M=0.006$ is to occur. Specifically, we first estimate each event study model 200 times by randomly reshuffling the commuting zones to which top inventors moved and then apply the sensitivity analysis proposed by Rambachan and Roth (2023) mentioned above to the estimates.

Table~\ref{tab:tabG1} shows the summary statistics of the simulated ``break-down'' values of $M$. In both Panels (a) and (b), the values are zero in 95\% of cases. These results indicate that the deviation from the linear trend up to $M=0.006$ occurs extremely rarely. Therefore, we may conclude that the parallel trends assumption is unlikely to be violated.

\medbreak
\begin{table}[h]
\caption{Summary statistics of the ``break-down'' values for the placebo simulations.}
\label{tab:tabG1}
\footnotesize{
\hskip 4cm
\begin{tabular}{lcc}
\hline\hline
&95th percentile
& 99th percentile \\ 
\hline
(a)  all local inventors & 0.0000
& 0.0010 \\
(b)  external inventors  & 0.0000
& 0.0010 \\ 
\hline\hline
\end{tabular}
}
  \begin{tablenotes}[flushleft]
\footnotesize
\item \hskip -.1cm
\emph{Notes:}
Panels (a) and (b) present the distribution of the  ``break-down'' values of $M$ obtained from the permutation-based placebo analysis for the event study model, where we consider all local inventors and external inventors, respectively. The ``break-down'' value is defined as the value of $M$ at which the null hypothesis that the first period effect is zero can no longer be rejected. The summary statistics are for 200 simulation results.
\end{tablenotes}
\vspace{-.4cm}
\end{table}

\renewcommand{\thesubsection}{G.\arabic{subsection}}
\renewcommand{\thefigure}{G\arabic{figure}}
\setcounter{figure}{0}
\renewcommand{\thetable}{G\arabic{table}}
\setcounter{table}{0}

\section{Falsification: The case of top baseball players}
\label{sec:H}

\vspace{-.1cm}
In this section, we first estimate the effect of state income tax differences on top MLB player migrations to quantify how location decisions of high-income individuals other than top inventors are affected by tax incentives. We then verify that the tax-induced migration of top MLB players does not affect local patent productivity. This falsification test allows us to isolate the causal effect of top inventor inflows on local patent productivity from other high-income migrants and mitigate potential threats to internal validity, thereby showing the robustness of our main findings.

\paragraph{Institutional Background.}

MLB players have limited opportunities for team selection. The most significant avenue to team selection for experienced players is free agency. After accruing six years of MLB service time, players become eligible for free agency, which allows them to maximize their market value and pursue optimal financial and career opportunities by negotiating contracts with any team. Free agency serves as a key mechanism for the migration of MLB players. We assume that MLB players are subject to the income tax rate of their home team's state. While this assumption does not fully capture the complexity of players' actual tax environment, it provides a reasonable approximation for examining tax differentials across team locations and their potential influence on players' migrations.\footnote{MLB players are in a complex tax environment, which is primarily determined by two key concepts: duty days and the jock tax. The former represent the total number of work days in a season, including games, practices, and travel. The latter allows states and municipalities to levy taxes on athletes for income earned from performing services within their jurisdictions. Thus, the players potentially owe taxes to multiple locations based on their play and travel schedule throughout the season. However, since the number of home games equals that of away games, the fraction of games played \mbox{at each away team's location is relatively small.}}

\paragraph{Data.}

Our analysis uses data from multiple sources to compile information on MLB players and team attributes. The primary source is the Sean Lahman Baseball Database, from which we extract data on players' positions, ages, salaries, past awards, and team affiliations, as well as on teams' annual performance metrics, home locations, and stadium capacities.\footnote{The Sean Lahman Baseball Database is a collection of baseball statistics covering MLB from 1871 to the present, which is accessible at
\url{https://cran.r-project.org/web/packages/Lahman/index.html}}
For more granular player performance data, we employ wins above replacement (WAR), which is a metric that captures each player's total contribution to his team.\footnote{Neil Paine provides the WAR from 1901, which is accessible at \url{https://github.com/Neil-Paine-1/MLB-WAR-data-historical}.}
To identify free agent declarations, we use the Retrosheet's transactions database, which provides detailed records of trades, contracts, and status changes.\footnote{Tom Ruane provides Retrosheet's transactions database, which covers player, manager, coach, and umpire IDs from 1873 to 2020, at \url{https://retrosheet.org/transactions/index.html}.}

\begin{figure}[h]
    \caption{Actual versus predicted salaries.}
    \label{fig:H1}
\hskip 3.75cm 
        \includegraphics[width=0.55\textwidth]{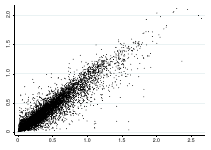} \\
 \footnotesize{\noindent
\emph{Notes:} The vertical and horizontal axes are predicted salaries and actual salaries in 10 million dollars, deflated by CPI using 2000 as the base year.}
\vspace{-1cm}
\end{figure}

While salary data are available from 1985 onwards, they are not exhaustive (since salary data are available for 62.43\% of player-year observations in MLB from 1985 to 2015). To address this limitation, we employ a machine learning approach to predict player salaries for this period. Specifically, we use a random forest algorithm to estimate salaries based on various player characteristics and performance.\footnote{Salaries used in the prediction model are adjusted to 2000 constant dollars to account for inflation. The model includes demographic variables, performance metrics, career milestones, as well as their lagged variables and interaction terms to capture potential non-linear relationships.}
The random forest model achieves strong predictive performance, with R-squared values of 0.975 for the training set and 0.813 for the test set. Figure~\ref{fig:H1} presents a scatter plot of actual versus predicted salaries, showing close alignment along the diagonal.

\vspace{-.05cm}

\clearpage
\newpage

\begin{figure}[h]
\caption{Relationship between top inventor flows, top player flows, and local patent productivity.}
\label{fig:H2}
\subfigure[Player vs. inventor flows]
{\includegraphics[clip, trim = 0 0 0 0, width=0.325\columnwidth]{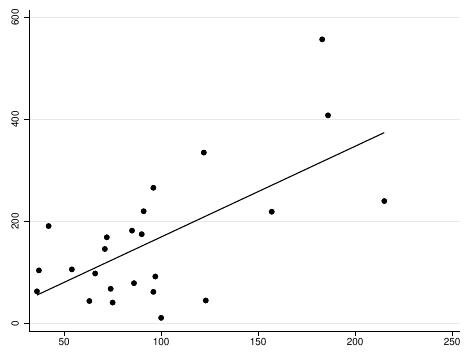}}
\subfigure[Inventor flows vs. log patent]
{\includegraphics[clip, trim = 0 0 0 0, width=0.325\columnwidth]{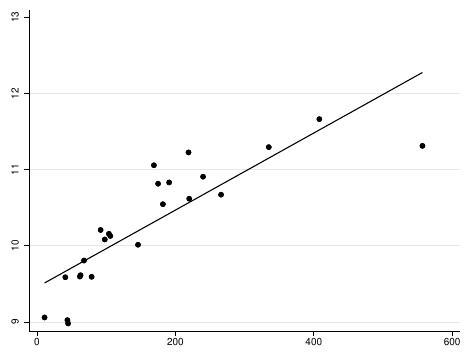}} 
\subfigure[Player flows vs. log patent]
{\includegraphics[clip, trim = 0 0 0 0, width=0.325\columnwidth]{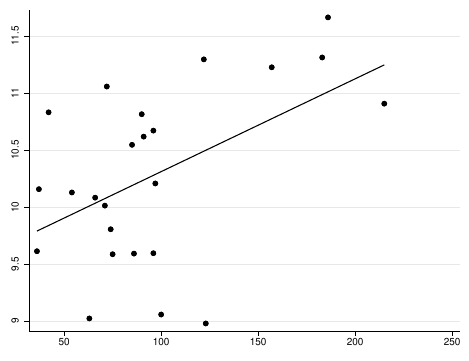}}  \\
\footnotesize{\noindent
\emph{Notes:} The vertical and horizontal axes in Panel (a) are top inventor inflows and top player inflows.
The vertical and horizontal axes in Panel (b) are the log number of patents and top inventor inflows.
The vertical and horizontal axes in Panel (c) are the log number of patents and top player inflows.}
\vspace{-.5cm}
\end{figure}

\paragraph{Summary statistics.}

To maintain consistency with our analysis on top inventor migrations, we restrict ourselves to the period from 1977 to 2009 and the teams in the U.S. by excluding Canadian teams.\footnote{Specifically, we exclude the Montreal Expos and their successor, the Washington Nationals, as well as the Toronto Blue Jays from our analysis.}
The resulting dataset consists of 28 teams located in 24 commuting zones, as well as 6,436 unique ``top players'' defined as players whose (predicted) salaries exceeded the ninety-fifth percentile of the U.S. income distribution in the year they declared free agency after six years of MLB service.\footnote{The data on the U.S. income distribution are obtained from the World Inequality Database: \url{https://wid.world/data/}.}
We examine the team choices of these top players. Of these top players, 42.6\% declared free agency at least once during their careers. To facilitate comparison with the top inventor analysis, we focus on the ATR at the ninety-fifth percentile of the U.S. income distribution. In our sample, 62.54\% of free agents meet the top player criterion. Panel (a) of Figure~\ref{fig:H2} illustrates the relationship between top player flows and top inventor flows to destination commuting zones. Panel (b) (resp., Panel (c)) depicts the relationship between top inventor (resp., player) inflows and local patent productivity in destination commuting zones. The correlation coefficients for Panels (a), (b), and (c) are 0.646, 0.847, and 0.495, respectively.

\begin{table}[h]
\caption{Top 10 commuting zones by top player inflows.}
\label{table:top_czs}
{\footnotesize
\hskip 1cm \begin{tabular}{rrlrcr}
\hline \hline
rank & cz number & counties &  state & team(s) & inflows \\
\hline
    1 & 19400 & Kings--Queens--New York & NY & NYA, NYN & 215 \\
    2 & 38300 & Los Angeles--Orange--San Bernardino & CA & CAL, LAN & 186 \\
    3 & 37800 & Alameda--Contra Costa--San Francisco & CA & OAK, SFN & 183 \\
    4 & 24300 & Cook--DuPage--Lake & IL & CHA, CHN & 157 \\
    5 & 33000 & Tarrant--Johnson--Parker & TX & TEX & 123 \\
    6 & 20500 & Middlesex--Worcester--Essex & MA & BOS & 122 \\
    7 & 29502 & Jackson--Johnson--Wyandotte & KS & KCA & 100 \\
    8 & 15200 & Cuyahoga--Summit--Lake & OH & CLE & 97 \\
    9 & 24701 & St. Louis--St. Clair & MO & SLN & 96 \\
    10 & 38000 & San Diego & CA & SDN & 96 \\
    \hline \hline
    \end{tabular}}
\begin{tablenotes}[flushleft]
\footnotesize
\item \hskip -.1cm \emph{Notes:} Inflows are defined as the number of top players who migrated into each commuting zone from 1977 to 2009. We adopt the abbreviations for the teams in the Sean Lahman Baseball Database, which are: BOS for Boston Red Sox, CAL for California Angels, CHA for Chicago White Sox, CHN for Chicago Cubs, CLE for Cleveland Guardians, KCA for Kansas City Royals, LAN for Los Angeles Dodgers, NYA for New York Yankees, NYN for New York Mets, OAK for Oakland Athletics, SDN for San Diego Padres, SFN for San Francisco Giants, SLN for St. Louis Cardinals, and TEX for Texas Rangers.
\end{tablenotes}    
\end{table}

Table~\ref{table:top_czs} summarizes top 10 commuting zones by top player inflows. Generally, commuting zones attracting more top players also tend to have greater top inventor inflows and more local innovations. However, this correlation does not imply causality between top player inflows and local patent productivity gains. Rather, it may reflect the tendency of larger commuting~zones (such as Kings--Queens--New York and Los Angeles--Orange--San Bernardino) to attract high-income, high-skill individuals in various sectors, including both top inventors and top players.

\paragraph{Counterfactual salaries.}
Following Kleven et al.~(2013), we consider three formulations for counterfactual salaries in the context of professional sports labor markets. We first adopt the ``perfect substitution technology'' assumption, i.e., perfect competition among players implies that pre-tax salary equals ability, i.e.,  $w_{ikt} = X_{it}^a$, where $X_{it}^a$ is the ability vector. In this case, the deterministic part of the utility can be rewritten as:
\vspace{-.1cm}
\begin{equation}
{\small
V_{i j k t} =
\alpha \ln [( 1 - \tau_{\sigma(k) t} ) X_{i t}^a ]
+ \gamma^h {\rm Home}_{i j k t-1}    + \gamma_k^x X_{i t} - \gamma^c C_{jk}  
   + Z_k.  
   }
   \label{eq:first}
   \vspace{-.1cm}
\end{equation}

The second formulation is a variant of the former, given by $w_{i k t} = X_{i t}^a \cdot \overline{w}_{k t}$, where $\overline{w}_{k t}$ is the overall salary level at team $k$ in year $t$ approximated by the estimated average salary from a random forest model. We further allow for the possibility that the coefficients on $\ln ( 1 - \tau_{\sigma(k) t})$, $\ln X_{i t}^a$, and $\ln  \overline{w}_{k t}$ may differ. In this case, $V_{ijkt}$ can be rewritten as:
\vspace{-.1cm}
\begin{equation}
{\small 
V_{i j k t} = \alpha^{t} \ln ( 1 - \tau_{\sigma(k) t} )  + \alpha^w \ln \overline{w}_{k t}
    + \gamma^h {\rm Home}_{i j k t-1}  + \alpha^a \ln X_{i t}^a  + \gamma_k^x X_{i t}        - \gamma^c C_{jk}  + Z_k.
   }
    \label{eq:second}
    \vspace{-.1cm}
\end{equation}

The last specification replaces counterfactual salaries with the random forest estimates $\widehat{w}^{rf}_{i k t}$, resulting in the deterministic utility function as follows:
\vspace{-.1cm}
\begin{equation}
{\small 
        V_{i j k t}= \alpha^t \ln ( 1 - \tau_{\sigma(k) t}
        ) +\alpha^w \ln \widehat{w}^{rf}_{i k t}  + \gamma^h {\rm Home}_{i j k t-1}
        + \gamma_k^x X_{i t} - \gamma^c C_{jk}  
   + Z_k.
        }
         \label{eq:last}
         \vspace{-.1cm}
\end{equation} 

\begin{table}[h]
    \caption{Multinomial logit regression results.}
    \label{tab:mlogit}
{\footnotesize
\hskip -.625cm
\begin{tabular}{lcccccccccc}
\hline\hline
        & (1) & (2) & (3) & (4) & (5) & (6) & (7) & (8) & (9) & (10) \\
        \hline
        $\ln (1-{\rm ATR})$ & 9.975 & 9.959 & 5.485 & 5.489 & 5.489 & 5.489 & 7.903 & 7.894 & 7.894 & 7.894 \\
                       & (2.124) & (2.122) & (2.349) & (2.347) & (2.347) & (2.347) & (2.212) & (2.210) & (2.210) & (2.210) \\
        $\ln ({\rm Salary})$ &        &        & 0.486  & 0.484  & 0.484  & 0.484  & 0.379  & 0.378  & 0.378  & 0.378  \\
                       &        &        & (0.100) & (0.100) & (0.100) & (0.100) & (0.117) & (0.117) & (0.117) & (0.117) \\
        Home (team)    & 1.899  & 1.748  & 1.907  & 1.805  & 1.805  & 1.805  & 1.899  & 1.751  & 1.751  & 1.751  \\
                       & (0.056) & (0.188) & (0.056) & (0.188) & (0.188) & (0.188) & (0.056) & (0.187) & (0.187) & (0.187) \\
       Home (league)  & 0.212  & 0.216  & 0.207  & 0.209  & 0.209  & 0.209  & 0.212  & 0.216  & 0.216  & 0.216  \\
                       & (0.043) & (0.043) & (0.043) & (0.043) & (0.043) & (0.043) & (0.043) & (0.043) & (0.043) & (0.043) \\
        Home (division) & 0.117  & 0.080  & 0.119  & 0.093  & 0.093  & 0.093  & 0.118  & 0.082  & 0.082  & 0.082  \\
                       & (0.045) & (0.052) & (0.045) & (0.053) & (0.053) & (0.053) & (0.045) & (0.052) & (0.052) & (0.052) \\[1mm]
        Player char.
        & Yes & Yes & Yes & Yes & Yes & Yes & Yes & Yes & Yes & Yes \\
        Player perf.  
        & Yes & Yes & Yes & Yes & Yes & Yes & Yes & Yes & Yes & Yes \\
        Team location          & No  & Yes & No  & Yes & Yes & Yes & No  & Yes & Yes & Yes \\[1mm]
        Observations & 95,028 & 95,028 & 94,503 & 94,503 & 94,503 & 94,503 & 95,028 & 95,028 & 95,028 & 95,028 \\
        \hline\hline
        \end{tabular}}
\begin{tablenotes}[flushleft]
\footnotesize
\item \hskip -.1cm \emph{Notes:} 
Coefficients are from multinomial logit regressions of top player $i$'s migration from team $j$ to team $k$ in year $t$. The baseline choice is retirement ($V_{ijkt}=0$ for $k=1$). ATR denotes the individual income average tax rate at the ninety-fifth percentile of the U.S. income distribution. Salary represents top players' pre-tax annual earnings. Columns 1-2 assume the perfect substitution technology in salary determination as in equation \eqref{eq:first}. Columns 3-6 use the estimated average pre-tax salaries for each team as in equation \eqref{eq:second}. Columns 7-10 adopt the counterfactual salary for each player-team-year combination based on the random forest model  as in equation \eqref{eq:last}. Columns 5 and 9 (resp., Columns 6 and 10) allow $\alpha^t$ (resp., $\alpha^w$) to vary across players. Home (team), Home (league), and Home (division) are dummy variables indicating whether players' choices match those of the previous year's team, league, or division. Additional controls include player characteristics (age, experience, their squares, and position) and player performance (log of 3-year WAR), as well as team location variables (same state, same census region, and log distance between team $j's$ and $k's$ locations). Team-specific coefficients $\gamma_{k}^{x}$ for player variables are estimated but not reported due to space constraints. All specifications include team fixed effects $Z_{k}$. Cluster-robust standard errors are in parentheses.
\end{tablenotes}
\vspace{-.5cm}
\end{table}

\paragraph{Estimation results.}

Table~\ref{tab:mlogit} presents the estimation results for $\alpha$, $\alpha^t$, $\alpha^w$, and $\gamma^h$ from the team choice model for top players. Columns 1-2 assume the perfect substitution technology in \eqref{eq:first}. Columns 3-6 use the estimated average pre-tax salaries for each team in \eqref{eq:second}. Columns 7-10 adopt the counterfactual salary for each player-team-year combination based on the random forest model in \eqref{eq:last}. To account for heterogeneity in team selection among top players, we also implement random coefficient specifications. Columns 5 and 9 (resp., Columns 6 and 10) allow $\alpha^t$ (resp., $\alpha^w$) to vary across top players. All models include team fixed effects $Z_k$ to capture unobservable characteristics of each team. The estimation results consistently show a significantly positive $\alpha^w$, indicating that top players tend to choose teams offering higher pre-tax salaries. The positive and significant $\alpha$ or $\alpha^t$ suggests that top players prefer teams in locations with lower income tax rates. Furthermore, the positive and significant $\gamma^h$ confirms the presence of top players' home preferences in team selection as in Kleven et al.~(2013). We use Column 8 of Table~\ref{tab:mlogit} to construct the Bartik instrument for the top player inflows $M_{dt}^{\rm ply}$.

The main IV regression results are reported in Table~\ref{tab:regression_results}. Given that the IV regressions involve two endogenous variables, we employ the Sanderson-Windmeijer tests to assess the relevance of our instruments (Sanderson and Windmeijer, 2016). The test results are presented in Table~\ref{table:sw_tests}. In all specifications, the test statistic for each endogenous variable exceeds 10, which is the commonly used rule-of-thumb value (Staiger and Stock, 1997).

\vspace{-.1cm}
\begin{table}[htbp]
    \caption{First-stage statistics for the falsification analysis.}
    \label{table:sw_tests}
    {\footnotesize
    \hskip 2cm
\begin{tabular}{lrrrrrr}
    \hline \hline 
                            & \multicolumn{1}{c}{(1)}     & \multicolumn{1}{c}{(2)}     & \multicolumn{1}{c}{(3)}     & \multicolumn{1}{c}{(4)}     & \multicolumn{1}{c}{(5)}     & \multicolumn{1}{c}{(6)}     \\
    \hline
    \multicolumn{7}{l}{(a) Sanderson-Windmeijer (under identification)}     \\
    Top inventor inflows     & 23.272  & 59.351  & 62.768  & 27.105  & 53.047  & 56.442  \\
    Top player inflows & 72.869  & 429.338  & 425.385  & 79.685  & 415.145  & 422.672  \\[2mm]
    \multicolumn{7}{l}{(b) Sanderson-Windmeijer (weak instruments)}      \\
    Top inventor inflows     & 23.206  & 29.590  & 20.861  & 25.323  & 24.779  & 17.567  \\
    Top player inflows & 72.663  & 214.052  & 141.378  & 74.447  & 193.918  & 131.552  \\
    \hline \hline
    \end{tabular}
    }
\begin{tablenotes}[flushleft]
\footnotesize
\item \hskip -.1cm \emph{Notes:}
The Sanderson-Windmeijer tests are used for assessing our instruments since the IV regression involves two endogenous variables (Sanderson and Windmeijer, 2016). Panel (a) presents the chi-squared test statistics for under-identification, while Panel (b) presents the $F$-test statistics for weak instruments. Columns~1~and~4 use $B_{dt}$ as an instrument for top inventor inflows $M_{dt}$. Columns 2 and 5 use $B_{dt}$ and $B_{dt}^{\sigma}$ as instruments for $M_{dt}$. Columns 3 and 6 use $B_{dt}$, $B_{dt}^{\sigma}$, and $B_{dt}^{\nu}$ as instruments for $M_{dt}$. In all cases, we use $B_{dt}^{\mathrm{ply}}$ as an instrument for top player inflows $M^{\mathrm{ply}}_{dt}$. In all specifications, the test statistic for each endogenous variable exceeds 10, which is the commonly used rule-of-thumb value (Staiger and Stock, 1997).
\end{tablenotes}
\vspace{-.6cm}
\end{table}

\renewcommand{\thesection}{Appendix \Alph{section}}
\renewcommand{\thesubsection}{H.\arabic{subsection}}
\renewcommand{\thefigure}{H\arabic{figure}}
\setcounter{figure}{0}
\renewcommand{\thetable}{H\arabic{table}}
\setcounter{table}{0}

\section{Shift exogeneity}
\label{sec:I}

Following Borusyak et al.~(2022), we exploit quasi-experimental variation in origin shocks by rewriting the destination-level structural equation \eqref{eq:iv} in terms of an origin-level equation as:
\vspace{-.5cm}
\begin{equation}
\ln \overline{Y}_{ot}^{\perp}=\phi^s \overline{M}_{ot}^{\perp}+\delta^{s} q_{ot} + \overline{\varepsilon}_{ot}^{\perp},
\label{eq:shock}
\vspace{-.1cm}
\end{equation}
where $\overline{v}_{o t}^{\perp}={\sum_{d} P_{o d t} v_{d}^{\perp}}/{\sum_{d} P_{o d t}}$ denotes the share-weighted average of a variable $v_{d t}^{\perp}$  across all destinations $d$ at time $t$, and $v_{dt}^{\perp}$ represents the residual from a sample projection of $v_{dt}$ on the control variable vector $X_{dt}$ in the destination-level equation \eqref{eq:iv}.\footnote{We will modify the control variable vector $X_{dt}$ to analyze the case with incomplete shares below.} The vector $q_{ot}$ includes appropriate origin-level control variables, which we detail below.

To establish shift-share IV consistency as in Borusyak~et~al.~(2022), we impose key assumptions. First, we assume that, conditional on some controls, the shifts are quasi-randomly assigned, i.e., $\mathrm{E}\left(I_{o t} | \bar{\varepsilon}, q, P\right)=\mu q_{o t}$, where $\bar{\varepsilon}=\left\{\bar{\varepsilon}_{o t}\right\}_{o t}, q=\left\{q_{o t}\right\}_{o t}$, and $P=\left\{P_{o d t}\right\}_{o d t}$ are vectors of origin-level shocks, observable commuting zone characteristics, and observable migration shares, respectively. We include origin commuting zone fixed effects and year fixed effects in $q_{ot}$ to control for time-invariant commuting-zone-level confounders and common time trends. Our identification strategy is that, after accounting for these factors, the residual variation in the number of top inventors in origin commuting zone $o$ is as-good-as-randomly assigned over time. We examine the validity of this assumption using Table~\ref{tab:fal} below.

Second, we impose two conditions: $\E\bigl(\sum_{{\rm cluster}}  \bigl( \sum_{o \in {\rm cluster}} \sum_{d \neq o}  {P}_{o d t} \bigr)^2 \bigr) \to 0$ for all $t$; and $\Cov\bigl(\widetilde{I}_{o t}, \widetilde{I}_{o' t}| \overline{\varepsilon}, q, {P} \bigr) = 0$ for all  $t$ and for all  $o$ and $o'$ such that ${\rm cluster} (o) \neq {\rm cluster}(o')$, where ${\rm cluster}$ stands for regions or states and $\widetilde{I}_{o t}=I_{o t}- \mu q_{ot}$ denotes the residual of the number of top inventors after controlling for origin commuting zone fixed effects and year fixed effects. The former implies that the effective sample size, determined by the inverse of the Herfindahl-Hirschman index of the shares, increases asymptotically. The latter ensures that the residuals are mutually uncorrelated between commuting zones such that ${\rm cluster} (o) \neq {\rm cluster}(o')$. We examine the validity of these assumptions using Tables~\ref{tab:shift_sum} and \ref{tab:icc} below.

Since we consider incomplete shares, i.e., $\sum_{o \neq d}P_{odt}\neq 1$ for each destination $d$ and year $t$, we include two additional controls in our destination-level specification: the sum of the shares $\sum_{o \neq d}P_{odt}$; and the exposure-weighted sum of the origin controls $\sum_{o \neq d} P_{odt}q_{ot}$ (see Section~4.2 in Borusyak et al., 2022). Thus, to rewrite the shift-level equation as the origin-level equation, we control for the destination-level characteristics $\{ X_{dt}, \sum_{o \neq d}P_{odt}, \sum_{o \neq d} P_{odt}q_{ot} \}$. This set of controls allows us to address both the incomplete shares and the shift exogeneity.

Table~\ref{tab:shift_sum} presents summary statistics for the shift-related variables. The first panel shows the distribution of the number of top inventors $I_{ot}$ and the distribution of the residual $\widetilde{I}_{ot}$ after controlling for origin commuting zone fixed effects and year fixed effects. Despite the difference in dispersion---the distribution of $I_{ot}$ exhibiting a larger spread than that of $\widetilde{I}_{ot}$---both demonstrate substantial variability.

\begin{table}[thbp]
    \caption{Summary statistics of shift variables.}
    \label{tab:shift_sum}
    {\footnotesize
\hskip 2.3cm 
\begin{tabular}{lccc}
    \hline\hline \\[-4mm]
        statistics & $I_{ot}$ & & $\widetilde{I}_{ot}$ \\
        \hline
        \multicolumn{1}{l}{Shift distribution} \\
        \quad \quad Mean & 47.509  & & \mbox{\White{0}0.000} \\
        \quad \quad Standard deviation  & 95.921 & & 57.757 \\
        \quad \quad Interquartile range & 43.000 & & 37.256 \\ 
        \multicolumn{1}{l}{Effective sample size ($1/{\rm HHI}$ of $P_{ot}$ weights)} \\
         \quad \quad Across CZs and years & &1180.018 &\\
         \quad \quad Across CZs & &\mbox{\White{0}298.514} & \\
        \multicolumn{1}{l}{Largest $P_{ot}$ weight} \\
         \quad \quad Across CZs and years & & 0.001 &\\
         \quad \quad Across CZs & &0.010 & \\
        \multicolumn{1}{l}{Observations} \\
         \quad \quad Number of CZ-year pairs & &{2,859} &\\
          \quad \quad Number of CZs & &{\mbox{\White{0,}391}} &\\
        \hline \hline
        \end{tabular}
        }
\begin{tablenotes}[flushleft]
\footnotesize
\item \hskip -.1cm \emph{Notes:} 
The first panel summarizes the distribution of the number of top inventors $I_{ot}$ and the distribution of the residual $\widetilde{I}_{ot}$ after controlling for origin commuting zone fixed effects and year fixed effects. The second and third panels report the effective sample size, calculated as the inverse of the Herfindahl-Hirschman index of $P_{ot} = \sum_{d \neq o} P_{odt}$ weights, and the largest $P_{ot}$ weight, respectively. The effective sample size and the largest $P_{ot}$ weight are computed for the full panel (across commuting zones and years) and cross-section (across commuting zones). The number of observations is provided for CZ-year pairs and CZs.
\end{tablenotes}
\vspace{-.1cm}
\end{table}

As in Borusyak et al.~(2022), we further examine the importance weight, $P_{ot} = \sum_{d \neq o} P_{odt}$. The second panel of Table~\ref{tab:shift_sum} reports the effective sample size, given by the inverse of the Herfindahl-Hirschman index (HHI) of $P_{ot}$ weights, which is sufficiently large across commuting zones and years. The largest $P_{ot}$ weight is small, which is consistent with the assumption that $\E\bigl(\sum_{{\rm cluster}}  \bigl( \sum_{o \in {\rm cluster}}  \sum_{d \neq o}  {P}_{o d t} \bigr)^2 \bigr) \to 0$. These results support the validity of the large-sample approximation and suggest favorable finite sample performance of the Bartik estimator.

To assess the correlation patterns of shocks and determine appropriate clustering for robust standard errors, we estimate intra-class correlation coefficients (ICCs) within U.S. census regions, states, and commuting zones using the random effects model:
\begin{eqnarray*}
I_{ot} &=& \iota_t + a_{\mathrm{region}(o)t} + b_{\sigma(o)t} + c_o + \epsilon_{ot} \\
\widetilde{I}_{ot} &=& \widetilde{\iota}_t + \widetilde a_{\mathrm{region}(o)t} + \widetilde b_{\sigma(o)t} + \widetilde c_o + \widetilde \epsilon_{ot},
\end{eqnarray*}
where $\iota_t$ is year fixed effects, $a_{\mathrm{region}(o)t}$ and $b_{\sigma(o)t}$ are time-varying regional and state random effects (to which commuting zone $o$ belongs), and $c_o$ is time-invariant random effects that are specific to commuting zone $o$.

\vspace{.2cm}
\begin{table}[H]
\caption{Intra-class correlation coefficients (ICCs).}
\label{tab:icc}
{\footnotesize
\hskip 4.5cm
\begin{tabular}{lcccc}
\hline \hline
        & region & state & CZ & observations \\
\hline
        $I_{ot}$ &  0.046  &  0.065 & 0.403 &  2,813  \\
        & (0.191) & (0.109)   & (0.073) & \\
        & [0.027] & [0.035]   & [0.105] & \\
        $\widetilde I_{ot}$ &  0.063  &  0.090 & 0.000 &  2,715  \\
        & (0.341) & (0.178)   & ($\cdot$)$^\dagger$ & \\
        & [0.052] & [0.074]   & [0.014] & \\
        \hline \hline
\end{tabular}
}
\begin{tablenotes}[flushleft]
\footnotesize
\item \hskip -.1cm \emph{Notes:} The reported intra-class correlation coefficients (ICCs) are estimated from the hierarchal model imposing an identify covariance structure for region random effects and that for state random effects. Robust standard errors are reported in parentheses, whereas bootstrapped standard errors are reported in square brackets. $\dagger$The estimated ICC for CZ lies at the lower bound of the possible range for ICC values, which is zero. Thus, we do not report the robust standard error for this estimate.
\end{tablenotes}
\vspace{-.5cm}
\end{table}

Table~\ref{tab:icc} presents the estimated ICCs, which shows limited evidence for clustering at the regional and state levels. It implies that the number of top inventors, which is the shift of our Bartik instrument, is uncorrelated across broader geographic units such as census regions and states. This aligns with our identifying assumptions based on Borusyak et al.~(2022).

To examine the validity of the assumption that $\E(I_{ot} | \overline{\varepsilon}, q, {P})=\mu q_{ot},$ we further conduct a regression-based falsification test proposed by Borusyak et al.~(2022). The test aim to corroborate the plausibility of the quasi-random assignment of the number of top inventors  across years within each origin commuting zone. For this falsification exercise, we use the lagged employment in four sectors as a potential confounder at the origin level, as we do in examining the share exogeneity at the destination level in \ref{app:rotem}.

Table~\ref{tab:fal} presents the results of an origin-level falsification test, which provides additional support for the validity of our identification strategy.  As in Borusyak et al. (2022), we regress the number of top inventors on the log of lagged sectoral employment, controlling for origin-commuting-zone fixed effects and year fixed effects and weighting by migration shares $P_{ot}$. The coefficients are not statistically significant at the conventional 5\% level.\footnote{Similar to Table~\ref{tab:tabF3}, the absence of significance at the 5\% level does not necessarily imply that the true coefficient is zero. Since our analysis involves multiple regressions, the rejection of the null hypothesis, if any, might be an artifact of multiple hypothesis testing rather than a genuine effect. To address this concern, we report p-values adjusted to control the family-wise error rate (FWER) using the Romano-Wolf multiple hypothesis correction method. We also compute the sharpened false discovery rate (FDR) q-values using the code provided by Anderson (2008).}
This falsification test complements our previous analysis and reinforces the exogeneity of our Bartik instruments.

\begin{table}[h!]
\caption{Regression coefficients.}
\label{tab:fal}
{\footnotesize
\hskip .65cm
    \begin{tabular}{p{4.0cm}cccc} 
    \hline
    \hline
    & \multicolumn{1}{c}{\scriptsize \begin{tabular}{c}
    Manufacturing
    \end{tabular}} & 
    \multicolumn{1}{c}{\scriptsize \begin{tabular}{c}
    Finance and \\ insurance
    \end{tabular}} & 
    \multicolumn{1}{c}{\scriptsize \begin{tabular}{c}
    Professional, \\ scientific, and \\ technical services
    \end{tabular}} & 
    \multicolumn{1}{c}{\scriptsize \begin{tabular}{c}
    Management of \\ companies and \\ enterprises
    \end{tabular}} \\
    \hline
    {\footnotesize Regression coefficients} & $-0.01017$ & $-0.00692$ & 0.00766 & 0.01184 \\
                                     & (0.00555) & (0.00586) & (0.00576) & (0.01135) \\
                                     & [0.06661] & [0.23826] & [0.18380] & [0.29673] \\
                                     & \{0.08890\} & \{0.23480\} & \{0.22880\} & \{0.23480\} \\
                                     & $<$0.36400$>$ & $<$0.36400$>$ & $<$0.36400$>$ & $<$0.36400$>$ \\
    \hline \hline
    \end{tabular}
    }
\begin{tablenotes}[flushleft]
\footnotesize
\item \hskip -.1cm \emph{Notes:} 
This table presents the coefficients obtained from regressing the number of top inventors on the lagged manufacturing employment, controlling for origin commuting zone fixed effects and year fixed effects and weighting by migration shares $P_{ot}$. The parentheses contain the conventional standard errors, while the square brackets show the p-values. The braces present the family-wise error rate (FWER) adjusted p-values, computed by using the Romano-Wolf multiple hypothesis correction method, and the angle brackets present the sharpened false discovery rate (FDR) q-values, calculated by using the code provided by Anderson (2008).
\end{tablenotes}
\vspace{-1cm}
\end{table}

\clearpage
\newpage

\section*{References for online appendix}
\begin{enumerate}
\item[] \hskip -.65cm [1] Anderson, Michael L. (2008).
Multiple Inference and Gender Differences in the Effects of Early Intervention: A Reevaluation of the Abecedarian, Perry Preschool, and Early Training Projects.
\emph{Journal of the American Statistical Association} 103(484): 1481--1495.
 
\item[] \hskip -.65cm [2] 
Chernozhukov, Victor, Christian Hansen (2008).
The Reduced Form: A Simple Approach to Inference with Weak Instruments.
\emph{Economics Letters} 100(1): 68--71.

\item[] \hskip -.65cm [3] 
Eckert, Fabian, Teresa C. Fort, Peter K. Schott,  Natalie J. Yang (2021).
Imputing Missing Values in the US Census Bureau's County Business Patterns.
\emph{NBER Working Paper} \#26632.

\item[] \hskip -.65cm [4] 
Kelejian, Harry H., Ingmar R. Prucha (1998).
A Generalized Spatial Two-Stage Least Squares Procedure for Estimating a Spatial Autoregressive Model with Autoregressive Disturbances.
\emph{Journal of Real Estate Finance and Economics} 17(1): 99--121.

\item[] \hskip -.65cm [5] 
Klarner, Carl (2015).
State Economic and Government Finance Data.
\emph{Harvard Dataverse}, V1.

\item[] \hskip -.65cm [6] 
Monath, Nicholas, Christina Jones, Sarvo Madhavan (2021).
Disambiguating Inventors, Assignees, and Locations.
American Institutes for Research.

\item[] \hskip -.65cm [7] 
Sanderson, Eleanor, Frank Windmeijer (2016). A Weak Instrument $F$-test in Linear IV Models with Multiple Endogenous Variables.
\emph{Journal of Econometrics} 190: 212--221.

\item[] \hskip -.65cm [8] 
Staiger, Douglas, James H. Stock (1997). Instrumental Variables Regression with Weak Instruments.
\emph{Econometrica} 65(3): 557--586.

\item[] \hskip -.65cm [9] 
Toole, Andrew A., Christina Jones,  Sarvothaman Madhavan (2021).
PatentsView: An Open Data Platform to Advance Science and Technology Policy.
\emph{USPTO Economic Working Paper} No. 2021-1.

\item[] \hskip -.85cm [10] 
Trajtenberg, Manuel, Gil Shiff, Ran Melamed (2006).
The ``Names Game'': Harnessing Inventors' Patent Data for Economic Research
\emph{NBER Working Paper} \#12479.

\end{enumerate}

\end{document}